\documentclass[a4paper, 11pt]{article}
\pdfoutput=1

\usepackage{jheppub}
\usepackage{caption}
\usepackage{subcaption}
\usepackage{graphicx}
\usepackage{xcolor}
\usepackage{amsmath,amssymb}
\usepackage{mathtools}

\title{Small $x$ Physics Beyond Eikonal Approximation: an Effective Hamiltonian Approach}
\author{Ming Li}
\affiliation{Department of Physics, The Ohio State University, Columbus, OH 43210, USA}

\emailAdd{li.13449@osu.edu}

\abstract{
Understanding the spin structure of hadrons in the small $x$ regime is an important direction to unravel the spin puzzle in hadronic physics. To include spin degrees of freedom in the small $x$ regime requires going beyond the usual eikonal approximation in high energy QCD. We developed an effective Hamiltonian approach to study spin related observables in the small $x$ regime using the shockwave formalism.  The small-$x$ effective Hamiltonian incorporates both quark and gluon propagators in the background fields and the background field induced interaction vertices up to next-to-eikonal order. A novel feature of sub-eikonal interactions is the background gluon field induced gluon radiation inside the shockwave. Its relation to chromo-electrically polarized Wilson line correlator is established both in small $x$ helicity evolution and in longitudinal double-spin asymmetry for gluon production.  
}

\begin{document}

\maketitle
\flushbottom
\newpage
\section{Introduction}
Understanding the spin structure of proton is one of the central problems in hadronic physics. Since the discovery by the European Muon Collaboration (EMC) \cite{EuropeanMuon:1987isl} showing that quark's intrinsic spin only contributes to a small portion of proton's spin, many experimental and theoretical efforts were devoted to understanding the proton spin puzzle \cite{Aidala:2012mv, Aschenauer:2015eha, AbdulKhalek:2021gbh,  Ji:2020ena}. Theoretical studies \cite{Jaffe:1989jz, Ji:1996ek} point out that besides quark's intrinsic spin, gluon's intrinsic spin (helicity), quark and gluon orbital angular momentum can contribute to proton spin. To study the fraction of proton's spin from gluon, significant advancement was made by the RHIC spin program \cite{Aschenauer:2015eha, RHICSPIN:2023zxx} at the Brookhaven National Laboratory measuring the double-spin asymmetry for particle and jet productions in longitudinally polarized proton-proton collisions. Including some of the experimental measurements into theoretical global analysis for extracting parton distribution functions $f(x, Q^2)$ found that gluons in the range $0.05<x<1$ constitute approximately $40\%$ of the proton's spin at $Q^2=10\,\rm{GeV}$ \cite{deFlorian:2008mr, deFlorian:2014yva, Nocera:2014gqa}. 
Estimating and constraining gluon helicity distribution at even smaller values of $x$ is currently under active theoretical study \cite{Cougoulic:2022gbk} and it is also one of the main goals of the future Electron-Ion Collider experiment \cite{AbdulKhalek:2021gbh}.

The collinear factorization formalism has been the cornerstone to study the double-spin asymmetry in longitudinally polarized proton-proton collisions, dating back to the tree-level partonic cross sections first incorporated in \cite{Babcock:1978yc,Babcock:1978gd}. More recently, global analysis based on generalization to next-to-leading order perturbative QCD contributions within the collinear factorization framework are carried out in \cite{deFlorian:2008mr, deFlorian:2014yva, Nocera:2014gqa}. This approach is particularly applicable when the produced particles and jets have large transverse momentum.  However, inclusive particle and jet productions  with large transverse momentum, especially in the midrapidity, are usually insensitive to  gluons at small $x$ whose typical transverse momentum are the gluon saturation scale $Q_s$ in the saturation regime \cite{Kovchegov:2012mbw,Iancu:2003xm, Albacete:2014fwa}.  To probe gluon helicity at smaller $x$, one needs to include effect of multiple scattering with small $x$ gluons and concentrates on particle/jet productions at moderate values of transverse momentum. Unfortunately, the collinear factorization formalism ceases to be applicable for particle and jet productions with transverse momentum around $Q_s$.  A more general transverse momentum dependent treatment beyond the collinear factorization formalism is desired.  

To faciliate calculating spin related observables in the small $x$ limit directly within the transverse momentum depdendent framework, we develop an effective Hamiltonian approach within the shockwave formalism. This approach is inspired by the seminal work \cite{Bjorken:1970ah}, in which the authors studied high energy QED in external fields.  We derived the small-$x$ effective Hamiltonian that describes high energy QCD processes up to sub-eikonal order. As is well known, leading order QCD processes in the high energy limit (eikonal approximation) are insensitive to spin degrees of freedom.  To probe the spin of quarks and gluons inside the proton, one has to go beyond the eikonal approximation.  We work in the shockwave formalism, treating the proton as background quark and gluon fields. The light-cone Hamiltonian for QCD in the background fields is then expanded in the eikonality parameter $\xi =e^{-\Delta Y}$ with $\Delta Y$ being the rapdity differerence between the projectile and target. The effective light-cone Hamiltonian up to linear order in $\xi$ is sufficient to calculate spin related observables at small $x$. This effective Hamiltonian contains both propagators and effective interaction vertices for quarks and gluons. The quadratic terms in the effective Hamiltonian automatically generate the single quark and the single gluon scattering amplitudes at small $x$, the so-called polarized Wilson lines that have already been obtained in the literature by several groups \cite{Cougoulic:2022gbk, Chirilli:2018kkw, Chirilli:2021lif,  Altinoluk:2020oyd, Altinoluk:2021lvu}.  There are three different interaction vertices in the effective Hamiltonian. At the order $\xi^{1/2}$, one has the background quark field induced quark-gluon conversion. At the order $\xi$, one has the background gluon field induced quark-antiquark-gluon vertex and gluon-gluon-gluon vertex. These three vertices are responsible for the additional complications and new features in spin related observables at small $x$. 

The three-particle interaction vertex induced by the background gluon field predicts that gluon could be emitted inside the shockwave at the sub-eikonal order. It is a new feature compared to the well-known physics at the eikonal order in which gluons are only allowed to be radiated either before or after interacting with the shockwave. This introduces additional contributions when calculating particle productions in polarized collisions and evaluating small $x$ rapidity evolutions of various transverse momentum dependent distribution functions \cite{Balitsky:2015qba, Balitsky:2016dgz}. To determine the significance of this phenomenon, we have performed explicit calculations of the process wherein a soft gluon is emitted inside a shockwave and have derived its contribution to the double-spin asymmetry for soft gluon production. Additionally, we have computed how the emission of gluons inside the shockwave affects the rapidity evolution of polarized Wilson line correlators. In both cases, we found that this effect is manifested in terms of the chromo-electrically polarized Wilson line correlator $\langle \mathrm{Tr}[U^{iG[2]}_{\mathbf{x}}U_{\mathbf{y}}^{\dagger}]\rangle$, which has been shown to be directly related to the small $x$ limit of gluon helicity TMD \cite{Cougoulic:2022gbk}.

The paper is organized as follows. In Sec.~\ref{sec:small_x_H}, the small-$x$ effective Hamiltonian of QCD together with the formalism to calculate scattering processes at the sub-eikonal order are developed. As an application of this formalism, the single quark/gluon scattering amplitudes at the sub-eikonal order are reproduced in Sec.~\ref{sec:single_parton_amplitude}. Sec.~\ref{sec:gluon_radiation_inside_shockwave} is devoted to study the significance of gluon radiation inside the shockwave. Discussions and conclusions are given in Sec.~\ref{sec:conclusion}.

\section{Small-$x$ Effective Hamiltonian}\label{sec:small_x_H}
There are several approaches to study QCD at small $x$. The most widely used approach is to start from the full QCD theory, calculate physical quantities and relevant Feynmann diagrams and finally take the small $x$ limit, typically by setting the center-of-mass collision energy $\sqrt{s}$ to be very large. However, we follow a different approach. Rather than using the complete QCD theory, we initially determine the effective QCD Hamiltonian, which is only applicable in the small $x$ limit. We then utilize this small-$x$ effective Hamiltonian to directly compute interesting physical quantities in the small $x$ limit. To study small $x$ physics, we adopt the shockwave formalism, which treats the target as background quark and gluon fields in the collision with the projectile. This enables us to describe the collision processes using QCD theory in background fields. This approach is the same as high-energy scatterings by external fields, wherein the external fields are highly Lorentz contracted.

\subsection{Light-cone Hamiltonian in the background fields}
Let the QCD Lagrangian density be 
\begin{equation}\label{eq:QCD_lagrangian}
\mathcal{L} = -\frac{1}{4} F_{\mu\nu}^a F^{a,\mu\nu} + \frac{1}{2}\bar{\Psi}i\gamma^{\mu} \overleftrightarrow{D_{\mu}}\Psi  - m\bar{\Psi}\Psi
\end{equation}
with the field strengh tensor  $F^{\mu\nu} = \partial^{\mu}A^{\nu} - \partial^{\nu}A^{\mu} + ig[A^{\mu}, A^{\nu}]$. The covariant derivatives are defined by $\overleftrightarrow{D_{\mu}} = \overrightarrow{D_{\mu}} - \overleftarrow{D_{\mu}}   $ with $\overleftarrow{D_{\mu}} = \overleftarrow{\partial_{\mu}} - ig A_{\mu}$ and $ \overrightarrow{D_{\mu}} = \partial_{\mu} + ig A_{\mu}$. Here $A_{\mu} = A_{\mu}^a t^a$ is defined in the fundamental representation of the $SU(3)$ color group. The fermion mass is denoted by $m$.

In the spirit of shockwave formalism, the nuclear target is characterized as classical gluon and quark fields in the small $x$ limit.
Denoting the background gluon fields and quark fields as $a_{\mu}^a$ and $\psi$ respectively, one makes the substitution 
\begin{equation}
A_{\mu}^e \rightarrow  A_{\mu}^a  + a_{\mu}^a,\qquad 
\Psi \rightarrow  \Psi + \psi
\end{equation}
into eq. \eqref{eq:QCD_lagrangian} to obtain the Lagrangian density in the background fields \cite{Peskin:1995ev}. 
\begin{equation}\label{eq:lagrangian_background}
\begin{split}
\mathcal{L} = &-\frac{1}{4} \mathcal{F}^e_{\mu\nu}\mathcal{F}^{e, \mu\nu}+\frac{1}{2}\bar{\Psi}i\gamma^{\mu} \overleftrightarrow{\mathcal{D}_{\mu}}  \Psi - g\bar{\Psi} A_{\mu} \gamma^{\mu} \Psi - m\bar{\Psi}\Psi\\
&- \frac{1}{2}ig f_{\mu\nu}^e [A^{\mu}, A^{\nu}]^e - g\bar{\Psi} \gamma^{\mu} A_{\mu} \psi- g\bar{\psi} \gamma^{\mu} A_{\mu} \Psi.\\
\end{split}
\end{equation}
The field strength tensor in the background field is defined as $\mathcal{F}_{\mu\nu} = \mathcal{D}_{\mu} A_{\nu} -\mathcal{D}_{\nu} A_{\mu} + ig[A_{\mu}, A_{\nu}]$. The covariant derivatives in the background field are $\overrightarrow{\mathcal{D}_{\mu}} = \partial_{\mu} + ig a_{\mu} $ and $ \overleftarrow{\mathcal{D}_{\mu}} = \overleftarrow{\partial_{\mu}} - ig a_{\mu}$. The background fields are assumed to satisfy classical equations of motion. Here $f_{\mu\nu} = \partial_{\mu} a_{\nu}-\partial_{\nu}a_{\mu} + ig[a_{\mu}, a_{\nu}]$. We use caligraphic letters to indicate expressions in which the ordinary derivative $\partial_{\mu}$ is replaced by covariant derivative $\mathcal{D}_{\mu}$ in the background field $a_{\mu}$ only.

We would like to obtain the corresponding Hamiltonian density from the Lagrangian density in eq. \eqref{eq:lagrangian_background} in the light-cone gauge $A^+=0$. Although the precise dynamics of the background fields themselves are not relevant to the current discussions, we also require that $a^+=0$ \footnote{A more general discussion in which $a^+$ is nonvanishing can be found in \cite{Chirilli:2018kkw}. On the other hand, terms containing $a^+$ are even higher orders in eikonality and will not contribute to effective Hamiltonin up to sub-eikonal order.}.  In the light-cone gauge, the field components $A^-, \Psi_B = \mathcal{P}_B\Psi$ are dependent fields and they can be expressed in terms of the independent fields $A^i, \Psi_G = \mathcal{P}_G\Psi$ \cite{Brodsky:1997de, Kogut:1969xa}. Here the spinor space projection operators are defined as $\mathcal{P}_G = \frac{1}{2}\gamma^-\gamma^+$, $\mathcal{P}_B = \frac{1}{2}\gamma^+\gamma^-$. One has the decomposition of quark field into good component and bad component $\Psi = \Psi_G + \Psi_B$. 

 To calculate the Hamiltonian density, one uses
\begin{equation}\label{eq:H_v1}
\mathcal{H} = \frac{\delta \mathcal{L}}{\delta (\partial_+ A^i) } \partial_+ A^i  + \frac{\delta \mathcal{L}}{\delta (\partial_+ \Psi_G)} \partial_+ \Psi_G + \partial_+ \Psi_G^{\dagger}\frac{\delta \mathcal{L}}{\delta (\partial_+\Psi_G^{\dagger})} - \mathcal{L},
\end{equation}
to obtain the light-cone Hamiltonian in the background fields
\begin{equation}\label{eq:LC_Hamiltonian_background}
\begin{split}
\mathcal{H}  =& \frac{1}{2} \mathcal{F}^{+-}_a\mathcal{F}_a^{+-}+\frac{1}{4} \mathcal{F}^{ij}_a\mathcal{F}_{a, ij} + \frac{1}{2}igf^a_{ij}[A^i, A^j]^a + a^-_b\left(- ig  [ A^{i}, \mathcal{F}^{+i}]_b + g\bar{\Psi}\gamma^+t^b\Psi\right) \\
&+\frac{1}{2}\bar{\Psi}_B i\gamma^- \overleftrightarrow{\partial_-} \Psi_B+ g\bar{\Psi}_G \gamma^i A_i \psi_B + g\bar{\psi}_B \gamma^i A_i \Psi_G .
\end{split}
\end{equation} 
It is supplemented by the constraint equations expressing the dependent fields $A^-, \Psi_B$ as
\begin{equation}\label{eq:auxiliary_psiB}
\Psi_B = \frac{\gamma^+ }{2i\partial_-}\Big[ (-i\gamma^i \mathcal{D}_i + g\gamma^i A_i + m )\Psi_G + g\gamma^i A_i \psi_G\Big]
\end{equation}
and 
\begin{equation}\label{eq:auxiliary_Aminus}
A^- = \frac{-1}{\partial_-} \left( \mathcal{D}_i A^i +\frac{1}{\partial_-}  J^{ +}\right) 
\end{equation}
with the light-cone time component of the color current $J^+ = J^+_0 + J^+_{\rm{int}}$ being
\begin{equation}
\begin{split}
J_0^+ &= -ig[F^{+i}, A_{i}]_b + g\sqrt{2}\Psi^{\dagger}_G t^b \Psi_G , \\
J^{+}_{\rm{int}} = &- 2ig[ f^{+i}, A_{i}]^b+ g\sqrt{2}\Psi^{\dagger}_G t^b \psi_{G}+ g\sqrt{2}\psi^{\dagger}_G  t^b \Psi_{G}.\\
\end{split}
\end{equation}
Here $J^+_0$ is independent of the background fields while $J_{\rm{int}}^+$ explicitly depends on the background fields. The inverse derivative is understood as 
$\frac{1}{\partial_-}\mathcal{F}^a(x^-) = \frac{1}{2} \int_{-\infty}^{+\infty} dz^- \epsilon(x^--z^-)\mathcal{F}^a(z^-)$
assuming antisymmetric boundary condition

Note that in eq. \eqref{eq:LC_Hamiltonian_background} the dependence on $A^-$ is only through $\mathcal{D}_-A^- \equiv \mathcal{F}^{+-}$, the chromoelectric fields. 
The various terms in eq. \eqref{eq:LC_Hamiltonian_background}  have clear physical meanings. The first two terms represent the energy density from chromoelectromagnetic fields in the background fields. The third term characterizes the background gluon fields induced mass term for the dynamical gluon fields. The fourth term is the ususal  coupling of the current $J^+a^-$ with the background fields.  The fifth term characterizes fermions' contribution to the energy density. The last two terms describe the conversion between quarks and gluons induced by background fermion fields.

Plugging eqs. \eqref{eq:auxiliary_psiB} and \eqref{eq:auxiliary_Aminus} into eq. \eqref{eq:LC_Hamiltonian_background}, the light-cone Hamiltonian density $\mathcal{H}= \mathcal{H}_0 +\mathcal{V}_0 + \mathcal{V}_B$ contains the free Hamiltonian density $\mathcal{H}_0$, the vacuum interaction $\mathcal{V}_0$ and the interaction with the background fields $\mathcal{V}_B$. 
\begin{equation}\label{eq:LC_Hamiltonian}
\begin{split}
\mathcal{H}_0 + \mathcal{V}_0 =& -\frac{1}{2} A_a^i \partial_l \partial^l A_i^a  + \left( \partial_i A^i +  \frac{1}{2\partial_-} J^+_0\right) \frac{1}{\partial_-} J_0^+\\
&+  ig[A_i, A_j]_b\partial^i A_b^j+ \frac{1}{4}(ig)^2 [A^i, A^j]_b [A_i, A_j]_b\\
&+\frac{i}{\sqrt{2}}\left(\Phi_B^{\dagger} \partial_-\Phi_B- \partial_-\Phi_B^{\dagger}\Phi_B    \right)\\
\end{split}
\end{equation}
with 
\begin{equation}
\Phi_B = \frac{\gamma^+}{2i\partial_-}  ( -i \gamma^i \partial_i + g\gamma^i A_i +m )\Psi_G.
\end{equation}
Eq.~\eqref{eq:LC_Hamiltonian} is the well-known light-cone Hamiltonian \cite{Brodsky:1997de} without background fields. 
The interaction with background fields has the following expression
\begin{equation}\label{eq:hamiltonian_density_int}
\begin{split}
\mathcal{V}_B= & -\frac{1}{2}A^i_a\Big( (\mathcal{D}_l\mathcal{D}^l -\partial_l\partial^l)^{ac} g_{ij} + 2ig (f_{ij})^{ac}  \Big)A_c^j+ \left( \partial_i A^i + \frac{1}{2\partial_-} J^+_0\right)\frac{1}{\mathcal{\partial}_-}J_{\rm{int}}^+\\
& + \left(ig[a_i, A^i]+  \frac{1}{2\mathcal{\partial}_-} J_{\rm{int}}^+ \right)\frac{1}{\partial_-} \left(J^+_0+J_{\rm{int}}^+ \right)-g^2\left[a^i, A^j\right] \left[A_i, A_j\right] +a^- J^+_0 \\
&+\frac{g}{2\sqrt{2}}\Big\{- \left( \Psi_G^{\dagger}\gamma^i a_i  + \psi_G^{\dagger}\gamma^i A_i\right)\gamma^-\Phi_B- \partial_-\Phi_B^{\dagger} \frac{\gamma^+}{\mathcal{\partial}_-} \left( \gamma^i a_i \Psi_G + \gamma^i A_i\psi_G\right)\\
&\quad+ig\left( \Psi_G^{\dagger}\gamma^i a_i  + \psi_G^{\dagger}\gamma^i A_i\right)\frac{1}{\mathcal{\partial}_-}\left( \gamma^i a_i \Psi_G + \gamma^i A_i\psi_G\right) + h.c.\Big\}\\
&+ g\bar{\Psi}_G \gamma^i A_i \psi_B + g\bar{\psi}_B \gamma^i A_i \Psi_G.
\end{split}
\end{equation}
Our focus lies in studying interactions that occur up to sub-eikonal order in high energy QCD. However, not all interaction terms in eq.~\eqref{eq:hamiltonian_density_int} contribute to sub-eikonal order. Hence, it becomes imperative to identify and isolate the sub-eikonal contributions. To achieve this, we introduce the eikonality parameter and proceed to expand the Hamiltonian as a power series expansion in terms of this parameter in the subsequent section.

\subsection{Expansion in eikonality}
The light-cone Hamiltonian obtained  in the previous section is
\begin{equation}\label{eq:full_hamiltonian}
H = \int d^2\mathbf{x} dx^- \mathcal{H} = H_0 + V = H_0 + V_0 + V_B.
\end{equation}
Recall the definition of $S$-matrix operator
\begin{equation}
\hat{S}\equiv S(+\infty, -\infty)=  \mathcal{P} \mathrm{exp}\left\{ -i\int_{-\infty}^{+\infty} dz^+ V_{\rm{I}}(z^+)\right\} .
\end{equation}
$S$-matrrix element is calculated by $S_{\rm{f}\rm{i}}= \langle \phi_{\rm{f}}|\hat{S}|\phi_{\rm{i}}\rangle$ with $|\phi_{\rm{i}}\rangle$ and $\langle \phi_{\rm{f}} |$ being the eigenstates of free Hamiltonian $H_0$ at asymptotic time $x^+=-\infty$ and $x^+=+\infty$ respectively. The interaction terms of the Hamiltonian in the interaction picture is defined by $V_{\rm{I}}(z^+) = e^{iH_0(z^+-z_0^+)}V(z_0^+) e^{-iH_0 (z^+-z_0^+)}$ with $z_0^+$ the reference time. We further assume the interaction with background fields only happen within the range $[x^+, x_0^+]$. The $S$-matrix operator thus has the factorized form
$\hat{S} = S(+\infty, x^+) S(x^+, x_0^+) S(x_0^+, -\infty) $ in which $V_B$ only contributes to $S(x^+, x_0^+)$.

We are particularly interested in states that have large longitudinal momentum. To obtain these states, we boost the states $|\phi_{\rm{i}}\rangle$ and $ \langle \phi_{\rm{f}} |$. Mathematically, it is implemented by
\begin{equation}
|\phi_{\rm{i}}\rangle_B = e^{-i\omega \hat{K}^3} |\phi_{\rm{i}}\rangle. 
\end{equation}
Here $\hat{K}^3$ is the Lorentz boost operator along the $z$ direction and the parameter $\omega$ characterizes the amount of boost. 
Noted that the boosted states are still eigenstates of the light-cone Hamiltonian because $\hat{H}_0 e^{-i\omega \hat{K}^3} |\phi_{\rm{i}}\rangle = e^{-\omega }e^{-i\omega \hat{K}^3} \hat{H}_0 |\phi_{\rm{i}}\rangle = (e^{-\omega} E_i)  e^{-i\omega \hat{K}^3} |\phi_{\rm{i}}\rangle$
with the help of $e^{i\omega \hat{K}^3} \hat{H}_0 e^{-i\omega \hat{K}^3} = e^{-\omega} \hat{H}_0$ \cite{Kogut:1969xa, Bjorken:1970ah}. 

To calculate $S$-matrix element between highly boosted states, instead of directly boosting the states, it is convenient to shift the boosting to the interactions \cite{Bjorken:1970ah}.  
\begin{equation}\label{eq:Smatrix_boosted_states}
\begin{split}
S_{\rm{f}\rm{i}}=&{}_{B}\langle \phi_{\rm{f}} |  \mathcal{P} \mathrm{exp}\left\{-i\int_{-\infty}^{+\infty} dz^+ V_{\rm{I}}(z^+) \right\} |\phi_{\rm{i}}\rangle_{B} \\
=&\langle \phi_{\rm{f}} |  \mathcal{P}\mathrm{exp}\left\{-i\int_{-\infty}^{+\infty} dz^+ e^{i\omega \hat{K}^3}V_{\rm{I}}(z^+) e^{-i\omega \hat{K}^3}\right\} |\phi_{\rm{i}}\rangle.\\
\end{split}
\end{equation}
The interaction term is transformed by boosting as 
\begin{equation}
e^{i\omega \hat{K}^3}V_{\rm{I}}(z^+) e^{-i\omega \hat{K}^3} =
 e^{iH_0e^{-\omega}(z^+-z_0^+)} \left[e^{i\omega \hat{K}^3}V(z_0^+)  e^{-i\omega \hat{K}^3}\right]e^{-iH_0e^{-\omega}(z^+-z_0^+)}
\end{equation}
To increase the collision energy in a scattering process, one can either boost the projectile or boost the target in the opposite direction. For the interaction with background fields, 
we find it convenient to boost the background fields instead of directly boosting the states.  For that, we will need to reverse the sign of the boost parameter in the above expressions $\omega\rightarrow -\omega$.  We also introduce the rescaled lightcone time $\tilde{x}^+ = e^{\omega} x^+$.  The $S$-matrix element  in eq. \eqref{eq:Smatrix_boosted_states} becomes
\begin{equation}\label{eq:s_matrix_final}
S_{\rm{f}\rm{i}}=\langle \phi_{\rm{f}} |  \mathcal{P}\mathrm{exp}\left\{-i\xi \int_{-\infty}^{+\infty} d\tilde{z}^+  \widetilde{V}_{\rm{I}}(\tilde{z}^+)\right\} |\phi_{\rm{i}}\rangle
\end{equation}
Here the interaction with background fields is first boosted and then transformed into the interaction picture by 
\begin{equation}\label{eq:boost_and_to_interaction_pic}
\begin{split}
&\widetilde{V}(\tilde{z}_0^+)  = e^{-i\omega K^3}V(z_0^+) e^{i\omega K^3},\\
&\widetilde{V}_{\rm{I}}(\tilde{z}^+) = e^{iH_0 (\tilde{z}^+-\tilde{z}_0^+)}\widetilde{V}(\tilde{z}_0^+)  e^{-iH_0(\tilde{z}^+-\tilde{z}_0^+)}. \\
\end{split}
\end{equation}
We have introduced the eikonality parameter $\xi = e^{-\omega}$ in eq.~\eqref{eq:s_matrix_final}.  Identifying $\xi = e^{-\Delta Y}$ with the rapidity difference between the projectile and target $\Delta Y = |Y_P-Y_T|$, the high energy limit $\Delta Y\rightarrow \infty$ corresponds to $\xi \rightarrow 0$.  In the case of deep inelastic scattering in which the Bjorken small-x parameter is defined by $x= \frac{Q^2}{2P\cdot q}$ with  $P^2 = m_N^2$ and $q^2 = -Q^2$, the eikonality parameter is found to be linearly related to the small-x Bjorken parameter $\xi = x e^{-\frac{m_N}{Q}} $. Therefore, the eikonality parameter is nothing but the small-$x$ parameter up to a positive constant multiplicative factor. 

Consequently, calculating the $S$-matrix element in the high energy limit is equivalent to expanding eq.~\eqref{eq:s_matrix_final} as power series expansion in $\xi$.  As we will explicitly demonstrate in the following, the background field boosted interaction term has the expansion
\begin{equation}\label{eq:V_eikonal_expansion}
\xi \widetilde{V}_{\rm{I}}(\tilde{z}^+) = \widetilde{V}_{\rm{I},(0)} (\tilde{z}^+) + \xi^{\frac{1}{2}}\, \widetilde{V}_{\rm{I},(\frac{1}{2})} (\tilde{z}^+)  + \xi \, \widetilde{V}_{\rm{I},(1)} (\tilde{z}^+) +\ldots
\end{equation}
Denoting the leading eikonal interaction operator as
\begin{equation}\label{eq:What_definition}
\hat{W} (\tilde{x}^+, \tilde{x}_0^+) = \mathcal{P}\mathrm{exp}\left\{-i\int_{\tilde{x}_0^+}^{\tilde{x}^+} d\tilde{z}^+  \widetilde{V}_{\rm{I},(0)}(\tilde{z}^+)\right\},  
\end{equation}
one can then expand the S-matrix operator up to first order in $\xi $ from eqs.\eqref{eq:s_matrix_final} and \eqref{eq:V_eikonal_expansion} 
\begin{equation}\label{eq:S_expanded_to_xi}
\begin{split}
&\hat{S}(\tilde{x}^+, \tilde{x}_0^+) =\mathcal{P}\mathrm{exp}\left\{-i\xi \int_{\tilde{x}_0^+}^{\tilde{x}^+} d\tilde{z}^+  \widetilde{V}_{\rm{I}}(\tilde{z}^+)\right\}\\
= &\hat{W} (\tilde{x}^+, \tilde{x}_0^+) -i \int_{\tilde{x}_0^+}^{\tilde{x}^+} d\tilde{w}^+ \hat{W} (\tilde{x}^+, \tilde{w}^+)\Big[ \xi^{\frac{1}{2}}\widetilde{V}_{\rm{I},(\frac{1}{2})} (\tilde{w}^+)  + \xi \widetilde{V}_{\rm{I},(1)} (\tilde{w}^+) \Big] \hat{W} (\tilde{w}^+, \tilde{x}_0^+)\\
& -\int_{\tilde{x}_0^+}^{\tilde{x}^+} d\tilde{w}_2^+ \int_{\tilde{x}_0^+}^{\tilde{w}_2^+} d\tilde{w}_1^+ \hat{W} (\tilde{x}^+, \tilde{w}_2^+) \Big[\xi^{\frac{1}{2}}\widetilde{V}_{\rm{I},(\frac{1}{2})} (\tilde{w}_2^+) \Big] \hat{W} (\tilde{w}_2^+, \tilde{w}_1^+)\Big[\xi^{\frac{1}{2}}\widetilde{V}_{\rm{I},(\frac{1}{2})} (\tilde{w}_1^+)\Big]\hat{W} (\tilde{w}_1^+, \tilde{x}_0^+)\\
&+\mathcal{O}(\xi^{\frac{3}{2}}).
\end{split}
\end{equation}
Eq. \eqref{eq:S_expanded_to_xi} is the main result of the this section. It is the starting point for calculating various scattering amplitudes up to next-to-eikonal order.  It should be pointed out that the Wilson line operator eq.~\eqref{eq:What_definition} contains sub-eikonal contributions due to the transformation to interaction picture given in eq.~\eqref{eq:boost_and_to_interaction_pic}. See detailed discussions in appendix \ref{app:sub-eikonal_transform_a-} in which these sub-eikonal contributions can be equivalently absorbed into $V_{(1)}$. In the following section, the expressions of $V_{(0)}, V_{(\frac{1}{2})}, V_{(1)}$ are derived.

\subsection{Effective light-cone Hamiltonian up to sub-eikonal order}\label{subsec:smallx_Hamiltonian}
The transformations of quark and gluon fields under Lorentz boost are (see appendix \ref{app:boost} and also \cite{Chirilli:2018kkw,Chirilli:2021lif})
\begin{equation}\label{eq:transform_one}
\begin{split}
&a^- \longrightarrow \widetilde{a}^- = e^{\omega}\,  a^-(e^{\omega} x^+, e^{-\omega} x^-, \mathbf{x}), \\
&a^i \longrightarrow \widetilde{a}^i = a^i (e^{\omega} x^+, e^{-\omega} x^-, \mathbf{x}), \\
&\psi_G \longrightarrow \widetilde{\psi}_G = e^{-\omega/2} \psi_G(e^{\omega} x^+, e^{-\omega} x^-, \mathbf{x}),\\
&\psi_B \longrightarrow \widetilde{\psi}_B = e^{\omega/2} \psi_B(e^{\omega} x^+, e^{-\omega} x^-, \mathbf{x}).\\
\end{split}
\end{equation}
The field strength tensor transforms as 
\begin{equation}\label{eq:transform_two}
\begin{split}
&f^{+i}\longrightarrow \widetilde{f}^{+i} = e^{-\omega} f^{+i} (e^{\omega} x^+, e^{-\omega} x^-, \mathbf{x}),\\
&f^{ij}\longrightarrow \widetilde{f}^{ij} = f^{ij} (e^{\omega} x^+, e^{-\omega} x^-, \mathbf{x}).\\
\end{split}
\end{equation}

We study the interaction with background fields given in eq.~\eqref{eq:hamiltonian_density_int} and examine how it transforms under the the transformations \eqref{eq:transform_one} and \eqref{eq:transform_two}. We will perform power series expansion in $\xi =e^{-\omega}$ and keep terms up to zeroth order in $\xi$. Note that we already have a factor $\xi$ in the exponential of eq.~\eqref{eq:s_matrix_final}.

Before analyzing each term, we first look at the factors involving inverse derivative and see how they transform by boosting the background fields. 
\begin{equation}\label{eq:inverse_d_current}
\begin{split}
&\left[\frac{1}{\partial_-}  J_{\rm{int}}^+\right]^a(x^-)
= \frac{1}{2} \int_{-\infty}^{\infty} dz^- \epsilon(x^- -z^-)  J^{+}_{\rm{int}}(z^-)\\
=&\frac{1}{2} \int_{-\infty}^{\infty} dz^- \epsilon(x^- -z^-)  \left(-2ig[f^{+i}, A_i] + g\sqrt{2} \Psi^{\dagger}_G t^c \psi_G + g\sqrt{2} \psi_G^{\dagger} t^c \Psi_G\right)\\
\Longrightarrow &\frac{1}{2} \int_{-\infty}^{\infty} dz^- \epsilon(x^- -z^-) \Big[ e^{-\omega}\Big(  -2ig[f^{+i}(\tilde{x}^+,\tilde{z}^-), A_i] \Big)\\
&\qquad+ e^{-\omega/2}\Big(\sqrt{2} \Psi^{\dagger}_G t^c \psi_G(\tilde{x}^+,\tilde{z}^-) + \sqrt{2} \psi_G^{\dagger} (\tilde{x}^+,\tilde{z}^-)t^c \Psi_G\Big)\Big]\\
\end{split}
\end{equation}
Here $\tilde{x}^+ = e^{\omega}x^+$ and $\tilde{z}^- = e^{-\omega} z^-$.  We use long right arrow to indicate expressions after boosting the background fields. Terms containing this factor eq.~\eqref{eq:inverse_d_current} do not contribute to interactions at sub-eikonal order as they are high powers in $\xi$.  As a result, the second and the third terms in eq.~\eqref{eq:hamiltonian_density_int} will not contribute at the sub-eikonal order except the term $ig[a_i, A^i] \frac{1}{\partial_-} J_0^+$.
 
The other factor containing inverse derivative is,
\begin{equation}
\begin{split}
&\frac{1}{\partial_-}\gamma^+ ( g\gamma^i a_i \Psi_G + g\gamma^i A_i\psi_G)
=\frac{1}{2} \int_{-\infty}^{\infty} dz^- \epsilon(x^- -z^-)  \gamma^+ ( g\gamma^i a_i \Psi_G + g\gamma^i A_i\psi_G)\\
\Longrightarrow & \frac{1}{2} \int_{-\infty}^{\infty} dz^- \epsilon(x^- -z^-)  \Big(  g \gamma^+ \gamma^i a_i(\tilde{x}^+, \tilde{z}^-) \Psi_G +e^{-\omega/2} g\gamma^+\gamma^i A_i \psi_G(\tilde{x}^+, \tilde{z}^-)\Big)\\
=&\frac{1}{2} \int_{-\infty}^{\infty} dz^- \epsilon(x^- -z^-)\Big(g \gamma^+ \gamma^i a_i(\tilde{x}^+, \tilde{z}^-) \Psi_G\Big)  + \mathcal{O}(\xi^{\frac{1}{2}}).
\end{split}
\end{equation}
In the last equality, we only kept the term contributing to interaction at the sub-eikonal order in the end.

We analyze the terms in eq. \eqref{eq:hamiltonian_density_int}. For notational simplicity, we suppress the transverse coordinates, which are not relevant to the analysis of eikonality expansion. The first two lines in eq. \eqref{eq:hamiltonian_density_int}
\begin{equation}
\begin{split}
&\int dx^+ dx^- \left(-\frac{1}{2}A^i_a\Big( (\mathcal{D}_l\mathcal{D}^l -\partial_l\partial^l)^{ac} g_{ij} + 2ig (f_{ij})^{ac}  \Big)A_c^j+ iga^i_b\big( ig[A^j, [A_i, A_j] ]_b+ [ A_i,\frac{1}{\partial_-} J^+_0]_b\big)\right)\\
\Longrightarrow & \int dx^+ dx^- \Big(-\frac{1}{2}A^i_a\Big( (\mathcal{D}_l\mathcal{D}^l -\partial_l\partial^l)^{ac} g_{ij} + 2ig (f_{ij})^{ac} \Big)(\tilde{x}^+, e^{-\omega} x^-) A_c^j\\
&\quad + iga^i_b (\tilde{x}^+, e^{-\omega}x^-)\big( ig[A^j, [A_i, A_j] ]_b+ [ A_i,\frac{1}{\partial_-} J^+_0]_b\big)\Big)\\
=&\xi \int d\tilde{x}^+ dx^- \Big(-\frac{1}{2}A^i_a\Big( (\mathcal{D}_l\mathcal{D}^l -\partial_l\partial^l)^{ac} g_{ij} + 2ig (f_{ij})^{ac} \Big)A_c^j\\
&\quad + iga^i_b\big( ig[A^j, [A_i, A_j] ]_b+ [ A_i,\frac{1}{\partial_-} J^+_0]_b\big)\Big) + \mathcal{O}(\xi^2)\\
\end{split}
\end{equation}
In the last line, we expanded the expression in powers of $\xi$ and only kept the leading order terms. The dynamical gluon fields have arguments $A_i \equiv  A_i(0, x^-)$ while for the background fields $a_i\equiv a_i(\tilde{x}^+, 0)$. 

The usual eikonal interaction term can be obtained by the last term in the second line of eq. \eqref{eq:hamiltonian_density_int}
\begin{equation}\label{eq:expand_arguments_aJ}
\begin{split}
 &\int dx^+ dx^-  a^-(x^+,x^-) J^+_{0}(x^+, x^-) 
 \Longrightarrow\int dx^+ dx^- e^{\omega} a^- (\tilde{x}^+, \tilde{x}^-)  J^+_{0}(x^+, x^-) \\
=&\int d\tilde{x}^+  a^- (\tilde{x}^+,0)\int dx^- J_0^+ (0, x^-) +\xi  \int d\tilde{x}^+  \partial_-a^- (\tilde{x}^+,0)\int dx^- x^- J_0^+ (0, x^-) \\
&+\xi \int d\tilde{x}^+ \tilde{x}^+ a^- (\tilde{x}^+,0)\int dx^- \partial_+ J_0^+ (0, x^-)  + \mathcal{O}(\xi^2).
\end{split}
\end{equation}
 In the last equality, we have performed Taylor expansion in powers of $\xi$. The first term is the well-known eikonal interaction. The other two terms are sub-eikonal interactions containing derivatives with repsect to the background fields and the dynamical fields. The second term characterizes longitudinal momentum exchange between projectile and the shockwave (see appendix~\ref{app:sub-eikonal_transform_a-} for its contribution to single particle scattering amplitude). Such process will not interfere with the eikonal order amplitude which on the other hand preserves longitudinal momentum of the projectile. The third term represents sub-eikonal contributions that will be equivalently included by the sub-eikonal order Wilson line operator transformations demonstrated in appendix~\ref{app:sub-eikonal_transform_a-}. 
 We therefore ignore these two terms in the following discussions.

The three instantaneous terms involving fermion fields in eq. \eqref{eq:hamiltonian_density_int} can be combined together.  The first one is transformed by
\begin{equation}\label{eq:fermion_term_one}
\begin{split}
&\int dx^+ dx^- \Big(- \frac{g}{\sqrt{2}}( \Psi_G^{\dagger}\gamma^i a_i  + \psi_G^{\dagger}\gamma^i A_i)\gamma^-\Phi_B\Big)\\
\Longrightarrow &-\frac{g}{\sqrt{2}}\int dx^+ dx^- \Big( \Psi_G^{\dagger}\gamma^i a_i (\tilde{x}^+, e^{-\omega }x^-) + e^{-\omega/2} \psi_G^{\dagger}(\tilde{x}^+, e^{-\omega }x^-)\gamma^i A_i\Big)\gamma^-\Phi_B(x^+, x^-)\\
= &-\xi \frac{g}{\sqrt{2}}  \int d\tilde{x}^+ a_i^b(\tilde{x}^+, 0 )\int dx^- \Psi_G^{\dagger}\gamma^i t^b \gamma^-\Phi_B(0, x^-)+\mathcal{O}(\xi^{\frac{3}{2}}).\\
\end{split}
\end{equation}
In the last equality, we have ignored the term containing $e^{-\omega/2}$ which contribute to order $\xi^{\frac{3}{2}}$. The next term is transformed as
\begin{equation}\label{eq:fermion_term_two}
\begin{split}
&-\frac{g}{\sqrt{2}} \int dx^+ dx^- \partial_-\Phi_B^{\dagger} \frac{1}{\partial_-} \gamma^+ ( \gamma^i a_i \Psi_G + \gamma^i A_i\psi_G)\\
\Longrightarrow &-\xi\frac{g}{\sqrt{2}}   \int d\tilde{x}^+ a_j^b(\tilde{x}^+, 0)  \int dx^-\partial_-\Phi_B^{\dagger}(0, x^-) \gamma^+\gamma^j  t^b \frac{1}{\partial_-}\Psi_G(0, x^-)+ \mathcal{O}(\xi^{\frac{3}{2}}).\\
\end{split}
\end{equation}
Similarly, the third term is transformed to be
\begin{equation}\label{eq:fermion_term_three}
\begin{split}
&\int dx^+ dx^-\frac{ig^2}{\sqrt{2}}( \Psi_G^{\dagger}\gamma^i a_i  + \psi_G^{\dagger}\gamma^i A_i)\frac{1}{\partial_-} ( g\gamma^i a_i \Psi_G + \gamma^i A_i\psi_G)\\
\Longrightarrow &\xi \frac{ig^2}{\sqrt{2}}\int d\tilde{x}^+ a^b_j(\tilde{x}^+, 0)a_i^c(\tilde{x}^+, 0) \int dx^-  \Psi_G^{\dagger} \gamma^j\gamma^i t^b t^c \frac{1}{\partial_-} \Psi_G+ \mathcal{O}(\xi^{\frac{3}{2}}).
\end{split}
\end{equation}
We need to combine the three expressions in eqs. \eqref{eq:fermion_term_one}, \eqref{eq:fermion_term_two}, \eqref{eq:fermion_term_three}. Keep in mind that these terms are accompanied by their complex conjuagate parts. We express the product of Dirac gamma matrices as 
\begin{equation}
\begin{split}
&\gamma^i\gamma^j = \frac{1}{2} \Big( [\gamma^i ,\gamma^j] + \{\gamma^i, \gamma^j\}\Big) = -2i S^{ij} -\delta^{ij},\\
&\gamma^j\gamma^i = \frac{1}{2} \Big(- [\gamma^i ,\gamma^j] + \{\gamma^i, \gamma^j\}\Big) = 2iS^{ij} -\delta^{ij}.\\
\end{split}
\end{equation}
We have used the generators for Lorentz transformation in spinor space $S^{\mu\nu} = \frac{i}{4} [\gamma^{\mu}, \gamma^{\nu}]$. 
In combining the three expressions, the terms that are quadratic in the fermion field are
\begin{equation}
\begin{split}
&\xi \frac{i}{\sqrt{2}}\int d\tilde{x}^+dx^-d^2\mathbf{x} \Big(a_i^b ( -ig \Psi_G^{\dagger} \gamma^i\gamma^j t^b \frac{1}{\partial_-} \partial_j \Psi_G + ig \partial_j \Psi_G^{\dagger}\gamma^j\gamma^i t^b \frac{1}{\partial_-} \Psi_G)+ a_j^ba_i^c g^2 \Psi_G^{\dagger} \gamma^j\gamma^i t^bt^c \frac{1}{\partial_-}\Psi_G\Big) \\
=&\xi \frac{i}{\sqrt{2}}\int d\tilde{x}^+dx^- d^2\mathbf{x}\Psi_G^{\dagger}  \Big(gf_{ji} S^{ij}  -(\mathcal{D}_l\mathcal{D}^l -\partial_l\partial^l)\Big)\frac{1}{\partial_-} \Psi_G \\
\end{split}
\end{equation}
We have used the identity $f_{ji}^d = \partial_j a_i^d -\partial_i a_j^d + ig (if^{bcd} t^d a_j^ba_i^c)$. and  that 
\begin{equation}
\mathcal{D}_l \mathcal{D}^l \Psi = \partial_l\partial^l \Psi + ig\partial_l a^l \Psi + 2ig a^l \partial_l \Psi + (ig)^2 a_l a^l \Psi.
\end{equation}
Integration by parts for transverse spatial derivatives are used throughout the derivations.
In eqs. \eqref{eq:fermion_term_one} and  \eqref{eq:fermion_term_two}, terms that contain the fermion mass cancel.  In combining the three expressions, the quark-antiquark-gluon interaction vertex is
\begin{equation}
\begin{split}
&\xi \frac{i}{\sqrt{2}} g^2\int d^2\mathbf{x} d\tilde{x}^+ a_i^b \int dx^- \Big(\Psi_G^{\dagger} t^b \gamma^i \frac{1}{\partial_-} (\gamma^j A_j \Psi_G) + \Psi_G^{\dagger} A_j \gamma^j \frac{1}{\partial_-} (\gamma^i t^b \Psi_G)\Big)\\
=&\xi \frac{i}{\sqrt{2}} g^2\int d^2\mathbf{x} d\tilde{x}^+dx^- a_i^b  A_j^c \Psi_G^{\dagger} \gamma^j \gamma^i t^ct^b \frac{1}{\partial_-} \Psi_G + h.c.
\end{split} 
\end{equation} 
Note that integration by parts is used and the boundary term $\int dx^- \partial_- (\frac{1}{\partial_-}\Psi_G^{\dagger} t^b \gamma^i \frac{1}{\partial_-} (\gamma^j A_j \Psi_G))$ is ignored. 

The last two terms in eq. \eqref{eq:hamiltonian_density_int} are transformed as
\begin{equation}
\begin{split}
&\int dx^+ dx^-\Big(g\bar{\Psi}_G \gamma^i A_i \psi_B + g\bar{\psi}_B \gamma^i A_i \Psi_G\Big)\\
\Longrightarrow&\int dx^+ dx^-  e^{\omega/2}\Big(g\bar{\Psi}_G \gamma^i A_i \psi_B(\tilde{x}^+, e^{-\omega}x^-) + g\bar{\psi}_B(\tilde{x}^+, e^{-\omega}x^-)  \gamma^i A_i \Psi_G\Big)\\
=&\xi^{1/2}\int d\tilde{x}^+ dx^- \Big(g\bar{\Psi}_G(0, x^-) \gamma^i A_i(0, x^-) \psi_B(\tilde{x}^+, 0)  + g\bar{\psi}_B(\tilde{x}^+, 0)  \gamma^i A_i(0, x^-) \Psi_G(0, x^-)\Big) + \mathcal{O}(\xi^{\frac{3}{2}})\\
\end{split}
\end{equation}
These two terms have power $\xi^{1/2}$.  It describes background fermion field induced conversion between quarks and gluons.
 
\begin{figure}[!t]
    \centering
    \includegraphics[width=0.6\textwidth]{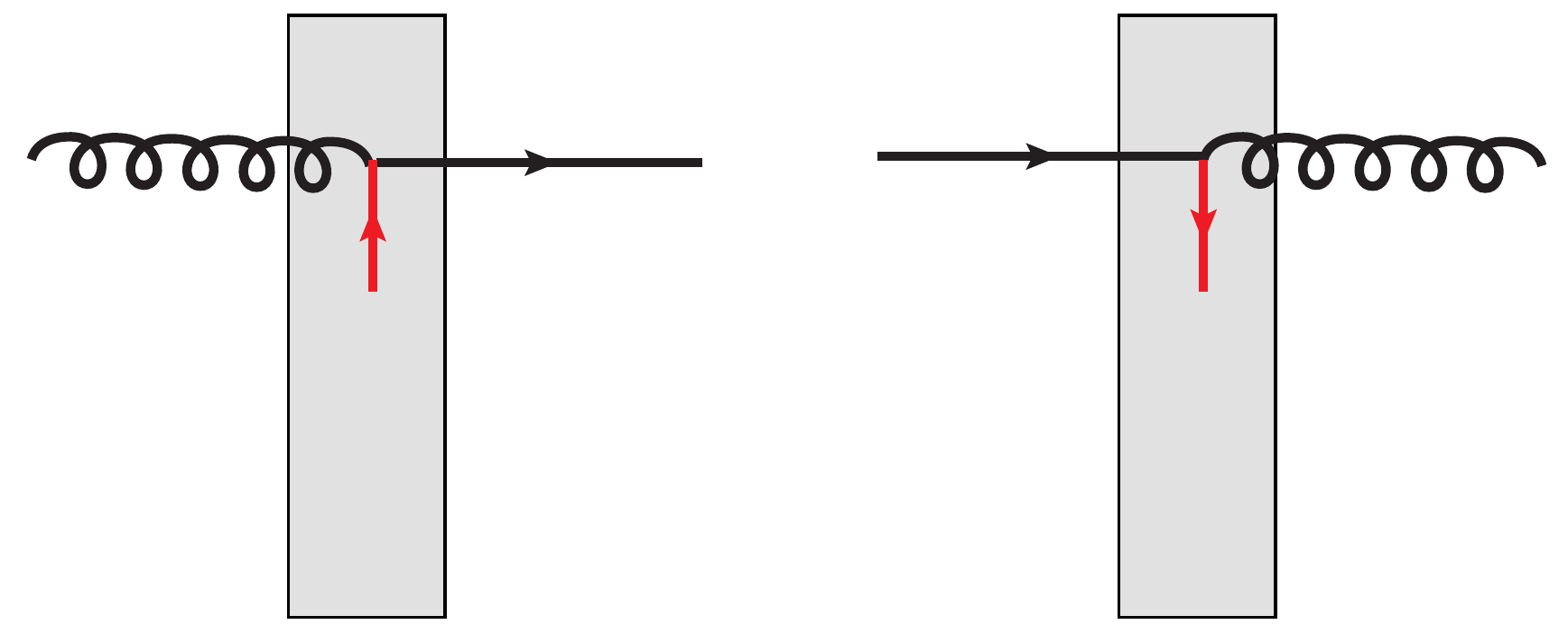}
    \caption{The order-$\xi^{\frac{1}{2}}$ sub-eikonal interaction representing background (anti)quark field induced conversion between quark and gluon. }
	\label{fig:vertex_1}
\end{figure}
Let us summarize the main results in this section.  
The eikonal interaction is
\begin{equation}\label{eq:V_0}
\begin{split}
V_{(0)} = a_b^- J_b^+ =& a_b^- \Big(g\bar{\Psi} \gamma^+ t^b \Psi - ig[A^i, F^{+i}]^b\Big).\\
\end{split}
\end{equation}
The order-$\xi^{\frac{1}{2}}$ sub-eikonal interaction as shown in Fig.~\ref{fig:vertex_1} is
\begin{equation}\label{eq:V_1/2}
\begin{split}
V_{(\frac{1}{2})}  =&g\bar{\Psi}_G \gamma^i A_i \psi_B + g\bar{\psi}_B \gamma^i A_i \Psi_G.\\
\end{split}
\end{equation}
It should be noted that only the bad component $\psi_B$ of the background fermion field is responsible for this sub-eikonal interaction.

\begin{figure}[!t]
    \centering
    \includegraphics[width=0.6\textwidth]{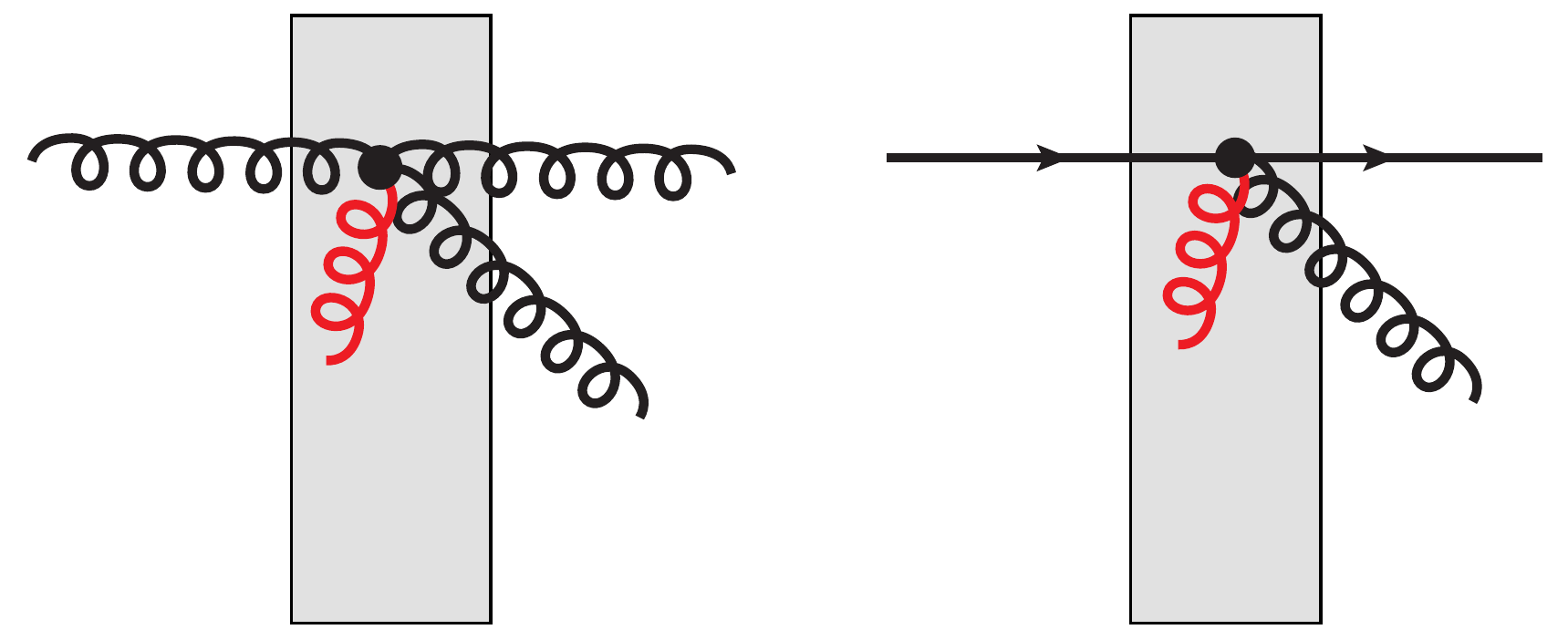}
    \caption{The order-$\xi$ sub-eikonal interaction representing background gluon field induced triple field vertices. }
	\label{fig:vertex_2}
\end{figure}
The order-$\xi$ sub-eikonal interactions due to the background  gluon and  quark fields has the expression
\begin{equation}\label{eq:V_1}
\begin{split}
V_{(1)} =& - \frac{1}{2} A_a^i \Big( (\mathcal{D}_l \mathcal{D}^l)^{ab} g_{ij} + 2ig (f_{ij})^{ab}\Big) A_b^j  + \frac{i}{\sqrt{2}} \Psi_G^{\dagger} \Big( gf_{ji} S^{ij} - \mathcal{D}_l \mathcal{D}^l \Big) \frac{1}{\partial_-} \Psi_G \\
&+ ig\left[A_i, A_j\right]_b (\mathcal{D}^i A^j )_b + (\mathcal{D}_i A^i)_b \frac{1}{\partial_-} \left( -ig\left[\partial_-A^j, A_j\right] + \sqrt{2}g\Psi_G^{\dagger} t^b \Psi_G\right)\\
&+ \frac{1}{\sqrt{2}} g \Psi_G^{\dagger} A_j \gamma^j \gamma^i \mathcal{D}_i \frac{1}{\partial_-}\Psi_G + h.c.\\
\end{split}
\end{equation}
It is interesting to note that at the sub-eikonal level,  the triple vertices, either three-gluon vertices or quark-antiquark-gluon vertices are induced only by the background tranverse gluon fields, see Fig.~\ref{fig:vertex_2}. In eq.~\eqref{eq:V_1}, we have
combined these background field induced triple interaction vertices with the corresponding vacuum triple interaction terms that contain ordinary spatial partial derivatives (given in eq. \eqref{eq:LC_Hamiltonian}). 
These combinations lead to interaction terms that depend on the covariant derivative $\mathcal{D}_i$ rather than simply the background gauge potential $a_i$.  When computing physical observables, gauge covariance becomes apparent with the help of these combinations. 

For the terms quadratic in the dynamical fields in eq.\eqref{eq:V_1}, we have also included the vacuum terms $-\frac{1}{2} A^i\partial_l \partial^l A^i$ and $ -\frac{i}{\sqrt{2}}\Psi_G^{\dagger}\partial_l\partial^l \frac{1}{\partial_-}\Psi_G$. These terms describe sub-eikonal order of the vacuum free propagator although they are independent of the background fields. In appendix~\ref{app:sub-eikonal_transform_a-}, it is shown that these terms are indeed sub-eikonal as they come from sub-eikonal order Wilson line operator transformation. More explanation will be given in the following section when computing single particle scattering amplitude. The upshot is that the dependence on the background gluon field $a_i$ is either through $f_{ij}$ or $\mathcal{D}_i$, maintaining explicit gauge covariance. 

The effective interaction is expressed purely in terms of independent dynamical fields $A^i, \Psi_G$. In the following, we will quantize the theory by substituting the modes expansion of these fields eqs.~\eqref{eq:mode_expansion}, see appendix~\ref{app:LCQ} for details. 
The quadratic terms provide the effective propagators for the gluon and the quark. 
The triple interaction terms represent background field induced three-field interaction.

\section{Single Particle Scattering Amplitude}\label{sec:single_parton_amplitude}
In this section, we calculate the various scattering amplitudes up to sub-eikonal order for single (anti) quark and gluon propagating through the background fields. The formula in eq.~\eqref{eq:S_expanded_to_xi} is our starting point.  Since the tilded coordinates in eq.~\eqref{eq:S_expanded_to_xi} are dummy variables and we have already performed Lorentz boosting on the background fields to obtain the interaction terms up to sub-eikonal order, we therefore ignore the tilde in all symbols in the following discussion for notational simplicity.

\subsection{Single (anti)quark scattering amplitude}
The scattering amplitude for a single quark propagating through the background fields is 
\begin{equation}\label{eq:quark_scattering_amplitude}
\langle q| \hat{S} |q\rangle\equiv M^{q\rightarrow q}( \{p^{\prime +}, \mathbf{x}', m', \sigma'\}; \{p^+, \mathbf{x}, m, \sigma\})
= \big\langle 0 \big | \hat{b}_{m', \sigma'}(p'^+, \mathbf{x}') \, \hat{S}\, \hat{b}^{\dagger}_{m, \sigma}(p^+, \mathbf{x}) \big|0 \big \rangle 
\end{equation}
The incoming quark has color index $m$ and spin index $\sigma$, longitudinal momentum $k^+$ and transverse coordinate $\mathbf{x}$. The corresponding primed quantities characterize the outgoing quark.  The $\hat{b}^{\dagger}$ is quark creation operator.  
Substituting eq.~\eqref{eq:S_expanded_to_xi} into eq.~\eqref{eq:quark_scattering_amplitude}, there are three terms in the eikonality expansion of $\hat{S}$ that contribute up to sub-eikonal order. 
The first term is the eikonal interaction with the background fields
\begin{equation}\label{eq:quark_M_one}
\begin{split}
& \big\langle 0 \big | \hat{b}_{m', \sigma'}(p'^+, \mathbf{x}') \, \hat{W} (x^+, x_0^+)\, \hat{b}^{\dagger}_{m, \sigma}(p^+, \mathbf{x}) \big|0 \big \rangle\\
=&(2\pi) 2p^+\delta(p^+-p'^+) \delta(\mathbf{x}-\mathbf{x}') \delta_{\sigma\sigma'}V^{m'm}_{\mathbf{x}}(x^+, x_0^+) 
\end{split}
\end{equation}
As expected, this is just the eikonal Wilson line in the fundamental representation for quark
\begin{equation}
V_{\mathbf{x}} (x^+, x_0^+) = \mathcal{P} \mathrm{exp} \left\{ -ig\int_{x_0^+}^{x^+} dz^+ a^-_b (z^+, \mathbf{x})  t^b \right\}.
\end{equation}
We have used the transformations $\hat{W}^{\dagger}\hat{b}_j\hat{W} = V_{ji}\hat{b}_i$ and $\hat{W}\hat{b}^{\dagger}_j\hat{W}^{\dagger} = \hat{b}^{\dagger}_iV_{ij}$, valid at eikonal order, see appendix \ref{app:sub-eikonal_transform_a-} for details. 
From eq.~\eqref{eq:S_expanded_to_xi}, the second contributing term is sub-eikonal
\begin{equation}\label{eq:quark_M_two}
\begin{split}
&-i\xi \int_{x_0^+}^{x^+} dw^+ \,  \big\langle 0 \big | \hat{b}_{m', \sigma'}(p'^+, \mathbf{x}') \hat{W} (x^+, w^+) V_{(1),\rm{I}} (w^+) \hat{W} (w^+, x_0^+)\hat{b}^{\dagger}_{m, \sigma}(p^+, \mathbf{x}) \big|0 \big \rangle\\
=&-i\xi\int_{x_0^+}^{x^+} dw^+ \, V^{m'n'}_{\mathbf{x}'} (x^+, w^+) \big\langle 0 \big | \hat{b}_{n', \sigma'}(p'^+, \mathbf{x}')  V_{(1),\rm{I}} (w^+) \hat{b}^{\dagger}_{n, \sigma}(p^+, \mathbf{x}) \big|0 \big \rangle V^{nm}_{\mathbf{x}}(w^+, x_0^+)\\
=& i \xi (2\pi) 2p^+\delta(p^+-p'^+) \delta_{\sigma\sigma'} \frac{1}{2p^+}\int_{x_0^+}^{x^+} dw^+ \, V^{m'n'}_{\mathbf{x}'} (x^+, w^+) \int_{\mathbf{z}} \delta(\mathbf{x}'-\mathbf{z}) \\
&\quad \times \Big[-(2\sigma)gf_{12}(w^+, \mathbf{z})  +\overleftarrow{\mathcal{D}}_l\overrightarrow{\mathcal{D}}^l(w^+, \mathbf{z})  \Big]^{n'n} \, \delta(\mathbf{x} -\mathbf{z}) V^{nm}_{\mathbf{x}}(w^+, x_0^+)
\end{split}
\end{equation}
We have substituted the portion of $V_{(1)}$ that is quadratic in quark fields from eq.~\eqref{eq:V_1} together with the mode expansion for quark field given in eq.~\eqref{eq:mode_expansion}. The transformation to interaction picture introduces higher order contributions in eikonality, so it is safe to set $V_{(1), \rm{I}} = V_{(1)}$ in the above calculations. 

From eq.~\eqref{eq:S_expanded_to_xi}, the third term contributing to quark scattering amplitude is 
\begin{equation}\label{eq:quark_M_three}
\begin{split}
&-\xi \int_{x_0^+}^{x^+} dw_2^+ \int_{x_0^+}^{w_2^+} dw_1^+ \big\langle 0 \big | \hat{b}_{m', \sigma'}(p'^+, \mathbf{x}') \hat{W} (x^+, w_2^+) V_{(\frac{1}{2}),\rm{I}} (w_2^+) \hat{W} (w_2^+, w_1^+)\\
&\qquad \qquad \times  V_{(\frac{1}{2}), \rm{I}} (w_1^+) \hat{W} (w_1^+, x_0^+)\hat{b}^{\dagger}_{m, \sigma}(p^+, \mathbf{x}) \big|0 \big \rangle \\
=&-\xi \int_{x_0^+}^{x^+} dw_2^+ \int_{x_0^+}^{w_2^+} dw_1^+V^{m'n'}_{\mathbf{x}'} (x^+, w_2^+) \big\langle 0 \big | \hat{b}_{n', \sigma'}(p'^+, \mathbf{y}_{\perp})V_{(\frac{1}{2}), \rm{I}} (w_2^+) \hat{W} (w_2^+, w_1^+)\\
&\qquad \qquad \times V_{(\frac{1}{2}), \rm{I}} (w_1^+)\hat{b}^{\dagger}_{n, \sigma}(p^+, \mathbf{x}_{\perp}) \big|0 \big \rangle V^{nm}_{\mathbf{x}}(w_1^+, x_0^+)\\
=&-\xi \frac{1}{2} g^2 (2\pi)\delta(p^+-p'^+)\delta(\mathbf{x}-\mathbf{x}')\int_{x_0^+}^{x^+} dw_2^+ \int_{x_0^+}^{w_2^+} dw_1^+V^{m'n'}_{\mathbf{x}} (x^+, w_2^+) t^{e'}_{n'l'} \psi^{\beta}_{B, l'}(w_2^+, \mathbf{x})\\
&\qquad \times \delta_{\sigma\sigma'} [ \gamma^- + 2\sigma \gamma^-\gamma^5]^{\alpha\beta} U_{\mathbf{x}}^{e'e}(w_2^+, w_1^+)\bar{\psi}_{B,l}^{\alpha}(w_1^+, \mathbf{x}_{\perp})t^e_{ln} V^{nm}_{\mathbf{x}}(w_1^+, x_0^+).\\
\end{split}
\end{equation}
We substituted the expression of $V_{(\frac{1}{2})}$ from eq.~\eqref{eq:V_1/2} together with the mode expansions for quark and gluon fields from eq.~\eqref{eq:mode_expansion}.  We have used the identity $\sum_{\lambda} \varepsilon^{i\ast}_{\lambda} \varepsilon^{i'}_{\lambda} = \delta^{ii'}$ and the eikonal transformation on the gluon creation operator $\hat{W}\hat{a}^{\dagger}_e \hat{W}^{\dagger} = \hat{a}^{\dagger}_{h} U^{he}$. 
Spinor space matrix identity (here $\alpha, \beta$ are indices in the spinor space with $\alpha$ the column index and $\beta$ the row index)
\begin{equation}
\Big[\gamma_i u_{G, \sigma}(p^+)\Big]^{\alpha}\Big[\bar{u}_{G, \sigma'}(p^+)\gamma_{i}\Big]^{\beta} =  p^+ \delta_{\sigma\sigma'} [ \gamma^- + 2\sigma \gamma^-\gamma^5]^{\alpha\beta}
\end{equation}
is also needed. 

Putting together the three terms from eqs.~\eqref{eq:quark_M_one}, \eqref{eq:quark_M_two} and \eqref{eq:quark_M_three},  the single quark scattering amplitude up to sub-eikonal order is 
\begin{equation}\label{eq:M_q_to_q}
\begin{split}
&M^{q\rightarrow q}(\{ p^{\prime +}, \mathbf{x}', m', \sigma'\}; \{p^+, \mathbf{x}, m, \sigma\})\\
=&(2\pi) 2p^+\delta(p^+-p^{\prime +}) \delta_{\sigma\sigma'} \Big[ \delta(\mathbf{x}-\mathbf{x}') V^{m'm}_{\mathbf{x}} + \xi \delta(\mathbf{x}-\mathbf{x}') 2\sigma \, V^{\rm{pol}[1]}_{\mathbf{x}}(p^+) + \xi V^{\rm{pol}[2]}_{\mathbf{x}', \mathbf{x}}(p^+)\Big]
\end{split}
\end{equation}
The polarized Wilson lines of type-one $ V^{\rm{pol}[1]}_{\mathbf{x}}(p^+)$ can be decomposed into $V^{\rm{pol}[1]}_{\mathbf{x}}(p^+) = V^{\rm{q}[1]}_{\mathbf{x}}(p^+) + V^{\rm{G}[1]}_{\mathbf{x}}(p^+)$  indicating whether the depedence is on the background quark field or the background gluon field. Their expressions are 
\begin{equation}\label{eq:polarized_wilson_V_one}
\begin{split}
V^{\rm{q}[1]}_{\mathbf{x}}(p^+) =&-  g^2 \frac{1}{2p^+}\int_{x_0^+}^{x^+} dw_2^+ \int_{x_0^+}^{w_2^+} dw_1^+V_{\mathbf{x}} (x^+, w_2^+) t^{e'} \psi^{\beta}_{B}(w_2^+,\mathbf{x})\left[ \frac{\gamma^-\gamma^5}{2}\right]^{\alpha\beta}\\
&\qquad \times  U_{\mathbf{x}}^{e'e}(w_2^+, w_1^+)\bar{\psi}_{B}^{\alpha}(w_1^+, \mathbf{x})t^e V_{\mathbf{x}}(w_1^+, x_0^+),\\
V^{\rm{G}[1]}_{\mathbf{x}}(p^+)=& -ig\frac{1}{2p^+}\int_{x_0^+}^{x^+} dw^+ \, V_{\mathbf{x}} (x^+, w^+)f_{12}(w^+, \mathbf{x})V_{\mathbf{x}}(w^+, x_0^+).\\
\end{split}
\end{equation}
The polarized Wilson lines of type-two $ V^{\rm{pol}[2]}_{\mathbf{x}', \mathbf{x}}(p^+)$ can also be decomposed into $V^{\rm{pol}[2]}_{\mathbf{x}', \mathbf{x}}(p^+) = \delta(\mathbf{x}-\mathbf{x}')V^{\rm{q}[2]}_{\mathbf{x}}(p^+) + V^{\rm{G}[2]}_{\mathbf{x}', \mathbf{x}}(p^+)$. Their explicit expressions are
\begin{equation}\label{eq:polarized_wilson_V_two}
\begin{split}
V^{\rm{q}[2]}_{\mathbf{x}}(p^+) =&-  \frac{g^2}{2p^+}\int_{x_0^+}^{x^+} dw_2^+ \int_{x_0^+}^{w_2^+} dw_1^+V_{\mathbf{x}} (x^+, w_2^+) t^{e'} \psi^{\beta}_{B}(w_2^+, \mathbf{x})\left[\frac{\gamma^-}{2}\right]^{\alpha\beta}   \\
&\qquad\qquad  \times U_{\mathbf{x}}^{e'e}(w_2^+, w_1^+)\bar{\psi}_{B}^{\alpha}(w_1^+, \mathbf{x})t^eV_{\mathbf{x}}(w_1^+, x_0^+).\\
V^{\rm{G}[2]}_{\mathbf{x}', \mathbf{x}}(p^+)
=&i \frac{1}{2p^+}\int_{x_0^+}^{x^+} dw^+ \, V_{\mathbf{x}'}^{m'n'} (x^+, w^+) \int_{\mathbf{z}} \delta(\mathbf{x}'-\mathbf{z}) \Big[\overleftarrow{\mathcal{D}}_l\overrightarrow{\mathcal{D}}^l (w^+, \mathbf{z})\Big]_{n'n} \, \delta(\mathbf{x}-\mathbf{z}) V^{nm}_{\mathbf{x}}(w^+, x_0^+).\\
\end{split}
\end{equation}
Eqs. \eqref{eq:polarized_wilson_V_one} and \eqref{eq:polarized_wilson_V_two} reproduce the polarized Wilson lines obtained in \cite{Cougoulic:2022gbk}. 

When the background fields are turned off by setting $\psi_B=0, a_i =a^-=0$, the single quark scattering amplitude as given in eq. \eqref{eq:M_q_to_q} does not vanish because of nonvanishing $V^{\rm{G}[2]}_{\mathbf{x}', \mathbf{x}}$.
\begin{equation}\label{eq:free_sub_eikonal}
V^{\rm{G}[2]}_{\mathbf{x}', \mathbf{x}}(p^+)
=i \frac{-\partial^2_{\mathbf{x}}}{2p^+} \delta(\mathbf{x}-\mathbf{x}') \left[x^+-x_0^+\right].
\end{equation}
It comes from the sub-eikonal order correction of free quark propagator
\begin{equation}
\int_{-\infty}^{\infty} \frac{dp^-}{2\pi} e^{ip^-(x^+-x_0^+)} \frac{i}{2p^+p^- -\mathbf{p}^2+i\epsilon} =\frac{1}{2p^+} e^{i\frac{\mathbf{p}^2}{2p^+}(x^+-x_0^+)}.
\end{equation}
Expanding the phase factor to linear order, one gets eq. \eqref{eq:free_sub_eikonal}. When computing cross section by squaring scattering amplitudes, these vacuum contributions should be subtracted. The lesson is that there are two sources of sub-eikonal physics. One is dynamical, genuinely related to the interactions with background fields  at the sub-eikonal order. The other is kinematic, which is just the sub-eikonal order expansion of the free propagator phase, having nothing to do with the background fields. In principle, one should replace $\overleftarrow{\mathcal{D}}_l\overrightarrow{\mathcal{D}}^l$ by $(\overleftarrow{\mathcal{D}}_l\overrightarrow{\mathcal{D}}^l - \overleftarrow{\partial}_l\overrightarrow{\partial}^l)$ for the interaction terms in eq.~\eqref{eq:V_1}. However, retaining the covariant derivatives $\overleftarrow{\mathcal{D}}_l\overrightarrow{\mathcal{D}}^l$ automatically keeps track of sub-eikonal contributions from free propagators. In appendix~\ref{app:sub-eikonal_transform_a-}, it is shown that the $\overleftarrow{\partial}_l\overrightarrow{\partial}^l$ term can be equivalently reproduced by the sub-eikonal order contributions to Wilson line operator transformation due to changing to the interaction picture. As a result, one can keep the $\overleftarrow{\mathcal{D}}_l\overrightarrow{\mathcal{D}}^l$ term as the sub-eikonal interaction and only use the eikonal order Wilson line operator transformation.

For single antiquark scattering amplitude, one can repeat the above calculations or making charge conjugation transformation on eq.~\eqref{eq:M_q_to_q}.  The fundamental representation color matrix changes as $t^e\rightarrow -t^{e\ast}$ so that the Wilson lines in fundamental representation changes as $V_{m'm} \rightarrow V^{\dagger}_{mm'}$. Under charge conjugation transformation, the Dirac bilinear terms change as 
\begin{equation}
\begin{split}
&\bar{\psi} \gamma^-\psi(y) \longrightarrow  - \bar{\psi}(y) \gamma^- \psi(x),\\
&\bar{\psi} \gamma^-\gamma^5\psi(y) \longrightarrow   \bar{\psi}(y) \gamma^-\gamma^5 \psi(x).\\
\end{split}
\end{equation}

\subsection{Single gluon scattering amplitude}
For the single gluon scattering amplitude up to sub-eikonal order, one can perform similar calculations as have been done in the above section, starting from eq. \eqref{eq:S_expanded_to_xi}. We only present the final result here. 
\begin{equation}\label{eq:gluonWL_sub-eikonal_decomposition}
\begin{split}
&M^{g\rightarrow g}(\{k^{\prime +}, \mathbf{x}', c', \lambda'\};\{k^+, \mathbf{x}, c, \lambda\} )\\
=&(2\pi) 2k^+\delta(k^+-k^{\prime +}) \delta_{\lambda\lambda'} \Big[\delta(\mathbf{x}-\mathbf{x}') U_{\mathbf{x}}+ \xi  \delta(\mathbf{x}-\mathbf{x}') \lambda U^{\rm{pol}[1]}_{\mathbf{x}}(k^+) + \xi U^{\rm{pol}[2]}_{\mathbf{x}', \mathbf{x}}(k^+)\Big]^{c'c}\\
\end{split}
\end{equation}
Again the polarized Wilson lines can be further decomposed as 
\begin{equation}
\begin{split}
&U^{\rm{pol}[1]}_{\mathbf{x}}(k^+)  = U^{\rm{q}[1]}_{\mathbf{x}} (k^+) + U^{\rm{G}[1]}_{\mathbf{x}}(k^+), \\
&U^{\rm{pol}[2]}_{\mathbf{x}', \mathbf{x}}(k^+) =\delta(\mathbf{x}-\mathbf{x}') U^{\rm{q}[2]}_{\mathbf{x}}(k^+) + U^{\rm{G}[2]}_{\mathbf{x}', \mathbf{x}}(k^+). \\
\end{split}
\end{equation}
Their explicit expressions are.
\begin{equation}\label{eq:U_pol_1}
\begin{split}
 U^{\rm{q}[1]}_{\mathbf{x}} (k^+)=&-\frac{g^2}{2k^+}\int_{x_0^+}^{x^+} dw_2^+ \int_{x_0^+}^{w_2^+} dw_1^+ U_{\mathbf{x}}^{c'h'}(x^+, w_2^+)\bar{\psi}_{B} (w_2^+, \mathbf{x})t^{h'} V_{\mathbf{x}}(w_2^+, w_1^+)t^h\\
 &\qquad\qquad \qquad  \times \left[ \frac{ \gamma^-\gamma^5}{2}\right] \psi_{B}(w_1^+,\mathbf{x})  U_{\mathbf{x}}^{hc}(w_1^+, x_0^+) +c.c.\\
 U^{\rm{G}[1]}_{\mathbf{x}}(k^+)=&-\frac{2ig}{2k^+}\int_{x_0^+}^{x^+} dw^+  U^{c'a}_{\mathbf{x}}(x^+, w^+)[f_{12}(w^+, \mathbf{x})]^{ab} U^{bc}_{\mathbf{x}}(w^+, x_0^+).\\ 
\end{split}
\end{equation}
\begin{equation}\label{eq:U_pol_2}
\begin{split}
U^{\rm{q}[2]}_{\mathbf{x}}(k^+)=&- \frac{g^2}{2k^+}\int_{x_0^+}^{x^+} dw_2^+ \int_{x_0^+}^{w_2^+} dw_1^+ U^{c'h'}_{\mathbf{x}}(x^+, w_2^+)\bar{\psi}_{B} (w_2^+, \mathbf{x})t^{h'} V_{\mathbf{x}}(w_2^+, w_1^+)t^h\\
&\qquad \times \left[ \frac{\gamma^-}{2} \right] \psi_{B}(w_1^+, \mathbf{x})  U_{\mathbf{x}}^{hc}(w_1^+, x_0^+)  + c.c.\\
U^{\rm{G}[2]}_{\mathbf{x}', \mathbf{x}}(k^+)=&\frac{i}{2k^+}\int_{x_0^+}^{x^+} dw^+  U_{\mathbf{x}'}^{c'a}(x^+, w^+)\int_{\mathbf{z}}\delta(\mathbf{x}'-\mathbf{z}) \Big[\overleftarrow{\mathcal{D}_l}\overrightarrow{ \mathcal{D}}^l (w^+, \mathbf{z}) \Big]^{ab}\delta(\mathbf{z}-\mathbf{x}) U_{\mathbf{x}}^{bc}(w^+, x_0^+).\\
\end{split}
\end{equation}
Note that ``$c.c.$'' represent the corresponding charge conjugation terms.

\subsection{Background field induced quark-gluon conversion}
When computing particle and jet productions in polarized collision, one has to consider background field induced quark-gluon converting processes like $g\leftrightarrow q$ and $g\leftrightarrow \bar{q}$, see Fig.~\ref{fig:vertex_1}. These subprocesses, representing order $\xi^{\frac{1}{2}}$ contribution,  can happen in pair flexibly in the scattering amplitude and the complex conjugate amplitude to make the final cross section be at order $\xi$. This flexibility typically increases the number of Feynman diagrams and introduces delicate cancellation among certain set of diagrams. This flexibility might also induce extra contributions to small $x$ rapidity evolution \cite{Chirilli:2021lif}. The background field induced quark-gluon conversion responsible for quark-gluon dijet production in deep inelastic electron-proton scatterings was recently investigated in \cite{Altinoluk:2023qfr}.  For future reference, we present the explicit expressions for these subprocesses in this section.

For gluon to quark conversion,
substituting the interaction $V_{(\frac{1}{2})}$ from eq.~\eqref{eq:V_1/2}, one obtains
\begin{equation}
\begin{split}
&M^{g\rightarrow q} (\{p^+, \mathbf{x}, c, \lambda\}, \{p^{\prime +}, \mathbf{z}, m, \sigma\})\\
= & \langle 0 | \hat{b}_{m,\sigma}(p^{\prime +}, \mathbf{z}) \hat{S}(x^+, x_0^+) \hat{a}^{\dagger}_{c, \lambda}(p^+, \mathbf{x})|0\rangle\\
=&-i \int_{x_0^+}^{x^+} dw^+   \langle 0 | \hat{b}_{m,\sigma}(p^{\prime +}, \mathbf{z}) \hat{W}(x^+, w^+) \hat{V}_{(1/2)}(w^+) \hat{W}(w^+, x_0^+) \hat{a}^{\dagger}_{c, \lambda}(p^+, \mathbf{x})|0\rangle\\
=&-i \int_{x_0^+}^{x^+} dw^+  V^{mm'}_{\mathbf{z}}(x^+, w^+) \langle 0 | \hat{b}_{m',\sigma}(p^{\prime +}, \mathbf{z}) \hat{V}_{(1/2)}(w^+)  \hat{a}^{\dagger}_{c', \lambda}(p^+, \mathbf{x})|0\rangle U^{c'c}_{\mathbf{x}}(w^+, x_0^+)\\
=&-i \int_{x_0^+}^{x^+} dw^+  V^{mm'}_{\mathbf{z}}(x^+, w^+) \langle 0 | \hat{b}_{m',\sigma}(p^{\prime +}, \mathbf{z}) \Big[g\int_{\mathbf{y}, q^+} \frac{1}{2q^+} \hat{b}^{\dagger}_{n, \rho}(q^+, \mathbf{x}) \bar{u}_{G, \rho}(q^+) \hat{a}_{e, \kappa}(q^+, \mathbf{y})\\
&\qquad \times  \varepsilon^{i}_{\kappa}\gamma_i t^e_{nn'} \psi_{B, n'}(w^+, \mathbf{y})  \Big] \hat{a}^{\dagger}_{c', \lambda}(p^+, \mathbf{x})|0\rangle U_{\mathbf{x}}^{c'c}(w^+, x_0^+)\\
=&(2\pi) 2p^+\delta(p^+-p^{\prime +}) \delta(\mathbf{x}-\mathbf{z}) \mathcal{M}^{g\rightarrow q}(x^+, x_0^+;p^+, \mathbf{x}, \{c, \lambda\}, \{m, \sigma\})
\end{split}
\end{equation}
The gluon-quark conversion process preserves the longitudinal momentum and transverse coordinates. It only changes the color and spin quantum numbers.  After factorizing out the Dirac delta functions, the conversion amplitude is 
\begin{equation}\label{eq:gluon_to_quark_amplitude}
\begin{split}
&\mathcal{M}^{g\rightarrow q}(x^+, x_0^+;p^+, \mathbf{x}, \{c, \lambda\}, \{m, \sigma\})\\
=& -ig \int_{x_0^+}^{x^+} V^{mm'}_{\mathbf{x}}(x^+, w^+) \frac{1}{2p^+} \bar{u}_{G, \sigma}(p^+) \varepsilon^i_{\lambda}\gamma_i t^{e}_{m'n} \psi_{B, n}(w^+, \mathbf{x}) U^{ec}_{\mathbf{x}}(w^+, x_0^+). \\
\end{split}
\end{equation}
Repeating the above calculations, one obtains the amplitude for gluon to antiquark conversion,
\begin{equation}
\begin{split}
& \mathcal{M}^{g\rightarrow \bar{q}}(x^+, x_0^+; p^+, \mathbf{x}, \{c, \lambda\}, \{m, \sigma\})\\
=& +ig\int_{x_0^+}^{x^+} dw^+ V^{\dagger m'm}_{\mathbf{x}}(x^+, w^+) \bar{\psi}_{B, n}(w^+;\mathbf{x}) \gamma_i \varepsilon^i_{\lambda} t^e_{nm'} \frac{1}{2p^+}v_{G, \sigma}(p^+) U_{\mathbf{x}}^{ec}(w^+, x_0^+).\\
\end{split}
\end{equation}
for quark to gluon inverse conversion, the amplitude is 
\begin{equation}\label{eq:quark_to_gluon_amplitude}
\begin{split}
& \mathcal{M}^{q\rightarrow g}(x^+, x_0^+; p^+, \mathbf{x}, \{m, \sigma\}, \{c, \lambda\})\\
=& -ig\int_{x_0^+}^{x^+}dw^+ U_{\mathbf{x}}^{ce}(x^+, w^+)\frac{1}{2p^+}\bar{\psi}_{B,n}(w^+, \mathbf{x}) \gamma_i t^e_{nm'}\varepsilon^{i\ast}_{\lambda} u_{G, \sigma}(p^+) V^{m'm}_{\mathbf{x}}(w^+, x_0^+).\\
\end{split}
\end{equation}
for antiquark to gluon conversion, 
\begin{equation}\label{eq:antiquark_to_gluon_amplitude}
\begin{split}
&\mathcal{M}^{\bar{q}\rightarrow g}(x^+, x_0^+; p^+, \mathbf{x}, \{m, \sigma\}, \{c, \lambda\}) \\
= &+ig\int_{x_0^+}^{x^+}dw^+ U^{ce}_{\mathbf{x}}(x^+, w^+) \frac{1}{2p^+} \bar{v}_{G, \sigma}(p^+)\varepsilon^{i\ast}_{\lambda}\gamma_i t^{e}_{m'n'}\psi_{B,n'}(w^+;\mathbf{x}) V^{\dagger mm'}_{\mathbf{x}}(w^+, x_0^+). \\
\end{split}
\end{equation}
Again, this expression can be obtained by directly applying charge conjugation on eq.~\eqref{eq:quark_to_gluon_amplitude}.

\subsection{Quark-antiquark pair converted to two gluons}
\begin{figure}[!t]
    \centering
    \includegraphics[width=0.6\textwidth]{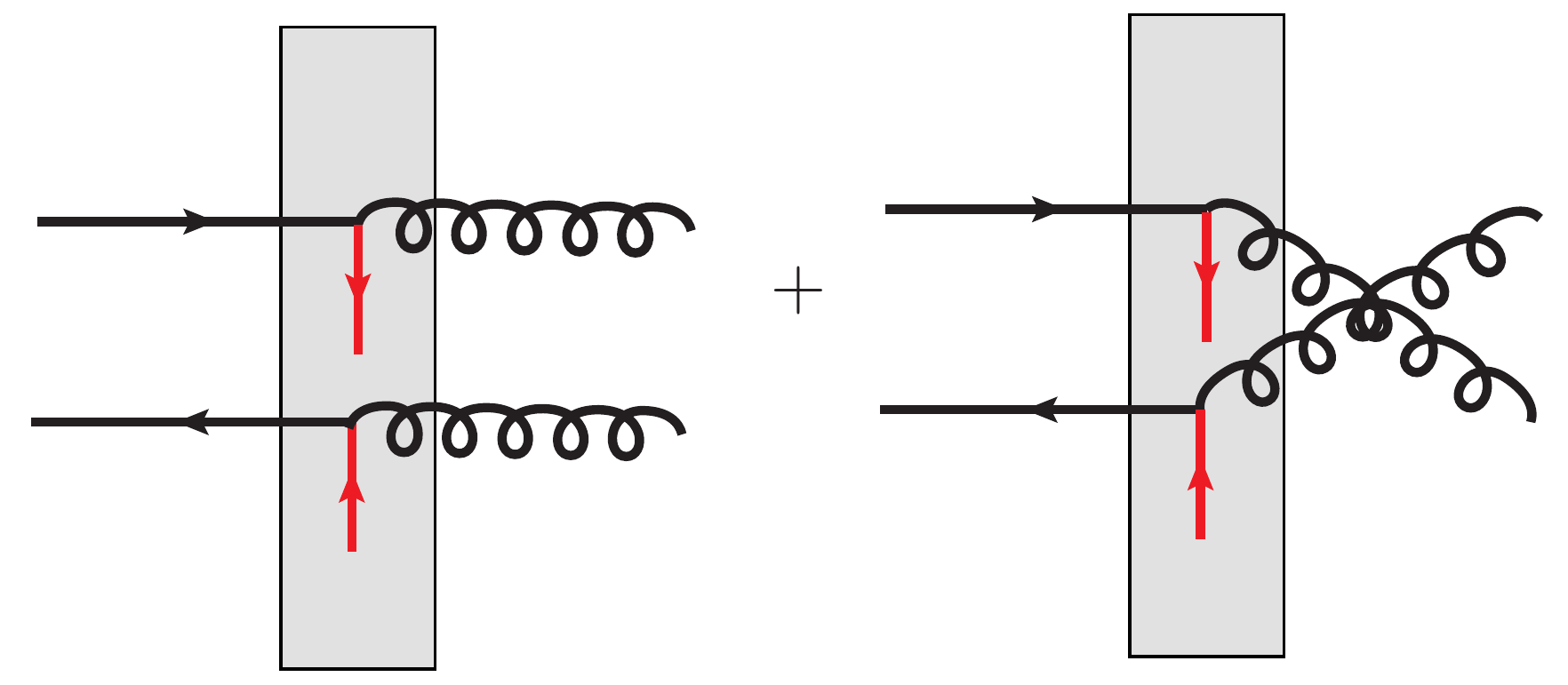}
    \caption{The process of $q\bar{q} \rightarrow gg$ induced by background quark fields. }
	\label{fig:qqbar_gg}
\end{figure}
Using the eikonality expansion of the S matrix operator from eq.~\eqref{eq:S_expanded_to_xi}, one can also compute the sub-eikonal process that a pair of quark and antiquark is converted into two gluons $\langle gg|\hat{S}|q\bar{q}\rangle $, see Fig.~\ref{fig:qqbar_gg}. In principle, one can obtain the amplitude $M^{q\bar{q} \rightarrow gg}$ as the product of two amplitudes computed in eqs.~\eqref{eq:quark_to_gluon_amplitude} and ~\eqref{eq:antiquark_to_gluon_amplitude}
\begin{equation}
M^{q\bar{q} \rightarrow gg} =M^{q\rightarrow g}M^{\bar{q}\rightarrow g}.
\end{equation}
However, we would like to demonstrate that this process can be directly computed from the eikonality expansion of the S matrix operator in eq. \eqref{eq:S_expanded_to_xi}, thus providing further evidence on the validity of the expansion. 
\begin{equation}\label{eq:M_qqbar_to_gg}
\begin{split}
&M^{q\bar{q} \rightarrow gg}(\{k_1^+, \mathbf{x}_1, m_1, \sigma_1\}, \{k_2^+, \mathbf{x}_2, m_2, \sigma_2\}; \{p_1^+, \mathbf{y}_1, c_1, \lambda_1\}, \{p_2^+, \mathbf{y}_2, c_2, \lambda_2\}) \\
=&\langle 0 | \hat{a}_{c_1, \lambda_1}(p_1^+, \mathbf{y}_1) \hat{a}_{c_2, \lambda_2}(p_2^+, \mathbf{y}_2) \, \hat{S}  \, \hat{b}^{\dagger}_{m_1, \sigma_1}(k_1^+, \mathbf{x}_1)\hat{d}^{\dagger}_{m_2, \sigma_2}(k_2^+, \mathbf{x}_2)|0\rangle \\
=&-\int_{x_0^+}^{x^+}dw_2^+ \int_{x_0^+}^{w^+_2}dw_1^+ \langle 0 | \hat{a}_{c_1, \lambda_1}(p_1^+, \mathbf{y}_1) \hat{a}_{c_2, \lambda_2}(p_2^+, \mathbf{y}_2) \Big[ \hat{W}(x^+, w_2^+)V_{(1/2)}(w_2^+)\hat{W}(w_2^+, w_1^+)\\
&\qquad \times V_{(1/2)}(w_1^+)\hat{W}(w_1^+, x_0^+) \Big]\hat{b}^{\dagger}_{m_1, \sigma_1}(k_1^+, \mathbf{x}_1)\hat{d}^{\dagger}_{m_2, \sigma_2}(k_2^+, \mathbf{x}_2)|0\rangle\\
=&-\int_{x_0^+}^{x^+}dw_2^+ \int_{x_0^+}^{w^+_2}dw_1^+ U^{c_1h_1}_{\mathbf{y}_1}(x^+, w_2^+)U^{c_2h_2}_{\mathbf{y}_2}(x^+, w_2^+) \langle 0 | \hat{a}_{h_1, \lambda_1}(p_1^+, \mathbf{y}_1) \hat{a}_{h_2, \lambda_2}(p_2^+, \mathbf{y}_2) \\
&\qquad \times \hat{V}_{(1/2)}(w_2^+)\hat{W}(w_2^+, w_1^+)\hat{V}_{(1/2)}(w_1^+)\hat{b}^{\dagger}_{n_1, \sigma_1}(k_1^+, \mathbf{x}_1)\hat{d}^{\dagger}_{n_2, \sigma_2}(k_2^+, \mathbf{x}_2)|0\rangle\\
&\qquad \times V^{n_1m_1}_{\mathbf{x}_1}(w_1^+, x_0^+) V^{\dagger m_2n_2}_{\mathbf{x}_2}(w_1^+, x_0^+). \\
\end{split}
\end{equation}
We have used the identities $\hat{W} \hat{b}^{\dagger}_j \hat{W}^{\dagger} = \hat{b}^{\dagger}_i V_{ij}$, $\hat{W} \hat{d}^{\dagger}_i \hat{W}^{\dagger} = V^{\dagger}_{ij}\hat{d}^{\dagger}_j$ and  $\hat{W}^{\dagger} \hat{a}_c \hat{W} = U^{ch}\hat{a}_h$ which are valid at the eikonal order.  Among the terms in $V_{(1/2)}(w_1^+)$ and $V_{(1/2)}(w_2^+)$, there are only two combinations that give nonvanishing contributions. One combination is 
\begin{equation}
\hat{V}_{(1/2)}(w_2^+) = g\int_{\mathbf{z}'}\int_{p^{\prime +}} \frac{1}{2p^{\prime +}} \bar{\psi}_{B, l'}(w_2^+, \mathbf{z}') \gamma_{i'}t^{e'}_{l'j'} \hat{a}^{\dagger}_{e', \lambda'}(p^{\prime +}, \mathbf{z}')\varepsilon^{i'\ast}_{\lambda'} \hat{b}_{j', s'}(p^{\prime +}, \mathbf{z}') u_{G, s'}(p^{\prime +}) 
\end{equation}
and 
\begin{equation}
\hat{V}_{(1/2)}(w_1^+) = g\int_{\mathbf{z}} \int_{p^+} \frac{1}{2p^+} \hat{d}_{j,s}(p^+, \mathbf{z})\bar{v}_{G, s}(p^+) \hat{a}^{\dagger}_{e, \lambda}(p^+, \mathbf{z})\varepsilon^{i\ast}_{\lambda}\gamma_i t^e_{jl}\psi_{B,l}(w_1^+, \mathbf{z}). 
\end{equation}
The other combination is to switch the expressions by $w_1^+\leftrightarrow w_2^+$. 
Using these expressions, 
eq.~\eqref{eq:M_qqbar_to_gg} can be expressed as 
\begin{equation}
\begin{split}
&\int_{x_0^+}^{x^+}dw_2^+ \int_{x_0^+}^{w^+_2}dw_1^+  \left[(2\pi)2k_1^+\delta(k_1^+-p_1^+)\delta(\mathbf{x}_1-\mathbf{y}_1) (2\pi)2k_2^+\delta(k_2^+-p_2^+)\delta(\mathbf{x}_2-\mathbf{y}_2)\right]\\
&\qquad \times g^2 \frac{1}{2k_1^+}\frac{1}{2k_2^+}\bar{\psi}_{B, l'}(w_2^+, \mathbf{x}_1)t^{h_1}_{l'n'_1}\gamma_{i'}\varepsilon^{i'\ast}_{\lambda_1} u_{G, \sigma_1}(k_1^+) \bar{v}_{G, \sigma_2}(k_2^+)\gamma_i \varepsilon^{i\ast}_{\lambda_2} t^e_{n_2l} \psi_{B, l}(w_1^+, \mathbf{x}_2)\\
&\qquad \times U^{c_1h_1}_{\mathbf{x}_1}(x^+, w_2^+)V_{\mathbf{x}_1}^{n'_1m_1}(w_2^+, x_0^+)U^{c_2e}_{\mathbf{x}_2}(x^+, w_1^+)V^{\dagger m_2n_2}_{\mathbf{x}_2}(w_1^+, x_0^+)\\
&\qquad +(w_1^+\leftrightarrow w_2^+). 
\end{split}
\end{equation}
(The extra minus sign comes from moving the annihilation operator $\hat{d}_{j,s}$ across $\psi_{B, l}$.)
Exchanging $w_1^+$ and $w_2^+$ is only for the integrand. It is equivalent to exchaning $w_1^+$ and $w_2^+$ in the integration measures while keeping the integrand unchanged. We will use
\begin{equation}
\int_{x_0^+}^{x^+}dw_2^+ \int_{x_0^+}^{w^+_2}dw_1^+   + \int_{x_0^+}^{x^+}dw_1^+ \int_{x_0^+}^{w^+_1}dw_2^+   = \int_{x_0^+}^{x^+}dw_2^+ \int_{x_0^+}^{x^+}dw_1^+.
\end{equation}
The final result for the scattering amplitude is 
\begin{equation}
\begin{split}
&M^{q\bar{q} \rightarrow gg}\\
=& \left[(2\pi)2k_1^+\delta(k_1^+-p_2^+)\delta(\mathbf{x}_1-\mathbf{y}_2) (2\pi)2k_2^+\delta(k_2^+-p_1^+)\delta(\mathbf{x}_2-\mathbf{y}_1)\right]\int_{x_0^+}^{x^+}dw_2^+ \int_{x_0^+}^{x^+}dw_1^+ \\
&\times g^2\frac{1}{4\sqrt{k_1^+k_2^+}}\bar{\psi}_{B, l'}(w_2^+, \mathbf{x}_1)\left[t^{h}V_{\mathbf{x}_1}(w_2^+, x_0^+)\right]_{l'm_1}[\gamma^- + \lambda_2 \gamma^-\gamma^5]\delta_{\lambda_1, -\lambda_2} \\
&\qquad \times \left[V^{\dagger}_{\mathbf{x}_2}(w_1^+, x_0^+)t^e\right]_{m_2l}\psi_{B, l}(w_1^+, \mathbf{x}_2) U_{\mathbf{x}_1}^{c_2h}(x^+, w_2^+) U_{\mathbf{x}_2}^{c_1e}(x^+, w_1^+)\\
&+\left[(2\pi)2k_1^+\delta(k_1^+-p_1^+)\delta(\mathbf{x}_1-\mathbf{y}_1) (2\pi)2k_2^+\delta(k_2^+-p_2^+)\delta(\mathbf{x}_2-\mathbf{y}_2)\right]\int_{x_0^+}^{x^+}dw_2^+ \int_{x_0^+}^{x^+}dw_1^+\\
&\times   g^2 \frac{1}{4\sqrt{k_1^+k_2^+}}\bar{\psi}_{B, l'}(w_2^+, \mathbf{x}_1)\left[t^{h} V_{\mathbf{x}_1}(w_2^+, x_0^+)\right]_{l'm_1} [\gamma^- + \lambda_1 \gamma^-\gamma^5]\delta_{-\lambda_1, \lambda_2}\\
&\qquad \times \left[V^{\dagger}_{\mathbf{x}_2}(w_1^+, x_0^+) t^e\right]_{m_2l} \psi_{B, l}(w_1^+, \mathbf{x}_2) U^{c_1h}_{\mathbf{x}_1}(x^+, w_2^+)U^{c_2e}_{\mathbf{x}_2}(x^+, w_1^+). \\
\end{split}
\end{equation}
This expression is exactly the same as $M^{q\rightarrow g}M^{\bar{q}\rightarrow g}$ using eq.~\eqref{eq:quark_to_gluon_amplitude} and eq.~\eqref{eq:antiquark_to_gluon_amplitude}.  The spinor space matrix elements have been further simplified by requiring the quark antiquark spin states satisfy $\delta_{\sigma_1,-\sigma_2}$. This is generally true when the quark antiquark pair comes from a photon/gluon splitting. We have used the spinor space identity
\begin{equation}
\begin{split}
&\gamma_{i'}\varepsilon^{i'\ast}_{\lambda_2}u_{G, \sigma_1}(k_1^+) \bar{v}_{G, \sigma_2}(k_2^+) \gamma_i \varepsilon^{i\ast}_{\lambda_1} \delta_{\sigma_1, -\sigma_2}\\
=&\sqrt{k_1^+k_2^+} [\gamma^- + \lambda_2 \gamma^-\gamma^5]\delta_{\lambda_1, -\lambda_2} 
\end{split}
\end{equation}
whose derivation can be found in the appendix~\ref{app:LCQ}. It is interesting to note that the polarizations of the two outging gluons are opposite.

\section{Gluon Radiation Inside the Shockwave}\label{sec:gluon_radiation_inside_shockwave}
The small $x$ effective Hamiltonian derived in Sec.~\ref{subsec:smallx_Hamiltonian} predicts that gluon radiation inside the shockwave, induced by background gluon fields, contributes to physical processes  at sub-eikonal order. In this section, we discuss two specific situations in which gluon radiation inside the shock contributes. One is about double-spin asymmetry for soft gluon production in longitudinally polarized collisions. The other is about small $x$ rapidity evolution of chrormo-magnetically polarized Wilson line correlator.  It is noted that gluon radiation inside the shockwave has been incorporated in studying gluon TMD evolution in \cite{Balitsky:2015qba, Balitsky:2016dgz}. In the context of jet quenching in heavy-ion collisions, medium induced gluon radiation has also been studied in \cite{Armesto:2011ir,Blaizot:2004wu}.

\subsection{Longitudinal double-spin asymmetry for soft gluon production}
For the incoming gluon we use $c, \lambda, p^+, \mathbf{p}$ to denote its color, polarization and momentum.  For the two outgoing gluons their color, polarization and momentum are $c_1, \lambda_1, p_1^+, \mathbf{p}_1$ and $c_2, \lambda_2, p_2^+, \mathbf{p}_2$, respectively. It is easier to do calculations in the mixed representation in which longitudinal momentum and transverse positions are used. We thus denote the corresponding transverse coordinates of the incoming and outgoing gluons as $\mathbf{x}_0, \mathbf{x}_1, \mathbf{x}_2$. We focus on the situation that one of the outgoing gluons is in the midrapidity region. The longitudinal momenta satisfy $p_1^+ \ll p_2^+\sim p^+$ or $p_1^+=zp^+$ with $z\rightarrow 0$. 

\begin{figure}[!t]
    \centering
    \includegraphics[width=0.8\textwidth]{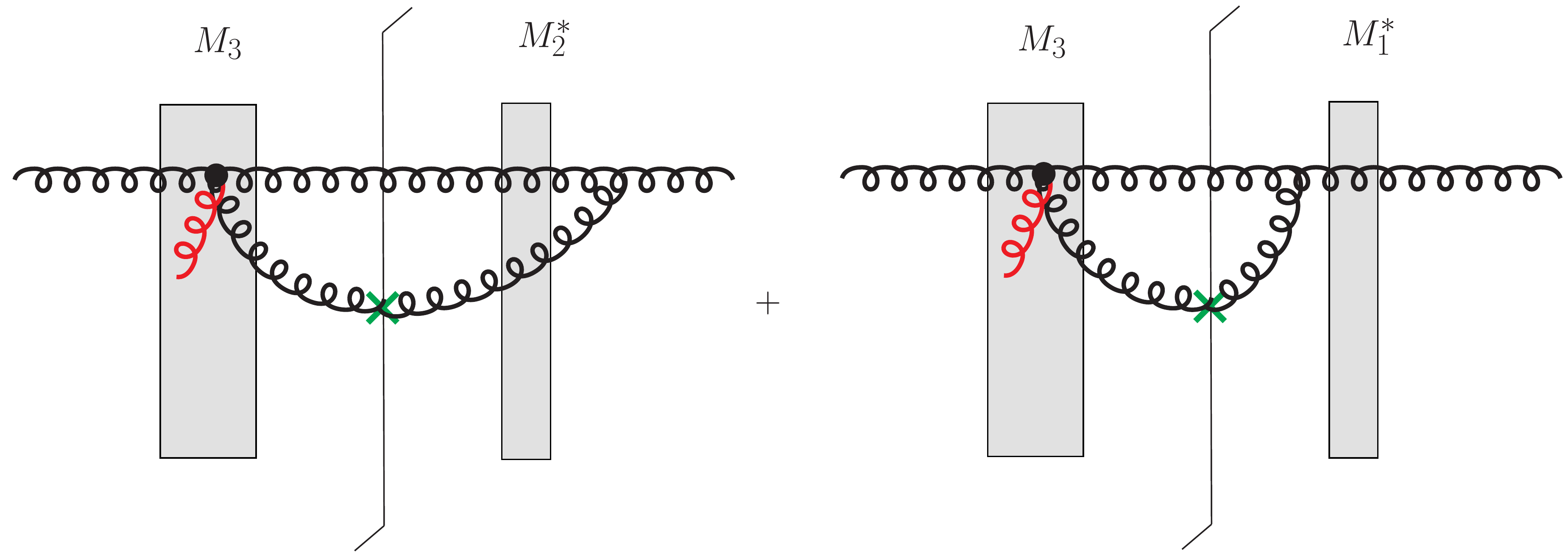}
    \caption{The interference terms involving gluon radiation inside the shock wave. The red gluon line inside the shockwave represents the background gluon fields. The green cross indicates the tagged soft gluon.}
	\label{fig:M6M2}
\end{figure}

We calculate the first diagram in Fig.~\ref{fig:M6M2}, representing the interference term between eikonal order initial state gluon radiation and the sub-eikonal order gluon radiation inside the shockwave. 
The eikonal order gluon radiation amplitude is computed as
\begin{equation}\label{eq:M2+M4}
\begin{split}
\mathcal{M}_2
= & \delta(\mathbf{x}_0-\mathbf{x}_2) (i gf^{ce_1e_2})
(2\delta_{\lambda\lambda_2} \varepsilon^{j\ast}_{\lambda_1} )\frac{i}{2\pi} \frac{(\mathbf{x}_1-\mathbf{x}_0)^j}{|\mathbf{x}_1-\mathbf{x}_0|^2} U^{c_1e_1}_{\mathbf{x}_1}U^{c_2e_2}_{\mathbf{x}_0}. \\
\end{split}
\end{equation}
For gluon radiation inside the shockwave, the amplitude is computed using the formula eq.~\eqref{eq:S_expanded_to_xi}.
\begin{equation}\label{eq:M6_exact}
\begin{split}
 M_3
=&\langle 0 | \hat{a}_{c_2, \lambda_2}(p_2^+, \mathbf{x}_2) \hat{a}_{c_1, \lambda_1}(p_1^+, \mathbf{x}_1) \hat{S}(x^+, x_0^+) \hat{a}^{\dagger}_{c, \lambda}(p^+, \mathbf{x}_0)|0\rangle\\
=&-i\xi \int_{x_0^+}^{x^+} dw^+ \langle 0 | \hat{a}_{c_2, \lambda_2}(p_2^+, \mathbf{x}_2) \hat{a}_{c_1, \lambda_1}(p_1^+, \mathbf{x}_1) \hat{W}(x^+, x_0^+)\hat{V}_{ggga}(w^+) \hat{W}(w^+, x_0^+) \hat{a}^{\dagger}_{c, \lambda}(p^+, \mathbf{x}_0)|0\rangle\\
=&-i\xi \int_{x_0^+}^{x^+} dw^+ U_{\mathbf{x}_1}^{c_1e_1}(x^+, w^+)U_{\mathbf{x}_2}^{c_2e_2}(x^+, w^+)U^{ec}_{\mathbf{x}_0}(w^+, x_0^+)\\
&\qquad \times \, \big\langle  0\big| \hat{a}_{e_1, \lambda_1}(p_1^+, \mathbf{x}_1) \hat{a}_{e_2, \lambda_2}(p_2^+, \mathbf{x}_2)   \hat{V}_{ggga}(w^+) \hat{a}^{\dagger}_{e, \lambda} (p^+, \mathbf{x}_0)\big|0\big\rangle \,  \\
=&(2\pi)2p^+\delta(-p^+ + p_1^+ + p_2^+)  \frac{1}{2p^+}\xi\int_{x_0^+}^{x^+} dw^+ U_{\mathbf{x}_1}^{c_1e_1}(x^+, w^+)U_{\mathbf{x}_2}^{c_2e_2}(x^+, w^+)U^{ec}_{\mathbf{x}_0}(w^+, x_0^+)\\
& \times \Big[ (igf^{dee_1} )[-\mathcal{D}_{\mathbf{x}_2}^i\delta(\mathbf{x}_2-\mathbf{x}_0)]_{de_2}\delta(\mathbf{x}_1-\mathbf{x}_0) \left(\varepsilon^i_{\lambda}\delta_{\lambda_2, -\lambda_1} -\varepsilon^{i\ast}_{\lambda_1}\delta_{\lambda_2\lambda}  - \frac{p^++p_1^+}{p_2^+}\delta_{\lambda_1\lambda} \varepsilon^{i\ast}_{\lambda_2}\right)\\
&\qquad +(igf^{dee_2})[-\mathcal{D}_{\mathbf{x}_1}^i\delta(\mathbf{x}_1-\mathbf{x}_0)]_{de_1}\delta(\mathbf{x}_2-\mathbf{x}_0)\left(\varepsilon^i_{\lambda}\delta_{\lambda_1, -\lambda_2}- \varepsilon^{i\ast}_{\lambda_2} \delta_{\lambda_1\lambda} - \frac{p^++p_2^+}{p_1^+} \delta_{\lambda\lambda_2} \varepsilon^{i\ast}_{\lambda_1}\right)\\
&\qquad +(igf^{de_1e_2})[-\mathcal{D}_{\mathbf{x}_0}^i\delta(\mathbf{x}_2-\mathbf{x}_0)]_{de} \delta(\mathbf{x}_1-\mathbf{x}_2)\left(\delta_{\lambda\lambda_2}\varepsilon_{\lambda_1}^{i\ast} - \delta_{\lambda \lambda_1}\varepsilon^{i\ast}_{\lambda_2} - \frac{p_1^+-p_2^+}{p^+} \varepsilon^i_{\lambda} \delta_{\lambda_1, -\lambda_2}\right)\Big]
\end{split}
\end{equation}
Here $\hat{V}_{ggga}$ represent the background field induced triple gluon interaction vertex given in eq.~\eqref{eq:V_1}. 
For the two contributing vertices in eq.~\eqref{eq:V_1}, one is local in longitudinal coordinate while the other one is nonlocal.

Eq.~\eqref{eq:M6_exact} needs to be simplified by taking the limit $z\rightarrow 0$  with $p_1^+=zp^+$.  In fact,  terms that contain the polarization factor $\delta_{\lambda\lambda_2} \varepsilon^{i\ast}_{\lambda_1}$ can be ignored because these terms do not communicate polarization of the incoming gluon to the outgoing gluons, leading to final result independent of the incoming gluon polarization, thus won't contribute to double-spin asymmetry. It is interesting to note that these terms happen to involve the factor $1/z$ in the soft gluon limit. 

There is the possible combination that the $1/z$ order gluon radiation inside the shockwave in the amplitude can combine with the complex conjugate amplitude in which the next-to-eikonal  (order $z$, see appendix~\ref{app:LCQ} for explicit expression) gluon is radiated either before or after eikonal scatterings. This combination turns out to give exactly the same result as the eq.~\eqref{eq:A_LL_result}.

Only keeping terms at $z^0$ order and excluding terms proportional to $\delta_{\lambda\lambda_2}\varepsilon^{i\ast}_{\lambda_1}$ ,  the simplified expression of $M_3$ contributing to double-spin asymmetry is 
\begin{equation}\label{eq:M3_gluon_inside_sw}
\begin{split}
\mathcal{M}_3
=& \frac{1}{2p^+}\int_{x_0^+}^{x^+} dw^+ U_{\mathbf{x}_1}^{c_1e_1}(x^+, w^+)U_{\mathbf{x}_2}^{c_2e_2}(x^+, w^+)U^{ec}_{\mathbf{x}_0}(w^+, x_0^+)\\
&\times\Big(  igf^{ee_1e_2}\left[-\partial^i_{\mathbf{x}_2} \delta(\mathbf{x}_2-\mathbf{x}_0) \delta(\mathbf{x}_1-\mathbf{x}_0) + \partial^i_{\mathbf{x}_1} \delta(\mathbf{x}_1-\mathbf{x}_0)\delta(\mathbf{x}_2-\mathbf{x}_0) -\partial^i_{\mathbf{x}_0} \delta(\mathbf{x}_2-\mathbf{x}_0) \delta(\mathbf{x}_1-\mathbf{x}_2)\right]\\
&\qquad +ig^2a_b^i(\mathbf{x}_1) \delta(\mathbf{x}_2-\mathbf{x}_0) \delta(\mathbf{x}_1-\mathbf{x}_0) 2(T^{e}T^{e_1})_{e_2b}\Big)\left(\varepsilon^i_{\lambda}\delta_{\lambda_2, -\lambda_1}  - \delta_{\lambda_1\lambda} \varepsilon^{i\ast}_{\lambda_2}\right)\\
\end{split}
\end{equation}
In the above expression, we have explicitly separated the spatial derivative terms and terms containing the background gluon field.
When computing the interfererence term $M_3M_2^{\ast}$,
the polarization factors involved is
\begin{equation}
\sum_{\lambda_1,\lambda_2}\left(\delta_{\lambda_2, -\lambda_1}  \varepsilon^{i}_{\lambda} - \delta_{\lambda\lambda_1}\varepsilon^{i\ast}_{\lambda_2}\right)\delta_{\lambda'\lambda_2}\varepsilon_{\lambda_1}^{j}  =\varepsilon^i_{\lambda} \varepsilon^{j\ast}_{\lambda'} -  \varepsilon^{j}_{\lambda}\varepsilon^{i\ast}_{\lambda'} = \delta_{\lambda\lambda'} (-i\lambda\epsilon^{ij}).
\end{equation}
The final result for the interference term  $M_3M_2^{\ast}$ is 
\begin{equation}\label{eq:A_LL_result}
\begin{split}
&\int_{\mathbf{x}_1, \mathbf{x}'_1}e^{-i\mathbf{p}_1\cdot(\mathbf{x}_1-\mathbf{x}'_1)}\sum_{c,c_1,c_2, \lambda_1, \lambda_2} \int_{\mathbf{x}_2, p_2^+, \mathbf{x}'_0, \mathbf{x}_0} \mathcal{M}_3\mathcal{M}_2^{\ast}(2\pi)2p^+\delta(p^+-p_1^+-p_2^+) + c.c.\\
=&-\lambda \delta_{\lambda \lambda'}2g^2N_c \int_{\mathbf{x}_1, \mathbf{x}'_1}e^{-i\mathbf{p}_1\cdot(\mathbf{x}_1-\mathbf{x}'_1)}\frac{1}{2\pi}\frac{\varepsilon^{ij}(\mathbf{x}'_1-\mathbf{x}_1)^j}{|\mathbf{x}'_1-\mathbf{x}_1|^2}\\
&\qquad \times \frac{1}{2p^+} \int_{x_0^+}^{x^+} dw^+\mathrm{Tr}\left[U_{\mathbf{x}_1}(x^+, w^+)\left(\overrightarrow{\mathcal{D}}_{\mathbf{x}_1}^i-\overleftarrow{\mathcal{D}}_{\mathbf{x}_1}^i\right) U_{\mathbf{x}_1}(w^+, x_0^+)U^{\dagger}_{\mathbf{x}'_1}\right] + c.c.\\
=&-\lambda \delta_{\lambda \lambda'} 4g^2 N_c\int_{\mathbf{x}_1, \mathbf{x}'_1}e^{-i\mathbf{p}_1\cdot(\mathbf{x}_1-\mathbf{x}'_1)}\frac{1}{2\pi}\frac{\varepsilon^{ij}(\mathbf{x}'_1-\mathbf{x}_1)^j}{|\mathbf{x}'_1-\mathbf{x}_1|^2} \mathrm{Tr}\left[U^{i\, G[2]}_{\mathbf{x}_1}U^{\dagger}_{\mathbf{x}'_1}\right] + c.c.\\
\end{split}
\end{equation}
In obtaining the first equality, we have simplified the Wilson line structures as follows
\begin{equation}
\begin{split}
&U_{\mathbf{x}_1}^{c_1c'_1}(x^+, w^+)U_{\mathbf{x}_1}^{c_2c'_2}(x^+, w^+)U^{c'c}_{\mathbf{x}_1}(w^+, x_0^+)\left[- U_{\mathbf{x}'_1}T^{c} U^{\dagger}_{\mathbf{x}_1}\right]^{c_1c_2}2(T^{c'}T^{c'_1})_{c'_2b}\\
=&-2N_c \mathrm{Tr}\left[U_{\mathbf{x}_1}(x^+, w^+)T^bU_{\mathbf{x}_1}(w^+, x_0^+)U_{\mathbf{x}'_1}^{\dagger}\right].\\
\end{split}
\end{equation}
\begin{equation}
\begin{split}
&\int_{\mathbf{x}_0} U_{\mathbf{x}_1}^{c_1c'_1}(x^+, w^+)U_{\mathbf{x}_1}^{c_2c'_2}(x^+, w^+)U^{c'c}_{\mathbf{x}_0}(w^+, x_0^+)\left[- U_{\mathbf{x}'_1}T^{c} U^{\dagger}_{\mathbf{x}_1}\right]^{c_1c_2} if^{c'c'_1c'_2} [-\partial^i_{\mathbf{x}_0} \delta(\mathbf{x}_1-\mathbf{x}_0)]\\
=&-N_c\mathrm{Tr}\left[U_{\mathbf{x}'_1}\partial_{\mathbf{x}_1}^iU^{\dagger}_{\mathbf{x}_1}(w^+, x_0^+) U^{\dagger}_{\mathbf{x}_1}(x^+, w^+) \right]\\
&\qquad -\mathrm{Tr}\left[U_{\mathbf{x}'_1}U^{\dagger}_{\mathbf{x}_1}(w^+, x_0^+)T^{c'}(\partial_{\mathbf{x}_1}^iU_{\mathbf{x}_1}(w^+, x_0^+))U^{\dagger}_{\mathbf{x}_1}(w^+, x_0^+)T^{c'} U^{\dagger}_{\mathbf{x}_1}(x^+, w^+) \right].
\end{split}
\end{equation}
\begin{equation}
\begin{split}
&\int_{\mathbf{x}_2} U_{\mathbf{x}_1}^{c_1c'_1}(x^+, w^+)U_{\mathbf{x}_2}^{c_2c'_2}(x^+, w^+)U^{c'c}_{\mathbf{x}_1}(w^+, x_0^+)\left[- U_{\mathbf{x}'_1}T^{c} U^{\dagger}_{\mathbf{x}_2}\right]^{c_1c_2} if^{c'c'_1c'_2}\left[-\partial^i_{\mathbf{x}_2} \delta(\mathbf{x}_2-\mathbf{x}_1) \right]\\
=&- \mathrm{Tr}\left[ U_{\mathbf{x}'_1}U^{\dagger}_{\mathbf{x}_1}(w^+, x_0^+)T^{c'}U_{\mathbf{x}_1}(w^+, x_0^+) \partial_{\mathbf{x}_1}^iU^{\dagger}_{\mathbf{x}_1}(w^+, x_0^+)T^{c'}U^{\dagger}_{\mathbf{x}_1}(x^+, w^+)\right].\\
\end{split}
\end{equation}
\begin{equation}
\begin{split}
&\int_{\mathbf{x}_0}U_{\mathbf{x}_1}^{c_1c'_1}(x^+, w^+)U_{\mathbf{x}_0}^{c_2c'_2}(x^+, w^+)U^{c'c}_{\mathbf{x}_0}(w^+, x_0^+)\left[- U_{\mathbf{x}'_1}T^{c} U^{\dagger}_{\mathbf{x}_0}\right]^{c_1c_2} if^{c'c'_1c'_2}\partial^i_{\mathbf{x}_1} \delta(\mathbf{x}_1-\mathbf{x}_0)\\
=&N_c \mathrm{Tr}[\partial^i_{\mathbf{x}_1}U_{\mathbf{x}_1}(x^+, w^+) U_{\mathbf{x}_1}(w^+, x_0^+)U^{\dagger}_{\mathbf{x}'_1}].
\end{split}
\end{equation}
In the last expression, we have used integration by parts for $\partial^i_{\mathbf{x}_1}$, noting that 
\begin{equation}
\epsilon^{ij}\mathbf{p}^i \int_{\mathbf{x}_1, \mathbf{x}'_1}e^{-i\mathbf{p}_1\cdot(\mathbf{x}_1-\mathbf{x}'_1)}\frac{(\mathbf{x}'_1-\mathbf{x}_1)^j}{|\mathbf{x}'_1-\mathbf{x}_1|^2} \langle \mathrm{Tr}[U_{\mathbf{x}_1}U^{\dagger}_{\mathbf{x}'_1}]\rangle =0.
\end{equation}

Eq.~\eqref{eq:A_LL_result} is expressed in terms of the polarized Wilson line with spatial index
\begin{equation}\label{eq:UiG2_def}
\begin{split}
U_{\mathbf{x}_1}^{iG[2]}(k^+) = &\frac{1}{2k^+} \int_{-\infty}^{+\infty} dx_1^+ U_{\mathbf{x}_1}[+\infty, x_1^+]\frac{1}{2}\Big[\overrightarrow{\mathcal{D}}_{\mathbf{x}_1}^i- \overleftarrow{\mathcal{D}}_{\mathbf{x}_1}^i\Big]U_{\mathbf{x}_1}[x_1^+, -\infty].\\
\end{split}
\end{equation}
Using the identity
\begin{equation}
\begin{split}
&\int_{x_0^+}^{x^+}dz^+ U_{\mathbf{x}}(x^+, z^+) f^{-i}(z^+, \mathbf{x}) U_{\mathbf{x}}(z^+, x_0^+)\\ =& a^i(x^+, \mathbf{x})U_{\mathbf{x}}(x^+, x_0^+) - U_{\mathbf{x}}(x^+, x_0^+) a^i(x^+_0, \mathbf{x}) + \frac{1}{ig}\partial^iU_{\mathbf{x}}(x^+,x_0^+),\\
\end{split}
\end{equation}
eq.~\eqref{eq:UiG2_def} can be reexpressed as \cite{Chirilli:2018kkw, Altinoluk:2020oyd}
\begin{equation}
 U^{i\rm{G}[2]}_{\mathbf{x}_1}(k^+)  =- \frac{ig}{2k^+} \int_{-\infty}^{+\infty}dx_1^+ x_1^+  U_{\mathbf{x}_1}(+\infty, x_1^+)f^{-i}(x_1^+, \mathbf{x}_1)U_{\mathbf{x}_1}(x_1^+, -\infty),
\end{equation}
which clearly shows that $U_{\mathbf{x}_1}^{i\rm{G}[2]}(k^+)$ is determined by the chormo-electric field $f^{-i}(x_1^+, \mathbf{x}_1)$. 

One can repeat the calculation for the second diagram in Fig.~\ref{fig:M6M2}, it vanishes because of $\mathrm{Tr}[U^{i\, G[2]}_{\mathbf{x}_1}U^{\dagger}_{\mathbf{x}_1}] = 0$.

Eq.~\eqref{eq:A_LL_result} is the main result of this section, which clearly shows that gluon radiation inside the shockwave contribute to longitudinal double-spin asymmetry in soft gluon production. Furthermore, its contribution is in the form of chromo-electrically polarized Wilson line correlator.

\subsection{Small $x$ evolution of polarized Wilson line correlator}
\begin{figure}[!t]
    \centering
    \includegraphics[width=0.8\textwidth]{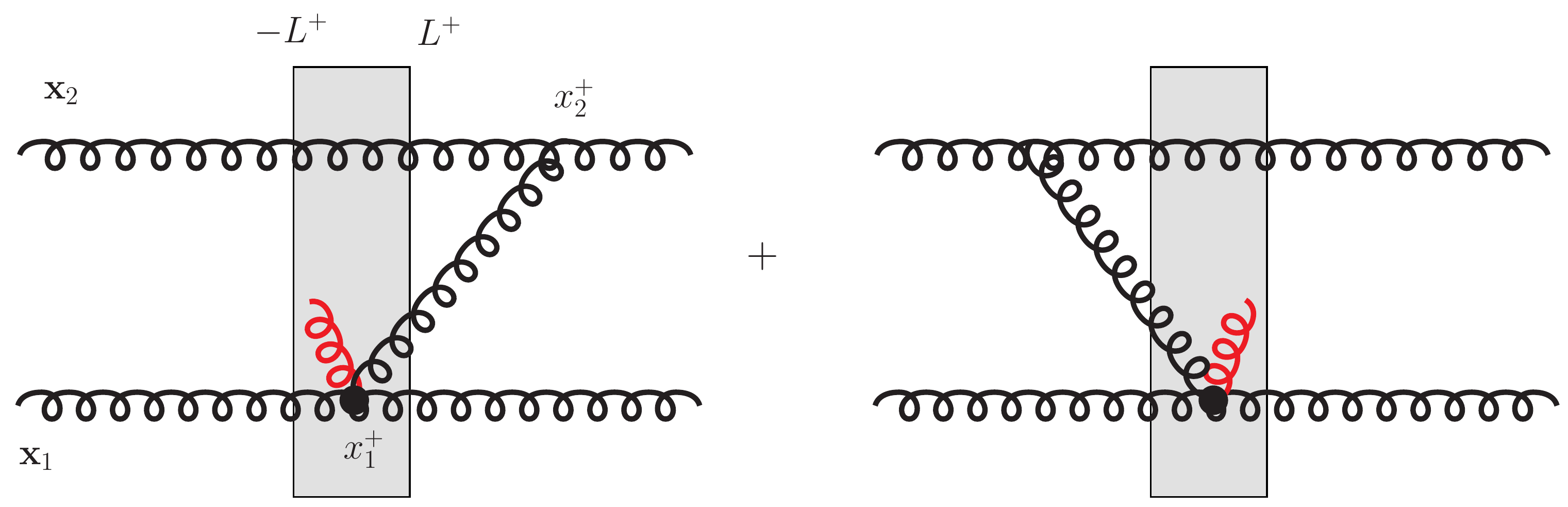}
    \caption{The single logarithmic contribution to small $x$ evolution of $\langle \mathrm{Tr}[U^{G[1]}_{\mathbf{x}_1}U^{\dagger}_{\mathbf{x}_2}]\rangle$. The red gluon lines represent the background gluon fields. }
	\label{fig:evolution_extra_one_new_labels}
\end{figure}

In this section, we calculate the amplitude shown in Fig.~\ref{fig:evolution_extra_one_new_labels}. They come from one step rapidity evolution of the chromo-magnetically polarized gluon dipole correlator $\langle \mathrm{Tr}[U^{G[1]}_{\mathbf{x}_1}U^{\dagger}_{\mathbf{x}_2}]\rangle $. The two diagrams represent contributions from gluon radiation inside the shockwave. We use two different methods to do the calculations. One is an operator treatment that was used by many groups \cite{Balitsky:2015qba, Kovchegov:2018znm, Cougoulic:2022gbk}. The other method is to directly compute the two diagrams. 

\subsubsection{Operator treatment}
We begin with the definition of the chromo-magnetically polarized gluon Wilson line correlator
\begin{equation}\label{eq:UG[1]U}
\begin{split}
\left\langle \mathrm{Tr}\left[U^{G[1]}_{\mathbf{x}_1} U^{\dagger}_{\mathbf{x}_2}\right] \right\rangle(k^+)  
= -\frac{2ig}{2k^+} \int_{-\infty}^{+\infty} dx_1^+ \left\langle \mathrm{Tr}\left[ U_{\mathbf{x}_1}[+\infty, x_1^+] f^{12}(x_1^+, 0^-, \mathbf{x}_1)U_{\mathbf{x}_1}[x_1^+, -\infty] U^{\dagger}_{\mathbf{x}_2}\right]\right\rangle 
\end{split}
\end{equation}
The field strength tensor $f^{12}$ carries longitudinal momentum $k^+$. We decomposed it into a quantum part and a classical part by $a^i(k^+) \rightarrow A^i(k^{\prime +}) +a^i(k^+-\delta k^+)$ with $ k^+-\delta k^+ < k^{\prime +} < k^+$. One then integrates out the quantum degrees of freedom $A^i(k^{\prime +})$ in the infinitesimal strip of longitudinal momentum $\delta k^+$. The separation into classical fields and quantum fluctuations is given by
\begin{equation}
\begin{split}
&f^{12} = \frac{1}{2}\epsilon^{ij}f^{ij} \\
= &\frac{1}{2}\epsilon^{ij}\Big(\partial^i a^j - \partial^j a^i + ig[a^i, a^j] + \partial^i A^j - \partial^j A^i + ig[A^i, A^j] +ig[A^i, a^j] +ig[a^i, A^j]\Big)\\
=&\frac{1}{2}\epsilon^{ij} \Big(f^{ij}+ig[A^i, A^j]\Big) + \epsilon^{ij}\Big(\partial^iA^j +ig[a^i, A^j]\Big). 
\end{split}
\end{equation}
We will focus on the piece $ \epsilon^{ij}(\partial^iA^j +ig[a^i, A^j])$ which is linear in quantum fluctuating fields.  Let the shockwave locate within the narrow range $[-L^+, L^+]$ around $0^+$. We further require that $
-L^+ < x_1^+ <L^+ $. In other words, the chromo-magnetic field lies inside the shockwave. The situations that the chromo-magnetic field lies before or after the shockwave have been studied in \cite{Cougoulic:2022gbk}.

The eikonal Wilson line $U_{\mathbf{x}_2}$ also needs to be expanded to linear order in quantum fluctuation field
\begin{equation}
U_{\mathbf{x}_2}^{mn} \simeq -ig \int_{-\infty}^{+\infty} dx_2^+  U_{\mathbf{x}_2}[+\infty, x_2^+] A^-(x_2^+, 0^-, \mathbf{x}_2) U_{\mathbf{x}_2}[x_2^+, -\infty]
\end{equation}
We consider two options for the ordering of $x_2^+$ with respect to $[-L^+, L^+]$. One is that $x_2^+ > L^+$ and the other is $x_2^+ <-L^+$, corresponding to the two diagrams in Fig.~\ref{fig:evolution_extra_one_new_labels}, respectively.

Then Eq. \eqref{eq:UG[1]U}, in reference to Fig.~\ref{fig:evolution_extra_one_new_labels}, can be written as the sum of the two cases
\begin{equation}
\begin{split}
\mathcal{M}_{\rm{I}}=&\frac{2g^2}{2k^+} \epsilon^{ij}\int_{-\infty}^{+\infty} dx_1^+\int_{L^+}^{+\infty} dx_2^+  \Big\langle  \mathrm{Tr}\Big[ U_{\mathbf{x}_1}[+\infty, x_1^+] \left(\partial^i A^j +ig[a^i, A^j]\right)U_{\mathbf{x}_1}[x_1^+, -\infty] \\
&\qquad \qquad\times U^{\dagger}_{\mathbf{x}_2}[x_2^+, -\infty] A^-(x_2^+, 0^-, \mathbf{x}_2)\Big]\Big\rangle\\
\end{split}
\end{equation}
and 
\begin{equation}
\begin{split}
\mathcal{M}_{\rm{II}}=&\frac{2g^2}{2k^+} \epsilon^{ij}\int_{-\infty}^{+\infty} dx_1^+\int_{-\infty}^{-L^+} dx_2^+  \Big\langle  \mathrm{Tr}\Big[ U_{\mathbf{x}_1}[+\infty, x_1^+] \left(\partial^iA^j +ig[a^i, A^j]\right)U_{\mathbf{x}_1}[x_1^+, -\infty]\\
&\qquad \qquad \times  A^-(x_2^+,0^-, \mathbf{x}_2)U^{\dagger}_{\mathbf{x}_2}[+\infty, x_2^+]\Big]\Big\rangle.
\end{split}
\end{equation}
The calculation of these two expressions follows very similar analysis.  We first calculate $\mathcal{M}_{\rm{I}}$, which can be written as
\begin{equation}\label{eq:M_1_explicit}
\begin{split}
\mathcal{M}_{\rm{I}}
=&\frac{2g^2}{2k^+} \epsilon^{ij}\int_{-\infty}^{+\infty} dx_1^+\int_{L^+}^{+\infty} dx_2^+  \Big\langle  \mathrm{Tr}\Big[ U_{\mathbf{x}_1}[+\infty, x_1^+] T^eU_{\mathbf{x}_1}[x_1^+, -\infty] U^{\dagger}_{\mathbf{x}_2}T^b\Big]\Big\rangle \\
&\qquad\times\left(\delta_{ed}\partial^i_{\mathbf{x}_1} -gf^{ecd}a_c^i(x_1^+, 0^-, \mathbf{x}_1)\right)\Big\langle  A_d^j(x_1^+, 0^-, \mathbf{x}_1) A^{-}_{b}(x_2^+, 0^-, \mathbf{x}_2)\Big\rangle .\\
\end{split}
\end{equation}
Because of $x_2^+>L^+$, we have set $U^{\dagger}_{\mathbf{x}_2}[x_2^+, -\infty] = U^{\dagger}_{\mathbf{x}_2}[+\infty, -\infty] \equiv U^{\dagger}_{\mathbf{x}_2}$.

The quantum flucations need to be averaged out by computing the two field correlation function in the background field
\begin{equation}\label{eq:prescription_corr}
\begin{split}
&\int_{L^+}^{+\infty} dx_2^+ \left\langle  A_d^j(x_1^+, 0^-, \mathbf{x}_1) A^{-}_{b}(x_2^+, 0^-, \mathbf{x}_2)\right\rangle\\
=&\int_{L^+}^{+\infty} dx_2^+\left\langle 0\left| A^{-}_{b}(x_2^+, 0^-, \mathbf{x}_2) \hat{S}(L^+, x_1^+) A_d^j(x_1^+, 0^-, \mathbf{x}_1) \right|0\right\rangle .\\
\end{split}
\end{equation}
The interaction with the shockwave only happens within the range $[L^+, x_1^+]$. 

Substituting the mode expansions for the fields $A^j$ and $A^+$ and using  
eikonal transformation of gluon creation operator
\begin{equation}
\hat{S}(L^+, x_1^+)\hat{a}^{\dagger}_{d, \lambda_1}(p_1, \mathbf{p}_1)\hat{S}^{\dagger}(L^+, x_1^+) = \int d^2\mathbf{w}_1 e^{i\mathbf{p}_1\cdot\mathbf{w}_1}  \hat{a}^{\dagger}_{h, \lambda_1}(p_1^-, \mathbf{w}_1)U^{hd}_{\mathbf{w}_1}(L^+, x_1^+), 
\end{equation}
 the two field correlator in eq. \eqref{eq:prescription_corr} can be computed as 
\begin{equation}\label{eq:two_field_correlator}
\begin{split}
&\int_{L^+}^{+\infty} dx_2^+ \Big\langle 0 \Big|A^{-}_{ b}(x_2^+, 0^-, \mathbf{x}_2) \hat{S}(L^+, x_1^+) A^{j}_d(x_1^+, 0^-, \mathbf{x}_1)\Big|0\Big\rangle \\
=&\int_{L^+}^{+\infty} dx_2^+ \sum_{\lambda_1, \lambda_2} \int_{p_1^+, p_2^+, \mathbf{p}_1, \mathbf{p}_2}  \Big[ e^{-i\frac{\mathbf{p}_2^2}{2p_2^+}x^+_2}e^{i\mathbf{p}_2\cdot\mathbf{x}_2}e^{i\frac{\mathbf{p}_1^2}{2p_1^+}x_1^+}e^{-i\mathbf{p}_1\cdot\mathbf{x}_1}\varepsilon^-_{\lambda_2}(p_2^+, \mathbf{p}_2)\varepsilon^{j\ast}_{\lambda_1}(p_1^+, \mathbf{p}_1)\\
&\times \int d\mathbf{w}_2 d\mathbf{w}_1 e^{-i\mathbf{p}_2\cdot\mathbf{w}_2} e^{i\mathbf{p}_1\cdot\mathbf{w}_1} U^{hd}_{\mathbf{w}_1}(L^+, x_1^+)\left[\delta_{bh}\, \delta_{\lambda_1\lambda_2} (2\pi)2p_1^+ \delta(p_1^+-p_2^+)  \delta^{(2)}(\mathbf{w}_1-\mathbf{w}_2)\right]\\
=&\int_{\mathbf{p}_2, \mathbf{p}_1, p_1^+} \Big[ e^{-i\frac{\mathbf{p}_2^2}{2p_1^+}L^+}e^{i\mathbf{p}_2\cdot\mathbf{x}_2}e^{i\frac{\mathbf{p}_1^2}{2p_1^+}x_1^+}e^{-i\mathbf{p}_1\cdot\mathbf{x}_1} (-i)\frac{2\mathbf{p}_2^j}{\mathbf{p}_2^2}\int  d\mathbf{w}_1 e^{-i\mathbf{p}_2\cdot\mathbf{w}_1} e^{i\mathbf{p}_1\cdot\mathbf{w}_1} U^{bd}_{\mathbf{w}_1}(L^+, x_1^+)\Big]\\
=&  \int_{p_1^+} e^{i\frac{\partial^2_{\mathbf{x}_2}}{2p_1^+}L^+} e^{-i\frac{\partial^2_{\mathbf{x}_1}}{2p_1^+}x^+_1} \frac{2}{2\pi}\frac{(\mathbf{x}_2-\mathbf{x}_1)^j}{|\mathbf{x}_2-\mathbf{x}_1|^2}  U^{bd}_{\mathbf{x}_1}(L^+, x_1^+)\\
\simeq &\int_{\delta k^+} \frac{dp_1^+}{p_1^+} \frac{1}{(2\pi)^2}\frac{(\mathbf{x}_2-\mathbf{x}_1)^j}{|\mathbf{x}_2-\mathbf{x}_1|^2}  U^{bd}_{\mathbf{x}_1}(L^+, x_1^+).
\end{split}
\end{equation}
In obtaining the last equality, we have ignored the phase factors as they will introduce higher order eikonality contributions. The integration over the longitudinal momentum is restricted within the infinitesimal range $\delta k^+$. In obtaining the second equality, we have carried out the integrals of $p_2^+$ and $x_2^+$. The integration over $x_2^+$ is given by 
\begin{equation}
\int_{L^+}^{+\infty}dx_2^+ e^{-i\frac{\mathbf{p}_2^2}{2p_2^+}x^+_2} =-i \frac{2p_2^+}{\mathbf{p}_2^2}e^{-i\frac{\mathbf{p}_2^2}{2p_2^+}L^+}.
\end{equation}
Taking into account of eq.~\eqref{eq:two_field_correlator}, the Wilson line structure in eq.~\eqref{eq:M_1_explicit} can be simplified as 
\begin{equation}
\begin{split}
&\mathrm{Tr}\Big[ T^bU_{\mathbf{x}_1}[+\infty, x_1^+] T^eU_{\mathbf{x}_1}[x_1^+, -\infty] U^{\dagger}_{\mathbf{x}_2}\Big] \partial_{\mathbf{x}_1}^i U^{be}_{\mathbf{x}_1}[+\infty, x_1^+]\\
=&N_c\mathrm{Tr}\Big[ \partial_{\mathbf{x}_1}^i U_{\mathbf{x}_1}[+\infty, x_1^+]U_{\mathbf{x}_1}[x_1^+, -\infty] U^{\dagger}_{\mathbf{x}_2}\Big]-\mathrm{Tr}\Big[ T^b\partial_{\mathbf{x}_1}^iU_{\mathbf{x}_1}[+\infty, x_1^+]  U_{\mathbf{x}_1}[ x_1^+, -\infty]T^d U^{\dagger}_{\mathbf{x}_2}\Big] U_{\mathbf{x}_1}^{bd}\\
=&\frac{1}{2}N_c\mathrm{Tr}\Big[ \partial_{\mathbf{x}_1}^i U_{\mathbf{x}_1}[+\infty, x_1^+]U_{\mathbf{x}_1}[x_1^+, -\infty] U^{\dagger}_{\mathbf{x}_2}\Big].
\end{split}
\end{equation}
We have used the color identity $T^dT^eT^d = \frac{1}{2}N_c T^e$ in obtaining the last equality.  On the other hand, one has 
\begin{equation}
\begin{split}
&\mathrm{Tr}\Big[T^b U_{\mathbf{x}_1}[+\infty, x_1^+] T^eU_{\mathbf{x}_1}[x_1^+, -\infty] U^{\dagger}_{\mathbf{x}_2}\Big]f^{ecd} U^{bd}_{\mathbf{x}_1}[+\infty, x_1^+]\\
=&\frac{1}{2} iN_c \mathrm{Tr}\Big[U_{\mathbf{x}_1}[+\infty, x_1^+]T^cU_{\mathbf{x}_1}[x_1^+, -\infty] U^{\dagger}_{\mathbf{x}_2}\Big].
\end{split}
\end{equation}
The final result for $\mathcal{M}_{\rm{I}}$ is 
\begin{equation}
\mathcal{M}_{\rm{I}}
=2g^2N_c\int_{\delta k^+} \frac{dp_1^+}{p_1^+} \frac{1}{(2\pi)^2}\frac{\epsilon^{ij}(\mathbf{x}_2-\mathbf{x}_1)^j}{|\mathbf{x}_2-\mathbf{x}_1|^2}\frac{1}{2k^+} \frac{1}{2} \int_{-\infty}^{+\infty} dx_1^+\mathrm{Tr}\Big[  U_{\mathbf{x}_1}[+\infty, x_1^+]\overleftarrow{\mathcal{D}}^i_{\mathbf{x}_1}U_{\mathbf{x}_1}[x_1^+, -\infty] U^{\dagger}_{\mathbf{x}_2}\Big].
\end{equation}

We carry out similar analysis for the second diagram in Fig.~\ref{fig:evolution_extra_one_new_labels}. Its expression is
\begin{equation}\label{eq:M_II_intermediate}
\begin{split}
\mathcal{M}_{\rm{II}}=&\frac{2g^2P^+}{2k^+} \epsilon^{ij}\int_{-\infty}^{+\infty} dx_1^+\int_{-\infty}^{-L^+} dx_2^+  \Big\langle  \mathrm{Tr}\Big[ U_{\mathbf{x}_1}[+\infty, x_1^+] T^eU_{\mathbf{x}_1}[x_1^+, -\infty]T^bU^{\dagger}_{\mathbf{x}_2}[+\infty, x_2^+]\Big]\Big\rangle\\
&\qquad \qquad \times \left(\partial^i_{\mathbf{x}_1}\delta^{de} -gf^{ecd}a^i_c(x_1^+, 0^-,\mathbf{x}_1)\right)\Big\langle A^-_b(x_2^+,0^-, \mathbf{x}_2)A^j_d(x_1^+, 0^-, \mathbf{x}_1)\Big\rangle.
\end{split}
\end{equation}
The two field correlation function with the ordering $x_2^+<x_1^+$ is computed by 
\begin{equation}\label{eq:two_field_corr_second}
\begin{split}
&\int_{-\infty}^{-L^+} dx_2^+ \Big\langle A^-_b(x_2^+,0^-, \mathbf{x}_2)A^j_d(x_1^+, 0^-, \mathbf{x}_1)\Big\rangle\\
=&\int_{-\infty}^{-L^+} dx_2^+ \int_{p_1^+, p_2^+, \mathbf{p}_1, \mathbf{p}_2} \Big[ e^{-i\frac{\mathbf{p}_1^2}{2p_1^+}x^+_1}e^{i\mathbf{p}_1\cdot\mathbf{x}_1}  e^{i\frac{\mathbf{p}_2^2}{2p_2^+}x_2^+}e^{-i\mathbf{p}_2\cdot\mathbf{x}_2}\frac{\mathbf{p}_2^j}{p_2^+} \int d\mathbf{w}_1 e^{-i\mathbf{p}_1\cdot\mathbf{w}_1} e^{i\mathbf{p}_2\cdot\mathbf{w}_1} \\
&\qquad \times (2\pi) 2p_1^+\delta(p_2^+-p_1^+)  U^{db}_{\mathbf{w}_1}(x_1^+, -L^+)\Big]\\
=& \int_{p_1^+} e^{i\frac{\partial^2_{\mathbf{x}_1}}{2p_1^+}x^+_1}e^{i\frac{\partial^2_{\mathbf{x}_2}}{2p_2^+}L^+} \frac{2}{2\pi} \frac{(\mathbf{x}_1-\mathbf{x}_2)^j}{|\mathbf{x}_1-\mathbf{x}_2|^2}  U^{db}_{\mathbf{x}_1}(x_1^+, -L^+)\\
\simeq & -\int_{\delta k^+}\frac{dp_1^+}{p_1^+}\frac{1}{(2\pi)^2} \frac{(\mathbf{x}_2-\mathbf{x}_1)^j}{|\mathbf{x}_2-\mathbf{x}_1|^2}  U^{db}_{\mathbf{x}_1}(x_1^+, -L^+).\\
\end{split}
\end{equation}
Compared to the corrrelation function obtained in eq. \eqref{eq:two_field_correlator}, an extra minus sign shows up, apart from the difference in Wilson line color indices. 
We used the integration over $x_2^+$, this time from $-\infty$ to $-L^+$. 
\begin{equation}
\int_{-\infty}^{-L^+} dx_2^+ e^{i\frac{\mathbf{p}_2^2}{2p_2^+}x_2^+} = -i\frac{2p_2^+}{\mathbf{p}_2^2} e^{-i\frac{\mathbf{p}_2^2}{2p_2^+}L^+}. 
\end{equation}
The two Wilson line structures in $\mathcal{M}_{\rm{II}}$ become 
\begin{equation}
\begin{split}
&\mathrm{Tr}\Big[ U_{\mathbf{x}_1}[+\infty, x_1^+] T^eU_{\mathbf{x}_1}[x_1^+, -\infty]T^bU^{\dagger}_{\mathbf{x}_2}\Big]\partial_{\mathbf{x}_1}^iU_{\mathbf{x}_1}^{eb}(x_1^+, -\infty)\\
=&N_c\mathrm{Tr}\Big[ U_{\mathbf{x}_1}[+\infty, x_1^+] \partial_{\mathbf{x}_1}^i U_{\mathbf{x}_1}[x_1^+,-\infty]U^{\dagger}_{\mathbf{x}_2}\Big]-\mathrm{Tr}\Big[ T^hU_{\mathbf{x}_1}[+\infty, x_1^+]\partial^i_{\mathbf{x}_1}U_{\mathbf{x}_1}[x_1^+, -\infty]T^bU^{\dagger}_{\mathbf{x}_2}\Big] U_{\mathbf{x}_1}^{hb}\\
=&\frac{1}{2}N_c\mathrm{Tr}\Big[ U_{\mathbf{x}_1}[+\infty, x_1^+] \partial_{\mathbf{x}_1}^i U_{\mathbf{x}_1}[x_1^+,-\infty]U^{\dagger}_{\mathbf{x}_2}\Big]
\end{split}
\end{equation}
and 
\begin{equation}
\begin{split}
&\mathrm{Tr}\Big[ U_{\mathbf{x}_1}[+\infty, x_1^+] T^eU_{\mathbf{x}_1}[x_1^+, -\infty]T^bU^{\dagger}_{\mathbf{x}_2}\Big]U_{\mathbf{x}_1}^{db}(x_1^+, -\infty) f^{ecd}\\
=&-\frac{1}{2} iN_c\mathrm{Tr}\Big[ U_{\mathbf{x}_1}[+\infty, x_1^+] T^c U_{\mathbf{x}_1}[x_1^+, -\infty]U^{\dagger}_{\mathbf{x}_2}\Big].
\end{split}
\end{equation}
The final result for $\mathcal{M}_{\rm{II}}$ is 
\begin{equation}
\mathcal{M}_{\rm{II}}
=-2g^2N_c\int_{\delta k^+} \frac{dp_1^+}{p_1^+} \frac{1}{(2\pi)^2}\frac{\epsilon^{ij}(\mathbf{x}_2-\mathbf{x}_1)^j}{|\mathbf{x}_2-\mathbf{x}_1|^2}\frac{1}{2k^+} \frac{1}{2} \int_{-\infty}^{+\infty} dx_1^+\mathrm{Tr}\Big[  U_{\mathbf{x}_1}[+\infty, x_1^+]\overrightarrow{\mathcal{D}}^i_{\mathbf{x}_1}U_{\mathbf{x}_1}[x_1^+, -\infty] U^{\dagger}_{\mathbf{x}_2}\Big]
\end{equation}

Summing up $\mathcal{M}_{\rm{I}}$ and $\mathcal{M}_{\rm{II}}$, one obtains 
\begin{equation}\label{eq:M1+M_2_final}
\mathcal{M}_{\rm{I}}+\mathcal{M}_{\rm{II}}=\frac{2\alpha_s N_c \Delta y}{\pi}\frac{\epsilon^{ij}(\mathbf{x}_1-\mathbf{x}_2)^j}{|\mathbf{x}_1-\mathbf{x}_2|^2}\Big\langle \mathrm{Tr}\left[  U^{i \rm{G}[2]}_{\mathbf{x}_1} U^{\dagger}_{\mathbf{x}_2}\right]\Big\rangle .
\end{equation}
Eq.~\eqref{eq:M1+M_2_final} is the main result of this section. It characterizes that gluon radiation inside the shockwave contributes to 
the rapidity evolution of chromo-magnetically polarized Wilson line correlator. Interestingly, the contribution is in the form of chromo-electrically polarized Wilson line correlator.  It is a single logarithmic contribution $\int_{\delta k^+} dp_1^+/p_1^+ = \Delta y = \ln \frac{1}{x}$.

\subsubsection{Directly calculating the diagrams}
In this section, we directly calculate the two diagrams in Fig.~\ref{fig:evolution_extra_one_new_labels}.
The scattering amplitude for two incoming gluons and two outgoing gluons is calculated by  
\begin{equation}\label{eq:direct_start}
\begin{split}
&\langle 0 | \hat{a}_{c', \lambda'}(p_2^{\prime +}, \mathbf{x}'_2) \hat{a}_{c', \lambda'}(p_1^{\prime +}, \mathbf{x}'_1) \hat{S}(+\infty, -\infty) \hat{a}_{c, \lambda_2}^{\dagger}(p_2^+, \mathbf{x}_2) \hat{a}_{c, \lambda_1}^{\dagger}(p_1^+, \mathbf{x}_1)|0\rangle \\
=&\langle 0 | \hat{a}_{c', \lambda'}(p_2^{\prime +}, \mathbf{x}'_2) \hat{a}_{c', \lambda'}(p_1^{\prime +}, \mathbf{x}'_1) \hat{S}(+\infty, L^+)\hat{S}(L^+, -L^+) \hat{a}_{c, \lambda_2}^{\dagger}(p_2^+, \mathbf{x}_2) \hat{a}_{c, \lambda_1}^{\dagger}(p_1^+, \mathbf{x}_1)|0\rangle \\
&+\langle 0 | \hat{a}_{c', \lambda'}(p_2^{\prime +}, \mathbf{x}'_2) \hat{a}_{c', \lambda'}(p_1^{\prime +}, \mathbf{x}'_1) \hat{S}(L^+, -L^+) \hat{S}(-L^+, -\infty)\hat{a}_{c, \lambda_2}^{\dagger}(p_2^+, \mathbf{x}_2) \hat{a}_{c, \lambda_1}^{\dagger}(p_1^+, \mathbf{x}_1)|0\rangle \\
\end{split}
\end{equation}
The incoming two gluons have the same color indices. The same is true for the two outgoing gluons. The two outgoing gluons also have the same polarization. Repeated indices are summed over. For generality, we keep all the longitudinal momentum and transverse coordinates different.

The first term in eq.~\eqref{eq:direct_start}, only considering the part corresponding to the first diagram in Fig.~\ref{fig:evolution_extra_one_new_labels}, is further expressed by
\begin{equation}
\begin{split}
&\langle 0 | \hat{a}_{c', \lambda'}(p_2^{\prime +}, \mathbf{x}'_2) \hat{a}_{c', \lambda'}(p_1^{\prime +}, \mathbf{x}'_1) \hat{S}(+\infty, L^+)\hat{S}(L^+, -L^+) \hat{a}_{c, \lambda_2}^{\dagger}(p_2^+, \mathbf{x}_2) \hat{a}_{c, \lambda_1}^{\dagger}(p_1^+, \mathbf{x}_1)|0\rangle\\
=&\sum_{e, e_2, \kappa, \kappa_2}\int_{q^+, q_2^+, \mathbf{y}_2, \mathbf{y}} \langle 0 | \hat{a}_{c', \lambda'}(p_2^{\prime +}, \mathbf{x}'_2) \hat{S}(+\infty, L^+) \hat{a}^{\dagger}_{e_2, \kappa_2}(q^+, \mathbf{y}_2) \hat{a}^{\dagger}_{e, \kappa}(q^+, \mathbf{y})|0\rangle\\
&\qquad \times  \langle 0| \hat{a}_{e_2, \kappa_2}(q_2^+, \mathbf{y}_2)\hat{W}(L^+, -L^+) \hat{a}_{c, \lambda_2}^{\dagger}(p_2^+, \mathbf{x}_2) |0\rangle\\
&\qquad \times  \langle 0| \hat{a}_{e, \kappa}(q^+, \mathbf{y}) \hat{a}_{c', \lambda'}(p_1^{\prime +}, \mathbf{x}'_1) \hat{S}(L^+, -L^+)  \hat{a}_{c, \lambda_1}^{\dagger}(p_1^+, \mathbf{x}_1)|0\rangle\\
=&\sum_{e,  \kappa, }\int_{q^+,  \mathbf{y}} e^{i(p_2^{\prime -} -p_2^- - q^-+i\epsilon)L^+} \Big[\psi^{g\rightarrow gg}_I(\{p_2^{\prime +}, \mathbf{x}'_2, c', \lambda'\}; \{q^+, \mathbf{y}, e, \kappa\}, \{p_2^+, \mathbf{x}_2, h, \lambda_2\}) \Big]^{\ast} \\
&\qquad \times U^{hc}_{\mathbf{x}_2}(L^+, -L^+)  M^{g\rightarrow gg}_3(\{p_1^+, \mathbf{x}_1, c, \lambda_1\}; \{q^+, \mathbf{y}, e, \kappa\}, \{p_1^{\prime +}, \mathbf{x}'_1, c', \lambda'\})
\end{split}
\end{equation}
We have expanded $\hat{S}(+\infty, L^+)$ to linear order in strong coupling constant.
The amplitude $M^{g\rightarrow gg}_3$ has been computed in eq.~\eqref{eq:M3_gluon_inside_sw},  representing background field induced gluon radiation. 
Note that we let the gluon $\{q^+, \mathbf{y}, e, \kappa\}$ be the soft gluon. In the limit that $q^+ \ll p_1^+$, the longitudinal momentum conservation leads to $p_1^+ = p_1^{\prime +}$ as expected for sub-eikonal order processes. We only kept the part of the polarization factor that will eventually give terms linear in the polarization of the incoming gluons.  The initial state gluon splitting wavefunction $\psi^{g\rightarrow gg}_I$ has also been computed in eq.~\eqref{eq:gluon_splitting_coordinate}. In the limit that $q^+\ll p_2^+$, the longitudinal momentum conservation enforces that $p_2^+ = p^{\prime +}_2$. 

Using these explicit expressions, $\mathcal{M}_{\rm{I}}$ becomes
\begin{equation}
\begin{split}
\mathcal{M}_{\rm{I}}
 =&\sum_{e,  \kappa, }\int_{q^+} e^{i(p_2^{\prime -} -p_2^- - q^-+i\epsilon)L^+} (-2\lambda_1 \delta_{\lambda_1\lambda_2})g^2 \frac{i}{2\pi} \frac{i\epsilon^{ij}(\mathbf{x}_1-\mathbf{x}_2)^j}{|\mathbf{x}_1-\mathbf{x}_2|^2} T^e_{c'h}U^{hc}_{\mathbf{x}_2}(L^+, -L^+)\\
&\times \frac{1}{2p_1^+}\int_{-L^+}^{L^+} dw^+ \Big( if^{de'h'}\left[ -2\partial_{\mathbf{x}_1}^iU_{\mathbf{x}_1}^{ee'}(L^+, w^+) U_{\mathbf{x}_1}^{c'h'}(L^+, w^+) U_{\mathbf{x}_1}^{dc}(w^+, -L^+) \right]\\
 & + iga^i_b (\mathbf{x}_1)U_{\mathbf{x}_1}^{ee'}(L^+, w^+) U_{\mathbf{x}_1}^{c'h'}(L^+, w^+) U_{\mathbf{x}_1}^{dc}(w^+, -L^+) 2(T^{d}T^{e'})_{h'b}\Big)\\
\end{split}
\end{equation}
We have discarded the factors characterizing longitudinal momentum conservation and transverse coordinate conservation. The Wilson line structures are further simplified 
\begin{equation}
\begin{split}
&T^e_{c'h}U^{hc}_{\mathbf{x}_2}(L^+, -L^+)if^{de'h'}\left[ -2\partial_{\mathbf{x}_1}^iU_{\mathbf{x}_1}^{ee'}(L^+, w^+) U_{\mathbf{x}_1}^{c'h'}(L^+, w^+) U_{\mathbf{x}_1}^{dc}(w^+, -L^+) \right]\\
=&-N_c\mathrm{Tr}\left[\partial^i_{\mathbf{x}_1}U_{\mathbf{x}_1}(L^+, w^+)U_{\mathbf{x}_1}(w^+, -L^+)U^{\dagger}_{\mathbf{x}_2}(L^+, -L^+)\right] \\
\end{split}
\end{equation}
and 
\begin{equation}
\begin{split}
& iga^i_b (\mathbf{x}_1) U_{\mathbf{x}_1}^{ee'}(L^+, w^+) U_{\mathbf{x}_1}^{c'h'}(L^+, w^+) U_{\mathbf{x}_1}^{dc}(w^+, -L^+) 2(T^{d}T^{e'})_{h'b} T^e_{c'h}U^{hc}_{\mathbf{x}_2}(L^+, -L^+)\\
=&N_c \mathrm{Tr}[U_{\mathbf{x}_1}(L^+, w^+) iga^i(\mathbf{x}_1) U_{\mathbf{x}_1}(w^+, -L^+)U^{\dagger}_{\mathbf{x}_2}(L^+, -L^+)].\\
\end{split}
\end{equation}
The final result for $\mathcal{M}_{\rm{I}}$ is 
\begin{equation}
\begin{split}
\mathcal{M}_{\rm{I}} =&\lambda_1\delta_{\lambda_1\lambda_2}  g^2N_c \frac{\Delta y}{2\pi}  \frac{i}{2\pi} \frac{i\epsilon^{ij}(\mathbf{x}_1-\mathbf{x}_2)^j}{|\mathbf{x}_1-\mathbf{x}_2|^2}\\
&\qquad \times \frac{1}{2p^+}\int_{-L^+}^{L^+}dw^+ \mathrm{Tr}\left[U_{\mathbf{x}_1}(L^+, w^+)\overleftarrow{\mathcal{D}}^i_{\mathbf{x}_1}U_{\mathbf{x}_1}(w^+, -L^+)U^{\dagger}_{\mathbf{x}_2}(L^+, -L^+)\right]. 
\end{split} 
\end{equation}
The integration over longitudinal momentum gives $\int dq^+/q^+ = \Delta y$. 

We analyze the second term in eq.~\eqref{eq:direct_start}, corresponding to the second diagram in  Fig.~\ref{fig:evolution_extra_one_new_labels}.
\begin{equation}
\begin{split}
&\langle 0 | \hat{a}_{c', \lambda'}(p_2^{\prime +}, \mathbf{x}'_2) \hat{a}_{c', \lambda'}(p_1^{\prime +}, \mathbf{x}'_1) \hat{S}(L^+, -L^+) \hat{S}(-L^+, -\infty)\hat{a}_{c, \lambda_2}^{\dagger}(p_2^+, \mathbf{x}_2) \hat{a}_{c, \lambda_1}^{\dagger}(p_1^+, \mathbf{x}_1)|0\rangle\\
=&\sum_{e, e_2, \kappa, \kappa_2}\int_{q^+, q_2^+, \mathbf{y}_2, \mathbf{y}}\langle 0 | \hat{a}_{c', \lambda'}(p_2^{\prime +}, \mathbf{x}'_2)\hat{W}(L^+, -L^+)  \hat{a}^{\dagger}_{e_2, \kappa_2}(q^+_2, \mathbf{y}_2)|0\rangle \\
&\qquad \times \langle 0| \hat{a}_{c', \lambda'}(p_1^{\prime +}, \mathbf{x}'_1) \hat{S}(L^+, -L^+) \hat{a}_{c, \lambda_1}^{\dagger}(p_1^+, \mathbf{x}_1)\hat{a}^{\dagger}_{e,\kappa}(q^+,\mathbf{y}) |0\rangle\\
&\qquad \times  \langle 0| \hat{a}_{e,\kappa}(q^+,\mathbf{y}) \hat{a}_{e_2, \kappa_2}(q^+_2, \mathbf{y}_2)\hat{S}(-L^+, -\infty)\hat{a}_{c, \lambda_2}^{\dagger}(p_2^+, \mathbf{x}_2) |0\rangle\\
=&\sum_{e, e_2, \kappa}\int_{q^+,\mathbf{y}}e^{i(p_2^- - q^- -p_2^{\prime -}+i\epsilon)L^+}U^{c'e_2}_{\mathbf{x}'_2}(L^+, -L^+) \psi^{g\rightarrow gg}_{I} (\{p_2^+, \mathbf{x}_2, c, \lambda_2\}; \{q^+, \mathbf{y}, e, \kappa\}, \{ p_2^{\prime +}, \mathbf{x}'_2, e_2, \lambda'\})\\
&\qquad \times \Big[M_3^{g\rightarrow gg}(\{p_1^{\prime +}, \mathbf{x}'_1, c', \lambda'\}; \{q^+, \mathbf{y}, e, \kappa\}, \{p_1^+, \mathbf{x}_1, c, \lambda_1\})\Big|_{L^+\leftrightarrow -L^+}\Big]^{\ast}
\end{split}
\end{equation}
Note that one has to exchange the role of $L^+$ and $-L^+$ when using the expression of $M_3^{g\rightarrow gg}$ calculated before. 
Using these explicit expressions, $\mathcal{M}_{\rm{II}}$ becomes
\begin{equation}
\begin{split}
\mathcal{M}_{\rm{II}} 
=&\sum_{e, e_2, \kappa}\int_{q^+}e^{i(p_2^- - q^- -p_2^{\prime -}+i\epsilon)L^+}  g^2(-2\lambda_1\delta_{\lambda_1\lambda_2}) \frac{i}{2\pi} \frac{i\epsilon^{ij}(\mathbf{x}_1-\mathbf{x}_2)^j}{|\mathbf{x}_1-\mathbf{x}_2|^2}U^{c'e_2}_{\mathbf{x}_2}(L^+, -L^+)T^{e}_{e_2c}\\
&\times \frac{1}{2p_1^{ +}} \int_{-L^+}^{L^+} dw^+ \Big(if^{h'e'h}\left[-2\partial^i_{\mathbf{x}_1}U_{\mathbf{x}_1}^{e'e}(w^+, -L^+) U_{\mathbf{x}_1}^{hc}(w^+, -L^+)U^{c'h'}_{\mathbf{x}_1} (L^+, w^+) \right]\\
&\qquad +iga_b^i(\mathbf{x}_1) U_{\mathbf{x}_1}^{e'e}(w^+, -L^+) U_{\mathbf{x}_1}^{hc}(w^+, -L^+)U^{c'h'}_{\mathbf{x}_1} (L^+, w^+)2(T^{h'}T^{e'})_{hb}\Big)
\end{split}
\end{equation}
Again, the Wilson line structures can be simplified as 
\begin{equation}
\begin{split}
&if^{h'e'h}\left[-2\partial^i_{\mathbf{x}_1}U_{\mathbf{x}_1}^{e'e}(w^+, -L^+) U_{\mathbf{x}_1}^{hc}(w^+, -L^+)U^{c'h'}_{\mathbf{x}_1} (L^+, w^+) \right]U^{c'e_2}_{\mathbf{x}_2}(L^+, -L^+)T^{e}_{e_2c}\\
=&N_c \mathrm{Tr}[U_{\mathbf{x}_1}(L^+, w^+)\partial^i_{\mathbf{x}_1}U_{\mathbf{x}_1}(w^+, -L^+)U^{\dagger}_{\mathbf{x}_2}] \\
\end{split}
\end{equation}
and
\begin{equation}
\begin{split}
&iga_b^i(\mathbf{x}_1) U_{\mathbf{x}_1}^{e'e}(w^+, -L^+) U_{\mathbf{x}_1}^{hc}(w^+, -L^+)U^{c'h'}_{\mathbf{x}_1} (L^+, w^+)2(T^{h'}T^{e'})_{hb}U^{c'e_2}_{\mathbf{x}_2}(L^+, -L^+)T^{e}_{e_2c}\\
=&N_c \mathrm{Tr}\left[U_{\mathbf{x}_1}(L^+, w^+)iga^i(\mathbf{x}_1) U_{\mathbf{x}_1}(w^+, -L^+)U^{\dagger}_{\mathbf{x}_2}(L^+, -L^+)\right]
\end{split}
\end{equation}
The final result for the second amplitude is 
\begin{equation}
\begin{split}
\mathcal{M}_{\rm{II}}& =-\lambda_1\delta_{\lambda_1\lambda_2} g^2N_c \frac{\Delta y}{2\pi}  \frac{i}{2\pi} \frac{i\epsilon^{ij}(\mathbf{x}_1-\mathbf{x}_2)^j}{|\mathbf{x}_1-\mathbf{x}_2|^2} \frac{1}{2p^+_1}\int_{-L^+}^{L^+} dw^+  \mathrm{Tr}[U_{\mathbf{x}_1}(L^+, w^+)\overrightarrow{\mathcal{D}}^i_{\mathbf{x}_1}U_{\mathbf{x}_1}(w^+, -L^+)U^{\dagger}_{\mathbf{x}_2}].
\end{split}
\end{equation}
Combining the two amplitudes 
\begin{equation}
\begin{split}
&\mathcal{M}_{\rm{I}} + \mathcal{M}_{\rm{II}}\\
=&-\lambda_1\delta_{\lambda_1\lambda_2} g^2N_c \frac{\Delta y}{2\pi}  \frac{i}{2\pi} \frac{i\epsilon^{ij}(\mathbf{x}_1-\mathbf{x}_2)^j}{|\mathbf{x}_1-\mathbf{x}_2|^2} \frac{1}{2p^+_1}\int_{-L^+}^{L^+} dw^+  \mathrm{Tr}\left[U_{\mathbf{x}_1}(L^+, w^+)(\overrightarrow{\mathcal{D}}^i_{\mathbf{x}_1}-\overleftarrow{\mathcal{D}}^i_{\mathbf{x}_1})U_{\mathbf{x}_1}(w^+, -L^+)U^{\dagger}_{\mathbf{x}_2}\right]\\
=&\lambda_1\delta_{\lambda_1\lambda_2} \frac{2\alpha_s N_c\Delta y}{\pi} \frac{\epsilon^{ij}(\mathbf{x}_1-\mathbf{x}_2)^j}{|\mathbf{x}_1-\mathbf{x}_2|^2}  \mathrm{Tr}[U^{i\rm{G}[2]}_{\mathbf{x}_1}U^{\dagger}_{\mathbf{x}_2}].\\
\end{split}
\end{equation}
The results of the two amplitudes coincide with eq.~\eqref{eq:M1+M_2_final}.
From the above expression, one can see that there is no tranvserse coordinate integration. Therefore, these two diagrams only contribute in the single logarithmic approximation \cite{Kovchegov:2021lvz}. 
\begin{figure}[!t]
    \centering
    \includegraphics[width=0.75\textwidth]{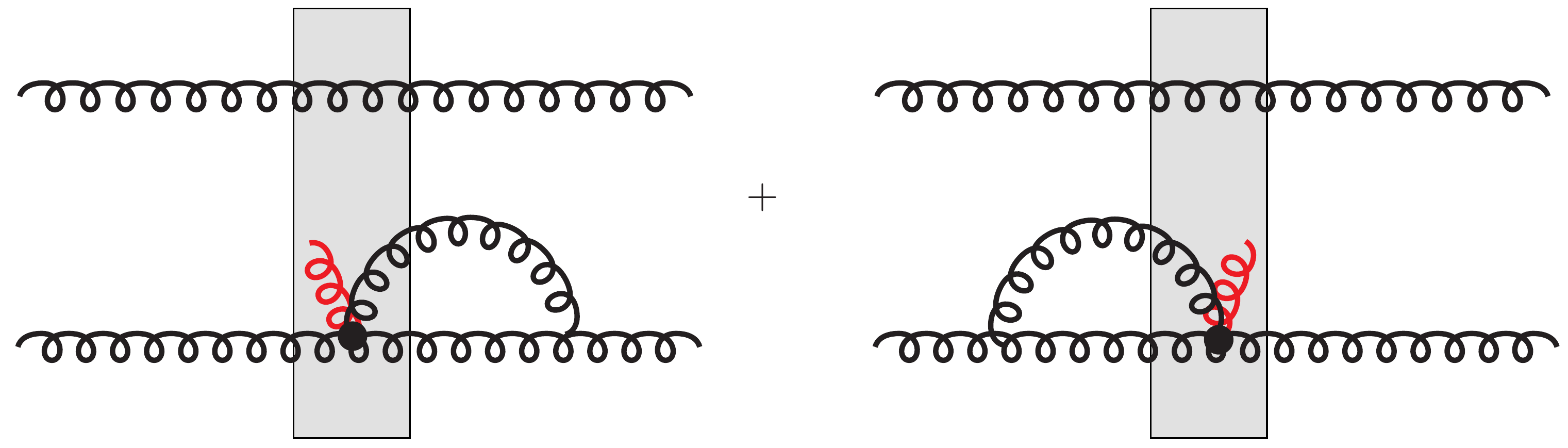}
    \caption{These two diagrams vanish. The red gluon lines represent the background gluon fields. }
	\label{fig:evolution_extra_two}
\end{figure}

One could also draw diagrams as shown in Fig.~\ref{fig:evolution_extra_two}. However, these two diagrams vanish because the background field induced triple gluon vertex is local in transverse coordinates while the soft gluon radiation is nonlocal in transverse coordinates.\\

We used two examples to show the importance of gluon radiation inside the shockwave. It contributes both to the longitudinal double-spin asymmetry of soft gluon production and to the rapidity evolution of polarized Wilson line correlator. It turns out that in both cases, the final result is related to chromo-electrically polarized Wilson line correlator $\langle \mathrm{Tr}[U^{iG[2]}_{\mathbf{x}}U^{\dagger}_{\mathbf{y}}]\rangle $. In \cite{Cougoulic:2022gbk}, it has been derived that the small $x$ limit of gluon helicity TMD is directly related to $\langle \mathrm{Tr}[U^{iG[2]}_{\mathbf{x}}U^{\dagger}_{\mathbf{y}}]\rangle$. 
\begin{equation}\label{eq:GluonHelicityTMD_UjG2}
\begin{split}
\Delta G_L(x, \mathbf{k}^2) 
=& \frac{4i}{g^2}\epsilon^{ij}\mathbf{k}^i\int d^2\mathbf{x} d^2\mathbf{y}  e^{-i\mathbf{k}_{\perp}\cdot(\mathbf{x}-\mathbf{y})}\Big\langle \mathrm{Tr}\left[U^{\dagger}_{\mathbf{y}}U^{j\rm{G}[2]}_{\mathbf{x}}(k^+)  - U^{j\rm{G}[2]\dagger }_{\mathbf{y}}(k^+) U_{\mathbf{x}}\right]\Big \rangle. \\
\end{split}
\end{equation}

One therefore conclude that the gluon radiation inside the shockwave manifesting itself in the form of the gluon polarized Wilson line correlator $\mathrm{Tr}[ U^{iG[2]}_{\mathbf{x}}U_{\mathbf{y}}^{\dagger}]$, being characterized by chromo-electric field $f^{-i}$, is related to the small $x$ limit of gluon helicity TMD .

\section{Summary}\label{sec:conclusion}
In this paper, we have derived the small-$x$ effective Hamiltonian for QCD in high energy within the shockwave formalism. The results are given in eqs.~\eqref{eq:V_0}, \eqref{eq:V_1/2}, \eqref{eq:V_1} valid up to sub-eikonal order. The straightforward continuation of the analysis to higher orders of eikonality is possible but probably very tedious. The use of a Hamiltonian approach to investigate high energy QCD at the eikonal order can be found in \cite{Kovner:2007zu,Kovner:2005pe}. 
We also established the approach to compute $S$-matrix elements up to sub-eikonal order in eq.~\eqref{eq:S_expanded_to_xi}.  As an application, various single quark/gluon scattering amplitudes, known as the polarized Wilson lines, are reproduced. This effective Hamiltonian approach, alternative to other approaches \cite{Balitsky:2015qba, Altinoluk:2020oyd, Chirilli:2018kkw, Cougoulic:2022gbk} has the advantage of directly isolating the relevant interactions up to sub-eikonal order and is particularly suitable to compute spin related observables at small $x$. 

The behavior of the eikonal interaction vertex eq.~\eqref{eq:V_0} under rapidity evolution had been carefully examined before, leading to the derivation of the JIMWLK renormalization group equation from a field theoretical approach in \cite{Jalilian-Marian:1997jhx, Jalilian-Marian:1997ubg, Iancu:2000hn, Ferreiro:2001qy}. The JIMWLK equation is general in the sense that it can be applied to any spin-independent observables at small $x$. For example, applying the JIMWLK equation on dipole correlator generates the Balitsky hierarchy and reproduces the BK equation in the large $N_c$ \cite{Balitsky:1995ub,Kovchegov:1999yj}. Currently, the small $x$ rapidity evolutions beyond eikonal order, particularly related to spin-dependent observables,  are analyzed in an observable-by-observable way. Since we have identified the relevant interaction vertices and propagators at the sub-eikonal order in the small $x$ effective Hamiltonian, it would be very interesting  though challenging to derive a general renormalization group equation that is valid at sub-eikonal order and automatically reproduces the evolution equations when being applied to different observables \cite{Cougoulic:2020tbc, Cougoulic:2019aja}. On the other hand, the small $x$ effective Hamiltonian approach offers the possibility to bridge high energy QCD to the general methodology in Hamiltonian formalism. As has been demonstrated that JIMWLK equation can be reproduced by the quantum Lindblad equation through quantum-classical correspondance \cite{Li:2020bys, Armesto:2019mna}. At the sub-eikonal order, the Hilbert space is enlarged as spin related degrees of freedoms start playing roles . It would also be very interesting to see if the Linblad formalism still applies to small $x$ helicity evolution using the small $x$ effective Hamiltonian.

One of the new features of small-$x$ effective Hamiltonian at sub-eikonal order is that gluons can be emitted inside the shockwave. This phenomenon has been barely discussed in the literature. We studied its effect in two situations: longitudinal double-spin asymmetry for soft gluon production in polarized collisions and rapidity evolution of polarized Wilson lines. In both cases, it is found that the contribution is given by the chromo-electrically polarized Wilson line correlator $\langle \mathrm{Tr}[U^{iG[2]}_{\mathbf{x}}U^{\dagger}_{\mathbf{y}}]\rangle $, which has been shown to be directly related to gluon helicity TMD in the small $x$ limit. It would be very interesting to see how gluon radiation inside the shockwave impacts the small $x$ rapidity evolutions and particle productions in phenomenological applications.

The small-$x$ effective Hamiltionian approach is developed within the light-cone quantization framework. As a result, it inevitably inherits the same zero mode problem \cite{Brodsky:1997de}. Recently, it was brought out that the chiral anomaly in polarized inclusive deep inelastic scatterings might be sensitive to zero modes \cite{Tarasov:2020cwl,Tarasov:2021yll} (see also \cite{Bhattacharya:2022xxw, Bhattacharya:2023wvy}), which could have been missed by the current approach. Nevertheless,
the small-$x$ effective Hamiltonian approach provides a systematic way of directly computing spin related observables including particle and jet productions in the small $x$ limit.  Of particular interest is the double-spin asymmetry for particle and jet productions in longitudinally polarized collisions relevant to experimental measurements at RHIC  \cite{STAR:2006opb, STAR:2007rjc, PHENIX:2008swq, STAR:2014wox, STAR:2013zyt, PHENIX:2014axc, PHENIX:2014gbf, PHENIX:2015fxo, STAR:2018iyz, PHENIX:2020trf, PHENIX:2016gzj, STAR:2018yxi, STAR:2019yqm, STAR:2021mfd, STAR:2021mqa, PHENIX:2022lgn}. This could include productions of neutral pion, $\eta$-meson, charged pions and $J/\psi$ at midrapidity, intermediate rapdity and forward rapidity, respectively. Double-spin asymmetry for direct photon production has also been measured at RHIC. Applications to deep inelastic scatterings relevant to future EIC would also be very intersting. We plan to study these observables in future works.

\acknowledgments
I thank Yuri Kovchegov for very helpful and inspiring discussions. I am grateful to Daniel Adamiak for discussions and checking many equations in the paper. I also appreciate very interesting conversations with Florian Cougoulic and Guillaume Beuf on related topics.   This material is based upon work supported by the U.S. Department of Energy, Office of Science, Office of Nuclear Physics under Award Number DE-SC0004286.

\appendix 
\section{Boost Transformations of Vector and Spinor Fields}\label{app:boost}
The vector fields and the spinor fields constitute different representations of the the Lorentz group. Their transformations under boost are obtained explicitly in this section.  The general Lorentz group representation is
\begin{equation}
U(\omega_{\mu\nu} ) = \mathrm{exp}\left\{ -\frac{i}{2} \omega_{\mu\nu} J^{\mu\nu}\right\}.
\end{equation}
Here $J^{\mu\nu} $ are the generators of the Lorentz group and $\omega_{\mu\nu}$ are the corresponding tranformation parameters. It is antisymmetric tensor. The generators satisfy the commutation relations
\begin{equation}
\Big[J^{\mu\nu}, J^{\rho\sigma}\Big] = i(g^{\nu\rho} J^{\mu\sigma} - g^{\mu\rho} J^{\nu\sigma} - g^{\nu\sigma}J^{\mu\rho} + g^{\mu\sigma} J^{\nu\rho}).
\end{equation}
For the vector representation, the generators have the following explicit expression
\begin{equation}
(J^{\mu\nu})^{\alpha}_{\,\,\beta} = i (g^{\mu\alpha} \delta^{\nu}_{\,\,\beta} - g^{\nu\alpha} \delta^{\mu}_{\,\, \beta}).
\end{equation}
We are interested in the boost along z-axis, the transformation is
\begin{equation}
U(\omega) = e^{-i\omega K^3}
\end{equation}
with 
\begin{equation}
K^3 = J^{03} = i\begin{pmatrix}
0 & 0 & 0 & 1\\
0 & 0 & 0 & 0\\
0 & 0 & 0 & 0 \\
1 & 0 & 0 &0\\
\end{pmatrix}.
\end{equation}
One obtains
\begin{equation}
U(\omega) \equiv \Lambda= \begin{pmatrix}
\cosh\omega & 0 & 0 & \sinh\omega \\
0 & 1 & 0 & 0 \\
0 & 0 & 1 & 0 \\
\sinh\omega & 0  & 0 & \cosh\omega \\
\end{pmatrix}.
\end{equation}
Under Lorentz boost, the gluon field transforms as $\widetilde{A}^{\mu} (x) = U(\omega) A^{\mu} (\Lambda^{-1} x)$. The explicit expressions for each component are
\begin{equation}\label{eq:Amu_boost_trans}
\begin{split}
&\widetilde{A}^+ = e^{\omega} A^+ (e^{-\omega} x^+, e^{\omega }x^-, \mathbf{x}_{\perp}),\\
&\widetilde{A}^- = e^{-\omega} A^- (e^{-\omega} x^+, e^{\omega }x^-, \mathbf{x}_{\perp}),\\
&\widetilde{A}^i = A^i( e^{-\omega} x^+, e^{\omega }x^-, \mathbf{x}_{\perp}).\\
\end{split}
\end{equation}
For the spinor representation
$U(\omega) = e^{-\frac{i}{2} \omega_{\mu\nu} S^{\mu\nu}}$
with the generators $
S^{\mu\nu} = \frac{i}{4} [\gamma^{\mu}, \gamma^{\nu}] $,
the boost operation has the explicit expression
\begin{equation}
U(\omega) = e^{-i\omega K^3}
\end{equation}
with 
\begin{equation}
K^3 = S^{03} = \frac{i}{2} \gamma^0\gamma^3.
\end{equation}
One can then obtain
\begin{equation}
U(\omega) = \sinh \frac{\omega}{2} \gamma^0\gamma^3 + \cosh\frac{\omega}{2}
= e^{\frac{\omega}{2}} \mathcal{P}_G + e^{-\frac{\omega}{2}} \mathcal{P}_B
\end{equation}
It is interesting to note that the good component  $\psi_G =\mathcal{P}_G\psi$ and the bad component $\psi_B = \mathcal{P}_B \psi$ transform differently under Lorentz boost. 
\begin{equation}
\begin{split}
&\widetilde{\psi}_G  = e^{\omega/2} \psi_G (e^{-\omega}x^+, e^{\omega} x^- ,\mathbf{x}_{\perp}) ,\\
&\widetilde{\psi}_B  = e^{-\omega/2} \psi_B (e^{-\omega}x^+, e^{\omega} x^- ,\mathbf{x}_{\perp}) .\\
\end{split}
\end{equation}
Using the above transformations, one can also compute the transformations of field strength tensor under Lorentz boost. 
For example, for $F^{+-} = \partial_- A^- - \partial_+ A^+ + ig[A^+, A^-]$,
one obtains 
\begin{equation}
\begin{split}
\widetilde{F}^{+-} =& \partial_- e^{-\omega} A^- (e^{-\omega}x^+, e^{\omega} x^-, \mathbf{x}_{\perp}) - \partial_+ e^{\omega} A^{+}(e^{-\omega}x^+, e^{\omega} x^-, \mathbf{x}_{\perp})\\
& + ig\Big[A^+(e^{-\omega}x^+, e^{\omega} x^-, \mathbf{x}_{\perp}), A^-(e^{-\omega}x^+, e^{\omega} x^-, \mathbf{x}_{\perp})\Big]\\
=&\tilde{\partial}_- A^{-}(\tilde{x}^+, \tilde{x}^-, \mathbf{x}_{\perp})-\tilde{\partial}_+ A^{+}(\tilde{x}^+, \tilde{x}^-, \mathbf{x}_{\perp}) + ig[ A^+(\tilde{x}^+, \tilde{x}^-, \mathbf{x}_{\perp}), A^-(\tilde{x}^+, \tilde{x}^-, \mathbf{x}_{\perp})]\\
=& F^{+-}(\tilde{x}^+, \tilde{x}^-, \mathbf{x}_{\perp})
\end{split}
\end{equation}
Similarly, one obtains
\begin{equation}
\begin{split}
&\widetilde{F}^{+-} = F^{+-}(\tilde{x}^+, \tilde{x}^-, \mathbf{x}_{\perp}),\\
&\widetilde{F}^{+i} = e^{\omega } F^{+i}(\tilde{x}^+, \tilde{x}^-, \mathbf{x}_{\perp}), \\
&\widetilde{F}^{-i} = e^{-\omega} F^{-i}(\tilde{x}^+, \tilde{x}^-, \mathbf{x}_{\perp}),\\
&\widetilde{F}^{ij} = F^{ij}(\tilde{x}^+, \tilde{x}^-, \mathbf{x}_{\perp}).
\end{split} 
\end{equation}

\section{Convention for Light-Cone Quantization}\label{app:LCQ}
The mode expansions for the dynamical fields expressed in terms of the corresponding creation and annihilation operators are
\begin{equation}\label{eq:mode_expansion}
\begin{split}
&A^{\mu}_a(x) = \int_0^{\infty} \frac{dp^+}{2p^+ (2\pi)} \int \frac{d^2\mathbf{p}}{(2\pi)^2} \sum_{\lambda} \left[e^{-ipx} \hat{a}_{a,\lambda}(p^+, \mathbf{p}) \varepsilon_{\lambda}^{\mu}(p^+, \mathbf{p})+ e^{ipx} \hat{a}^{\dagger}_{a,\lambda}(p^+, \mathbf{p}) \varepsilon_{\lambda}^{\ast\mu}(p^+, \mathbf{p})\right],\\
&\Psi_{i}(x) =  \int_0^{\infty} \frac{dk^+}{2k^+ (2\pi)} \int \frac{d^2\mathbf{k}}{(2\pi)^2}\sum_{\sigma}\left[e^{-ikx} \hat{b}_{i,\sigma}(k^+, \mathbf{k}) u_{\sigma}(k^+, \mathbf{k})+ e^{+ikx} \hat{d}^{\dagger}_{i,\sigma}(k^+, \mathbf{k}) v_{\sigma}(k^+, \mathbf{k})\right],\\
\end{split}
\end{equation}
with the commuation relations being 
\begin{equation}
\begin{split}
&\left[\hat{a}_{\lambda_1, c_1} (p^+_1, \mathbf{p}_1), \hat{a}^{\dagger}_{\lambda_2, c_2}(p^+_2, \mathbf{p}_2)\right] = (2p^+_1) (2\pi)^3\delta(p^+_1 - p^+_2) \delta^{(2)}(\mathbf{p}_1-\mathbf{p}_2) \delta_{\lambda_1\lambda_2}\delta_{c_1c_2}\, ,\\
&\left\{\hat{b}_{\sigma_1, i_1} (k^+_1, \mathbf{k}_1), \hat{b}^{\dagger}_{i_2, \alpha_2}(k^+_2, \mathbf{k}_2)\right\} = (2k^+_1) (2\pi)^3\delta(k^+_1 - k^+_2) \delta^{(2)}(\mathbf{k}_1-\mathbf{k}_2) \delta_{\sigma_1\sigma_2}\delta_{i_1i_2}\,,\\
&\left\{\hat{d}_{\sigma_1, i_1} (k^+_1, \mathbf{k}_1), \hat{d}^{\dagger}_{i_2, \alpha_2}(k^+_2, \mathbf{k}_2)\right\} = (2k^+_1) (2\pi)^3\delta(k^+_1 - k^+_2) \delta^{(2)}(\mathbf{k}_1-\mathbf{k}_2) \delta_{\sigma_1\sigma_2}\delta_{i_1i_2}\,.\\
\end{split}
\end{equation}
We will use the shorthand notations
\begin{equation}
\int_{p^+}\equiv \int_0^{\infty} \frac{dp^+}{2p^+ (2\pi)}, \quad \int_{\mathbf{p}} \equiv\int \frac{d^2\mathbf{p}}{(2\pi)^2}, \quad \int_{\mathbf{x}} \equiv \int d^2\mathbf{x}.
\end{equation}
It should be noted that in the free field expansions given in eq.~\eqref{eq:mode_expansion}, $A^+ =0$ and  
$A^- = -\frac{\partial_i}{\partial_-}A^i$ are implementd by the requirements on the polarization vector
\begin{equation}
\varepsilon_{\lambda}^+(p^+, \mathbf{p}) =0, \qquad \varepsilon_{\lambda}^- (p^+, \mathbf{p}) = \frac{\mathbf{p}^i \varepsilon_{\lambda}^i(p^+, \mathbf{p})}{p^+}
\end{equation}
as the independent field components are $A^i$. Here $\varepsilon^i_{\lambda} = \frac{1}{\sqrt{2}}(1, i\lambda)$.  Similarly, for the fermion fields, not all the components of the spinors are independent 
\begin{equation}
\begin{split}
&u_{B, \sigma}(k)  = \frac{\gamma^+}{2k^+} (\mathbf{k}^j \gamma^j + m) u_{G,\sigma}(k),\\
&v_{B, \sigma}(k) = \frac{\gamma^+}{2k^+}(\mathbf{k}^j \gamma^j -m)  u_{G,\sigma}(k).\\
\end{split}
\end{equation}
For the explicit expressions of the spinor $u(k), v(k)$, we use the Kogut-Soper convention \cite{Kogut:1969xa, Brodsky:1997de} (see also \cite{Beuf:2016wdz}).
\begin{equation}
\begin{split}
&u(k_1, \frac{1}{2}) = \frac{1}{2^{1/4} \sqrt{k_1^+}} \begin{pmatrix} \sqrt{2}k_1^+\\
k_1^{x} + ik_1^y\\ m\\ 0\\ \end{pmatrix}\,, \qquad u(k_1, -\frac{1}{2}) = \frac{1}{2^{1/4} \sqrt{k_1^+}} \begin{pmatrix} 0\\m \\ -k_1^x+ik_1^y\\ \sqrt{2}k_1^+\\ \end{pmatrix},\\
&v(k_2,\frac{1}{2}) =\frac{1}{2^{1/4} \sqrt{k_2^+}} \begin{pmatrix} 0\\-m \\ -k_2^x+ik_2^y\\ \sqrt{2}k_2^+\\ \end{pmatrix}\, , \qquad v(k_2,-\frac{1}{2}) =\frac{1}{2^{1/4} \sqrt{k_2^+}}\begin{pmatrix} \sqrt{2}k_2^+\\
k_2^{x} + ik_2^y\\- m\\ 0\\ \end{pmatrix}\,.
\end{split}
\end{equation}
In addition, the gamma matrices are taken in the chiral representation
\begin{equation}
\gamma^0 = \begin{pmatrix} 0 & 1 \\ 1 & 0\\ \end{pmatrix}\,,\qquad \gamma^{i} =\begin{pmatrix} 0 & -\sigma^i\\ \sigma^i & 0 \\ \end{pmatrix} 
\end{equation}
Here $\sigma^i$ are Pauli matrices and $\gamma^5 = i \gamma^0\gamma^1\gamma^2\gamma^3$. 
For any matrix in spinor space, it can be decomposed as 
\begin{equation}
M = A I + B_{\mu} \gamma^{\mu} + C_{\mu\nu} \sigma^{\mu\nu} + D_{\mu} \gamma^{\mu}\gamma^5 + E \gamma^5
\end{equation}
with the coefficients being
\begin{equation}
A = \frac{1}{4}\mathrm{Tr}[M], \quad B^{\mu} = \frac{1}{4} \mathrm{Tr}[M\gamma^{\mu}], \quad C^{\mu\nu} = \frac{1}{8} \mathrm{Tr}[ M\sigma^{\mu\nu}] , \quad D^{\mu} = -\frac{1}{4} \mathrm{Tr}[M\gamma^{\mu}\gamma^5], \quad E = \frac{1}{4} \mathrm{Tr}[M\gamma^5]
\end{equation}
By explicit matrix algebra, one can verify the following identities that have been used in the main content of the paper. 
\begin{equation}
 \Big[\gamma_i u_{G, \sigma}(p^+)\bar{u}_{G, \sigma'}(p^+)\gamma_{i}\Big] = p^+ \delta_{\sigma\sigma'} [ \gamma^- + 2\sigma \gamma^-\gamma^5].
 \end{equation}
\begin{equation}
\left[\gamma_{i'}  \varepsilon_{\lambda'}^{i'\ast} u_{G, \sigma}(k^+)\bar{u}_{G, \sigma}(k^+)  \varepsilon_{\lambda}^i\gamma_i\right] = k^+\delta_{\lambda\lambda'} [ \gamma^- + \lambda \gamma^-\gamma^5].
\end{equation}

To characterize gluon radiation, either before or after scattering with the shockwave, one needs the light-cone wavefunction for gluon splitting.  This is textbook knowledge, we reproduce the result in this appendix for reference. The gluon splitting wave function from initial state has opposite sign compared to that from final state. We calculate the initial state gluon splitting.  
\begin{figure}[!t]
    \centering
    \includegraphics[width=0.4\textwidth]{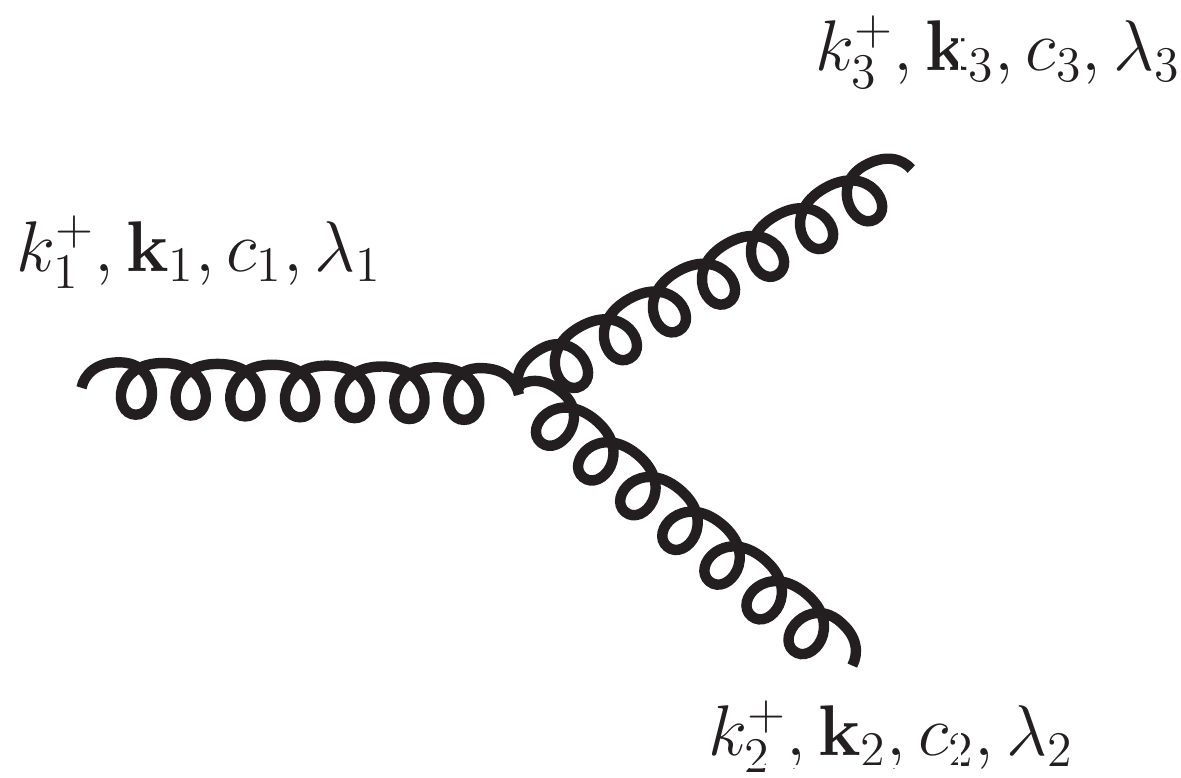}
    \caption{Gluon splitting vertex. }
	\label{fig:gluon_splitting}
\end{figure}
\begin{equation}
\begin{split}
&\Psi^{g\rightarrow gg}(k_1^+, \mathbf{k}_1, c_1, \lambda_1; k_2^+, \mathbf{k}_2, c_2, \lambda_2, k_3^+, \mathbf{k}_3, c_3, \lambda_3)\\
=&\frac{1}{k_1^- -k_2^- - k_3^-+i\epsilon} \langle 0 |\hat{a}_{c_3, \lambda_3}(k_3^+, \mathbf{k}_3) \hat{a}_{c_2, \lambda_2}(k_2^+, \mathbf{k}_2) V_{ggg}\hat{a}^{\dagger}_{c_1, \lambda_1}(k_1^+, \mathbf{k}_1) |0\rangle \\
=&-(2\pi)^3 \delta(k_1^+-k_2^+-k_3^+)\delta(\mathbf{k}_1-\mathbf{k}_2-\mathbf{k}_3)gf^{c_1c_2c_3} \left(\frac{\mathbf{k}_1^2}{2k_1^+} - \frac{\mathbf{k}_2^2}{2zk_1^+}  - \frac{(\mathbf{k}_1-\mathbf{k}_2)^2}{2(1-z)k_1^+}\right)^{-1}\\
&\times \left[ \delta_{\lambda_2, -\lambda_3} i\varepsilon^j_{\lambda_1}(2z\mathbf{k}_1 -2\mathbf{k}_2)^j + \delta_{\lambda_1\lambda_2} i\varepsilon^{j\ast}_{\lambda_3}\frac{(-2z\mathbf{k}_1+2\mathbf{k}_2)^j}{1-z} - \delta_{\lambda_1\lambda_3}i\varepsilon^{j\ast}_{\lambda_2} \frac{(-2\mathbf{k}_2 + 2z\mathbf{k}_1)^j}{z} \right]\\
=&-gf^{c_1c_2c_3} (2\pi)^3 2k_1^+\delta(k_1^+-k_2^+-k_3^+)\delta(\mathbf{k}_1-\mathbf{k}_2-\mathbf{k}_3)  \\
&\quad \times  \Big[z(1-z)\delta_{\lambda_2, -\lambda_3} i\varepsilon^j_{\lambda_1} -z\delta_{\lambda_1\lambda_2} i\varepsilon^{j\ast}_{\lambda_3}  -(1-z)\delta_{\lambda_1\lambda_3} i\varepsilon^{j\ast}_{\lambda_2} \Big]\frac{2(\mathbf{k}_2-z\mathbf{k}_1)^j}{(\mathbf{k}_2-z\mathbf{k}_1)^2}
\end{split}
\end{equation}
The three-gluon vertex $V_{ggg}$ from eq. \eqref{eq:LC_Hamiltonian} has been substituted in the second equality.  The delta function enforces $k_2^+ = zk_1^+$ and $k_3^+=(1-z)k_1^+$ and $\mathbf{k}_3 =\mathbf{k}_1-\mathbf{k}_2$. It is more useful to have an expression in transverse coordinate space
\begin{equation}
\begin{split}
&\Psi^{g\rightarrow gg}(k_1^+, \mathbf{y}_1, c_1, \lambda_1; k_2^+, \mathbf{y}_2, c_2, \lambda_2, k_3^+, \mathbf{y}_3, c_3, \lambda_3)\\
=&\int_{\mathbf{k}_1, \mathbf{k}_2, \mathbf{k}_3} e^{-i\mathbf{k}_1\cdot\mathbf{y}_1}e^{i\mathbf{k}_2\cdot\mathbf{y}_2}e^{i\mathbf{k}_3\cdot\mathbf{y}_3}\,\, \Psi^{g\rightarrow gg}(k_1^+, \mathbf{k}_1, c_1, \lambda_1; k_2^+, \mathbf{k}_2, c_2, \lambda_2, k_3^+, \mathbf{k}_3, c_3, \lambda_3)\\
=&-gf^{c_1c_2c_3} (2\pi) 2k_1^+\delta(k_1^+-k_2^+-k_3^+) \Big[z(1-z)\delta_{\lambda_2, -\lambda_3} i\varepsilon^j_{\lambda_1} -z\delta_{\lambda_1\lambda_2} i\varepsilon^{j\ast}_{\lambda_3}  -(1-z)\delta_{\lambda_1\lambda_3} i\varepsilon^{j\ast}_{\lambda_2} \Big] \\
&\quad \times \int_{\mathbf{k}_1,\mathbf{k}_2} e^{-i\mathbf{k}_1\cdot(\mathbf{y}_1-\mathbf{y}_3)}e^{i\mathbf{k}_2\cdot(\mathbf{y}_2-\mathbf{y}_3)} \frac{2(\mathbf{k}_2-z\mathbf{k}_1)^j}{(\mathbf{k}_2-z\mathbf{k}_1)^2}\\
=&-gf^{c_1c_2c_3} (2\pi) 2k_1^+\delta(k_1^+-k_2^+-k_3^+)  \Big[z(1-z)\delta_{\lambda_2, -\lambda_3} i\varepsilon^j_{\lambda_1} -z \delta_{\lambda_1\lambda_2} i\varepsilon^{j\ast}_{\lambda_3}  -(1-z)\delta_{\lambda_1\lambda_3} i\varepsilon^{j\ast}_{\lambda_2} \Big]\\
&\quad \times \int_{\mathbf{k}_3,\mathbf{k}_2} e^{i\mathbf{k}_3\cdot(\mathbf{y}_3-\mathbf{y}_1)}e^{i\mathbf{k}_2\cdot(\mathbf{y}_2-\mathbf{y}_1)}\frac{2[(1-z)\mathbf{k}_2-z\mathbf{k}_3]^j}{[(1-z)\mathbf{k}_2-z\mathbf{k}_3]^2}.
\end{split}
\end{equation}
In the last equality, we have the expression from integrating out the momentum $\mathbf{k}_1$ rather than the momentum $\mathbf{k}_3$. 

In the situation when the gluon radiated is soft $z\rightarrow 0$, keeping terms up to linear order in $z$, one obtains 
\begin{equation}\label{eq:gluon_splitting_coordinate}
\begin{split}
&\Psi^{g\rightarrow gg}(k_1^+, \mathbf{y}_1, c_1, \lambda_1; k_2^+, \mathbf{y}_2, c_2, \lambda_2, k_3^+, \mathbf{y}_3, c_3, \lambda_3)\Big|_{z\rightarrow 0}\\
=&-gf^{c_1c_2c_3} (2\pi) 2k_1^+\delta(k_1^+-k_2^+-k_3^+)  \delta(\mathbf{y}_3-\mathbf{y}_1) 2 \frac{i}{2\pi} \frac{(\mathbf{y}_2-\mathbf{y}_1)^j}{|\mathbf{y}_2-\mathbf{y}_1|^2} \\
&\qquad \times \Big[ -\delta_{\lambda_1\lambda_3} i\varepsilon^{j\ast}_{\lambda_2}+z\Big(\delta_{\lambda_2, -\lambda_3} i\varepsilon^j_{\lambda_1} - \delta_{\lambda_1\lambda_2} i\varepsilon^{j\ast}_{\lambda_3}  \Big)\Big].
\end{split}
\end{equation}
In the sub-eikonal order, we only kept terms that transfer polarization information to the softer gluons.

\section{Sub-eikonal Transformations Related to $a^-$ Field}\label{app:sub-eikonal_transform_a-}
From the interaction term at the eikonal order 
\begin{equation}
\begin{split}
V_0 =& \int dx^-d^2\mathbf{x} a_b^-(x^+, 0, \mathbf{x}) J_b^+(0, x^-, \mathbf{x}) \\
= &\int dx^-d^2\mathbf{x}a_b^-(x^+, 0, \mathbf{x}) \left( g\bar{\Psi} \gamma^+ t^b \Psi - ig[A^i, F^{+i}]^b\right)\\
\end{split}
\end{equation}
and the definition of Wilson line operator 
\begin{equation}\label{eq:def_wl_operator}
\hat{W}(x_f^+, x_i^+) = \mathcal{P} \mathrm{Exp}\left\{-i\int^{x_f^+}_{x_i^+} dz^+ V_{(0), \rm{I}}(z^+)\right\}
\end{equation}
If one ignores the transformation to interaction picture in $V_{(0), \rm{I}}(z^+) = e^{iH_0z^+}V_{(0)}e^{-iH_0z^+}$, one obtains the well-known transformations for creation operators at the eikonal order. 
\begin{equation}\label{eq:creation_operators_eikonal}
\begin{split}
&\hat{W}(x_f^+, x_i^+) \hat{a}^{\dagger}_{h, \lambda}(p^+, \mathbf{y}) \hat{W}^{\dagger}(x_f^+, x_i^+) = \hat{a}^{\dagger}_{c, \lambda}(p^+, \mathbf{y}) U^{ch}_{\mathbf{y}}(x_f^+, x_i^+),\\
&\hat{W}(x^+_f, x_i^+) \hat{b}^{\dagger}_{i, \rho}(p^+, \mathbf{y}) \hat{W}^{\dagger}(x_f^+, x_i^+) = \hat{b}^{\dagger}_{j, \rho}(p^+, \mathbf{y}) V^{ji}_{\mathbf{y}}(x_f^+, x_i^+),\\
&\hat{W}(x_f^+, x_i^+) \hat{d}^{\dagger}_{i, \rho}(p^+, \mathbf{y}) \hat{W}^{\dagger}(x_f^+, x_i^+) =  V^{\dagger, ij}_{\mathbf{y}}(x_f^+, x_i^+)\hat{d}^{\dagger}_{j, \rho}(p^+, \mathbf{y}).\\
\end{split}
\end{equation}
Here the Wilson lines in the adjoint representation and the fundamental representation are
\begin{equation}
U_{\mathbf{y}} (x_f^+, x_i^+) = \mathcal{P} \mathrm{exp} \left\{ -ig\int_{x_i^+}^{x_f^+} dz^+ a^-_b (z^+, \mathbf{y}) T^b \right\}.
\end{equation}
\begin{equation}
V_{\mathbf{y}} (x_f^+, x_i^+) = \mathcal{P} \mathrm{exp} \left\{ -ig\int_{x_i^+}^{x_f^+} dz^+ a^-_b (z^+, \mathbf{y})  t^b \right\}.
\end{equation}
Our goal is to obtain sub-eikonal corrections to the transformations in eq.~\eqref{eq:creation_operators_eikonal}. 

Recall that the free Hamiltonian 
\begin{equation}
H_0 = \frac{1}{2} \int dx^- d^2\mathbf{x} \left[\bar{\Psi} \frac{m^2 + \partial_l \partial^l}{i\partial_-} \gamma^+ \Psi - A^i_a \partial_l \partial^l A_i^a\right]
\end{equation}
can be expressed in terms of creation and annihilation operators as
\begin{equation}
H_0 =\int_{p^+, \mathbf{p}}  E_{p}\Big[ \hat{b}_{p, \sigma}^{\dagger} \hat{b}_{p, \sigma} - \hat{d}_{p,\sigma}\hat{d}^{\dagger}_{p,\sigma}+ \hat{a}_{p,\lambda}^{\dagger}\hat{a}_{p,\lambda}\Big] 
\end{equation}
The light-cone energy is $E_p =p^-= \frac{ \mathbf{p}^2}{2p^+}$ in which we have ignored the mass of quarks and gluons. From this explicit expression, the transformation to the \textit{interaction picture} is calculated to be
\begin{equation}
e^{iH_0z^+} \hat{a}_{\lambda}^{\dagger}(p^+, \mathbf{x}) e^{-iH_0 z^+} = e^{i\frac{-\partial_{\mathbf{x}}^2}{2p^+} z^+} \hat{a}_{\lambda}^{\dagger}(p^+, \mathbf{x})
\end{equation}
similar transformations hold for $\hat{b}^{\dagger}_{p, \sigma}$ and $\hat{d}^{\dagger}_{p, \sigma}$. 
The color current has the explicit expression
\begin{equation}
\begin{split}
&\hat{J}^+_b(\mathbf{x}) = \int dx^- J^+_b(0, x^-, \mathbf{x})\\
=& g\int_{k^+} \Big[\hat{b}^{\dagger}_{i,\sigma}(k^+, \mathbf{x}) t^b_{ij} \hat{b}_{j, \sigma} (k^+, \mathbf{x}) + \hat{d}_{i, \sigma} (k^+, \mathbf{x}) t^b_{ij} \hat{d}^{\dagger}_{j, \sigma} (k^+, \mathbf{x}) + \hat{a}^{\dagger}_{c, \lambda}(k^+, \mathbf{x}) T^b_{ce} \hat{a}_{e, \lambda}(k^+, \mathbf{x})\Big]
\end{split}
\end{equation}
Using this expression, one can obtain the commutation relations
\begin{equation}
\begin{split}
&[ J^+_b(\mathbf{x}), \hat{a}^{\dagger}_{h, \lambda}(p^+, \mathbf{y})] = g\hat{a}^{\dagger}_{c, \lambda}(p^+, \mathbf{x}) T^b_{ch} \delta(\mathbf{x}-\mathbf{y}), \\
&[J^+_b(\mathbf{x}), \hat{b}^{\dagger}_{j, \rho}(p^+, \mathbf{y})] = g \hat{b}_{i, \rho}^{\dagger}(p^+, \mathbf{x}) t^b_{ij} \delta(\mathbf{x}-\mathbf{y}), \\
&[J^+_b(\mathbf{x}), \hat{d}_{i, \rho}^{\dagger}(p^+, \mathbf{y})] = -g t^b_{ij} \hat{d}^{\dagger}_{j, \rho}(p^+, \mathbf{x}) \delta(\mathbf{x}-\mathbf{y}).
\end{split}
\end{equation}
The eikonal Wilson line operator eq.~\eqref{eq:def_wl_operator}
contains sub-eikonal contribution due to the transformation to interaction picture $V_{0, I}(z^+) = e^{iH_0z^+}V_{(0)}e^{-iH_0z^+}$.
We need to compute its ation on creation operators up to sub-eikonal order
\begin{equation}\label{eq:adagger_eikonal_transform}
\begin{split}
& \hat{W}(z_N^+, z_0^+) \hat{a}^{\dagger}_{h, \lambda}(p^+, \mathbf{y}) \hat{W}^{\dagger}(z_N^+, z_0^+)\\
=&\lim\limits_{\substack{\Delta x^+\rightarrow 0,\\N\Delta x^+=z_N^+-z_0^+}}\prod_{j=0}^{N-1}\hat{W}(z_{j+1}^+, z_j^+) \hat{a}^{\dagger}_{h, \lambda}(p^+, \mathbf{y}) \hat{W}^{\dagger}(z_{j+1}^+, z_j^+).
\end{split}
\end{equation}
We use gluon creation operator as an example to demonstrate the derivations. 
In eq.~\eqref{eq:adagger_eikonal_transform}, it is computed by dividing the time interval $[x_f^+, x_i^+]$ into $N$ pieces $\Delta x^+ = \frac{x_f^+ -x_i^+}{N}$ and in the end taking $N\rightarrow \infty$ and $\Delta x^+\rightarrow 0$ limit with $N\Delta x^+ = x_f^+-x_i^+$ fixed.  We use the denotations $z_0^+=x_i^+$, $z_i^+ =z_0^+ + i \Delta x^+$, $z_N^+ = x_f^+$. 

For the general expression, it is unclear how to get a closed form expression by taking the limits directly. However, one can obtain closed form expression up to sub-eikonal order. The final result for the transformation of gluon creation operator by eikonal Wilson line operator up to eikonal order is
\begin{equation}
\begin{split}
&\hat{W}(z_N^+, z_0^+) \hat{a}^{\dagger}_{h, \lambda}(p^+, \mathbf{y}) \hat{W}^{\dagger}(z_N^+, z_0^+)\\
 =& \hat{a}^{\dagger}_{c, \lambda}(p^+, \mathbf{y}) U_{\mathbf{y}}^{ch}(z_N^+, z_0^+) +  \frac{i}{2p^+}\int_{z_0^+}^{z_N^+} dz^+\hat{a}_{c,\lambda}^{\dagger}(p^+,\mathbf{ y})\partial_{\mathbf{y}}^2U^{ch'}_{\mathbf{y}}(z_N^+, z^+) U^{h'h}_{\mathbf{y}}(z^+, z_0^+)\\
&\qquad \qquad +2\partial^i_{\mathbf{y}}\hat{a}_{c,\lambda}^{\dagger}(p^+,\mathbf{ y})\partial^i_{\mathbf{y}}U^{ch'}_{\mathbf{y}}(z_N^+, z^+)U_{\mathbf{y}}^{h'h} (z^+, z_0^+)\\
\end{split}
\end{equation}
The first term recovers the well-known eikonal transformation and the second terms represents the sub-eikonal correction. 
Repeating the above analysis for the gluon annihilation operator, one gets
\begin{equation}
\begin{split}
&\hat{W}^{\dagger}(z^+_N, z_{0}^+) \hat{a}_{h, \lambda}(p^+, \mathbf{y}) \hat{W}(z_N^+, z_{0}^+)\\
=&U^{hc}_{\mathbf{y}}(z_N^+, z_0^+) \hat{a}_{c, \lambda}(p^+, \mathbf{y})+\frac{i}{2p^+} \int_{z_0^+}^{z^+_N} dz^+ U^{hd}_{\mathbf{y}}(z_N^+, z^+)\partial^2_{\mathbf{y}}U^{dc}_{\mathbf{y}}(z^+, z_0^+)\hat{a}_{c, \lambda}(p^+, \mathbf{y}) \\
&\qquad \qquad + 2U_{\mathbf{y}}^{hd}(z_N^+, z^+)\partial^i_{\mathbf{y}}U^{dc}_{\mathbf{y}}(z^+, z_0^+) \partial^i_{\mathbf{y}}\hat{a}_{c, \lambda}(p^+, \mathbf{y}).\\
\end{split}
\end{equation}
Using the above transformations, one can calculate the single gluon scattering amplitude,
\begin{equation}\label{eq:single_amplitude_sub-eikonal_a-}
\begin{split}
&\langle 0 | \hat{a}_{c', \lambda'}(p^{\prime +},\mathbf{x}') \hat{W}(x_f^+,x_i^+) \hat{a}^{\dagger}_{c, \lambda}(p^+, \mathbf{x})|0\rangle\\
=&\frac{1}{2} \Big[\langle 0 | W^{\dagger}(x_f^+, x_i^+)\hat{a}_{c', \lambda'}(p^{\prime +},\mathbf{x}') \hat{W}(x_f^+,x_i^+) \hat{a}^{\dagger}_{c, \lambda}(p^+, \mathbf{x})|0\rangle \\
&\qquad + \langle 0 | \hat{a}_{c', \lambda'}(p^{\prime +},\mathbf{x}') \hat{W}(x_f^+,x_i^+) \hat{a}^{\dagger}_{c, \lambda}(p^+, \mathbf{x})W^{\dagger}(x_f^+, x_i^+)|0\rangle\Big]\\
=&(2\pi)2p^+\delta(p^+-p^{\prime +})\delta_{\lambda\lambda'} \Big\{\delta(\mathbf{x}-\mathbf{x}') U^{c'c}_{\mathbf{x}}(x_f^+, x_i^+)\\
&+\frac{1}{2} \frac{i}{2p^+}\int_{x_i^+}^{x_f^+} dz^+ \Big[U^{c'd}_{\mathbf{x}'}(x_f^+, z^+)\partial^2_{\mathbf{x}'}U^{dc}_{\mathbf{x}'}(z^+, z_i^+) \delta(\mathbf{x}-\mathbf{x}') + 2U^{c'd}_{\mathbf{x}'}(x_f^+, z^+)\partial^i_{\mathbf{x}'}U^{dc}_{\mathbf{x}'}(z^+, x_i^+)\partial^i_{\mathbf{x}'}\delta(\mathbf{x}-\mathbf{x}')\Big]\\
&+\frac{1}{2} \frac{i}{2p^+} \int_{x_i^+}^{x_f^+}dz^+\Big[ \delta(\mathbf{x}'-\mathbf{x})\partial^2_{\mathbf{x}}U^{cd}_{\mathbf{x}}(x_f^+, z^+) U^{dc}_{\mathbf{x}}(z^+, x_i^+) + 2\partial^i_{\mathbf{x}}\delta(\mathbf{x}'-\mathbf{x})\partial^i_{\mathbf{x}}U^{c'd}_{\mathbf{x}}(x_f^+, z^+)U^{dc}_{\mathbf{x}}(z^+, x_i^+)\Big]\Big\}\\
=&(2\pi)2p^+\delta(p^+-p^{\prime +})\delta_{\lambda\lambda'} \Big\{\delta(\mathbf{x}-\mathbf{x}') U^{c'c}_{\mathbf{x}}(x_f^+, x_i^+)\\
&-\frac{i}{2p^+} \int_{x_i^+}^{x_f^+} dz^+ U^{c'd}_{\mathbf{x}'}(x_f^+, z^+) \int_{\mathbf{z}}\left[\partial^i_{\mathbf{z}}\delta(\mathbf{x}'-\mathbf{z})\partial^i_{\mathbf{z}}\delta(\mathbf{x}-\mathbf{z})\right]U^{dc}_{\mathbf{x}}(z^+, x_i^+) \\
&- \frac{i}{2p^+}(x_f^+-x_i^+)\frac{1}{2} \left[\partial^2_{\mathbf{x}'}\delta(\mathbf{x}-\mathbf{x}')U^{c'c}_{\mathbf{x}'}(x_f^+, x_i^+) +\partial^2_{\mathbf{x}}\delta(\mathbf{x}'-\mathbf{x}) U^{c'c}_{\mathbf{x}}(x_f^+, x_i^+) \right]\Big\}.\\
\end{split}
\end{equation}
In obtaining the last equality, we have repeatedly utilized integration by parts to move the partial derivatives to act on the Dirac delta functions instead of the Wilson lines.

For the two terms at sub-eikonal order, the first term recovers the corresponding non local terms in the single gluon scattering amplitude in eq.~\eqref{eq:U_pol_2} when setting the background field $a^i=0, \psi_B =0$. The second term is a boundary term (boundary in the longitudinal direction, not the transverse direction) which is proportional to the width of the shockwave $x_f^+ -x_i^+$. It is precisely the phase factor from the free propagator given in Eq.~\eqref{eq:free_sub_eikonal}.  If one turns off the background fields $a^-=0$, These two terms vanish as expected
\begin{equation}
- \frac{i}{2p^+}(x_F^+-x_I^+)\frac{1}{2}  \delta^{c'c} \int_{\mathbf{z}}\partial^2_{\mathbf{z}}\left[\delta(\mathbf{x}'-\mathbf{z}) \delta(\mathbf{x}-\mathbf{z})\right] = 0.
\end{equation}\\

We would like to calculate the contributions of sub-eikonal Taylor expansion of $a^-_bJ^+_b$ to the single particle scatering amplitude. 
The two sub-eikonal terms arer 
\begin{equation}\label{eq:two_terms_taylor_eikonal}
\int d^2\mathbf{z} dz^+ dz^- \Big[ z^- \partial_-a^-_b(z^+, 0, \mathbf{z}) J^+_b(0, z^-, \mathbf{z}) + z^+ a^-_b(z^+, 0, \mathbf{z})\partial_+ J^+_b(0, z^-, \mathbf{z})\Big]
\end{equation}
For the second term. 
The time dependence $z^+ \partial_+ J^+(0, z^-, \mathbf{z})$ is introduced by the lowest order expansion of  $e^{iH_0 z^+} J^+_b(0, z^-, \mathbf{z}) e^{-iH_0z^+}$. In general the time dependence is generated by the full Hamiltonian $e^{iH z^+} J^+_b(0, z^-, \mathbf{z}) e^{-iHz^+}$. In the case when $H=H_0$, it is part of the sub-eikonal transformation that has already been included in obtaining eq.~\eqref{eq:single_amplitude_sub-eikonal_a-}. One should not double count its contribution.  

For the first term, we utilized the gluonic part to demonstrate the derivations. The gluons' contribution to the time dependent color current is 
\begin{equation}
\begin{split}
J_0^+(z^+, z^-, \mathbf{z}) =&gf^{bcd} \int_{p^+, q^+} e^{i(p^--q^-)z^+}e^{i(p^+-q^+)z^-} \hat{a}^{\dagger}_{c, \lambda}(p^+, \mathbf{z})\hat{a}_{d, \lambda}(q^+, \mathbf{z}) (-iq^+)\\
&+e^{-i(p^--q^-)z^+}e^{-i(p^+-q^+)z^-} \hat{a}^{\dagger}_{d, \lambda}(q^+, \mathbf{z})\hat{a}_{c, \lambda}(p^+, \mathbf{z}) (iq^+). 
\end{split}
\end{equation}
Terms containing $\hat{a}\hat{a}$ and $\hat{a}^{\dagger}\hat{a}^{\dagger}$ will not contribute to single gluon scattering amplitude and we ignore them. 

The first term in eq.~\eqref{eq:two_terms_taylor_eikonal} contains
\begin{equation}
\int dz^- z^- J^+_b(0, z^-, \mathbf{z})
=-gf^{bcd} \frac{1}{2}\int_{p^+} \Big[\hat{a}^{\dagger}_{c, \lambda}(p^+, \mathbf{z})(\overrightarrow{\partial_{p^+}}- \overleftarrow{\partial_{p^+}})\hat{a}_{d, \lambda}(p^+, \mathbf{z}) \Big].
\end{equation}
Using this explicit expression, we now compute its  contribution to the single gluon scattering amplitude at sub-eikonal order. 
\begin{equation}
\begin{split}
&-i\int_{x_i^+}^{x_f^+} dz^+\langle 0 | \hat{a}_{c', \lambda'}(p^{\prime +},\mathbf{x}') \hat{W}(x_F^+,z^+)
 V_{(1), L}(z^+)\hat{W}(z^+,x_I^+) \hat{a}^{\dagger}_{c, \lambda}(p^+, \mathbf{x})|0\rangle\\
= &-gf^{bed}\frac{1}{2} \int d^2\mathbf{z}\int dz^+ f_b^{+-}(z^+, 0, \mathbf{z})\langle 0 | \hat{a}_{c', \lambda'}(p^{\prime +},\mathbf{x}') \hat{W}(x_f^+,z^+)\\
 &\qquad \times  \int_{q^+} \Big[\hat{a}^{\dagger}_{e, \kappa}(q^+, \mathbf{z})(\overrightarrow{\partial_{q^+}}- \overleftarrow{\partial_{q^+}})\hat{a}_{d, \kappa}(q^+, \mathbf{z}) \Big]\hat{W}(z^+,x_i^+) \hat{a}^{\dagger}_{c, \lambda}(p^+, \mathbf{x})|0\rangle\\
 =&ig \delta_{\lambda\lambda'} \delta(\mathbf{x}-\mathbf{x}')\Big[(2\pi) (p^+ + p^{\prime +}) \partial_{p^+}\delta(p^{\prime +}-p^+)\Big]\int dz^+ \left[U_{\mathbf{x}}(x_f^+, z^+)f^{+-}(z^+, 0, \mathbf{x}) U_{\mathbf{x}}(z^+, x_i^+)\right]^{c'c}.\\
\end{split}
\end{equation}
It is apparent that this sub-eikonal interaction involves longitudinal momentum exchange between the projectile and the shockwave.  Transverse coordinates are preserved as well as the polarization.  The polarized Wilson line has the insertion of longitudinal chromoelectric field $f^{+-} = \partial_- a^-$. When calculating spin related observables at small $x$, they are represented as the interference terms between the eikonal order amplitude and sub-eikonal order amplitude. The eikonal order amplitudes preserve longitudional momentum conservation and  therefore they will not interfere with sub-eikonal interactions that involve longitudional momentum exchange with the shockwave.


\bibliography{double_spin_asymmetry}
\end{document}